\documentclass[11pt]{article}
\usepackage[utf8]{inputenc}

\usepackage{parskip}
\usepackage[a4paper, total={6in, 9.5in}]{geometry}
\usepackage{graphicx}
\usepackage{amsmath, amsfonts}
\usepackage[bottom]{footmisc}
\usepackage{float}

\newfloat{experiment}{htbp}{loe}
\floatname{experiment}{Experiment}

\newcounter{definition}
\renewcommand{\thedefinition}{\arabic{definition}}
\newenvironment{definition}{%
  \refstepcounter{definition}
  \par\medskip
  \noindent
  \textbf{Definition \thedefinition.}\quad
}{
  \par\medskip
}

\newcounter{theorem}
\renewcommand{\thetheorem}{\arabic{theorem}}
\newenvironment{theorem}{%
  \refstepcounter{definition}
  \par\medskip
  \noindent
  \textbf{Theorem \thetheorem.}\quad
}{
  \par\medskip
}

\usepackage[style=numeric-comp, sorting=none]{biblatex}
\title{On Post-Quantum Cryptography Authentication for Quantum Key Distribution\vspace{6mm}}
\author{Juan Antonio Vieira Giestinhas$^{1}$\footnote{Correspondence: qph504@york.ac.uk} , Timothy Spiller$^1$
\\[3mm]{\centering $^1$ University of York, YO10 5DD, York, United Kingdom}}
\date{}

\begin{document}

\maketitle

\begin{abstract}
The traditional way for a Quantum Key Distribution (QKD) user to join a quantum network is by authenticating themselves using pre-shared key material. While this approach is sufficient for small-scale networks, it becomes impractical as the network grows, due to the total quadratic increase in the number of pre-shared keys required. To address this scalability issue, Public Key Infrastructure (PKI) combined with Post-Quantum Cryptography (PQC) offers a more scalable solution, allowing users to authenticate the QKD traffic remotely to obtain information-theoretical secure (ITS) keys under the presented assumptions. Unlike traditional PKI, which relies on classical cryptographic algorithms such as RSA, the approach presented in this paper leverages PQC algorithms that are believed to be resistant to quantum attacks. Similarly to the SIGMA or TLS protocols, authentication, confidentiality, and integrity are achievable against bounded adversaries to ensure secure and scalable quantum networks.
\end{abstract}

\section{Introduction}\label{intro}

The development of quantum computers is becoming an increasing threat to the most commonly used asymmetric algorithms, such as Rivest-Shamir-Adleman (RSA) or Elliptic-Curve Cryptography (ECC)~\cite{Del2018HSP}. With the advent of Shor's algorithm~\cite{Shor}, the feasibility of retrospective decryption (harvest now, decrypt later) could become possible sooner than expected, posing risks to sensitive data.

On one hand, Quantum Cryptography, notably Quantum Key Distribution (QKD)~\cite{GisinQC2002, QuantumCryptographyPirandola_2020, SecureQKDrealisticDevices2020}, can be used to distribute secret keys in an Information-Theoretical Secure (ITS) manner, with the condition that an authenticated channel is available. The classical channel for QKD is typically authenticated with pre-shared keys and ITS Message Authentication Codes (MAC)~\cite{ITS-MAC-WEGMAN1981265, ITS-MAC-secret-key-rate}. However, this approach is not convenient as it forces honest parties to physically meet each other.

On the other hand, Post-Quantum Cryptography (PQC) is another alternative to asymmetric cryptography that is believed to be secure against quantum and classical algorithms. By combining both technologies, ITS keys given by QKD are obtainable through a remote classical channel authenticated by PQC. Once the first QKD iteration authenticated by PQC is effected, the shared keys thus obtained may be used to perform ITS authentication in further QKD iterations.

This paper focuses on non-ITS authentication for QKD, specifically relying on PQC and Public Key Infrastructure (PKI), notably with certificates given by a Certificate Authority (CA), where parties do not have pre-shared key material to authenticate the classical channel, which is required to distill an authenticated shared secret key from a QKD protocol. A general case of authentication for QKD is presented. Two examples of PQC protocols are presented, a signature-based authentication protocol, and an alternative scheme using Key Encapsulation Mechanisms (KEM) for authentication without signatures. Furthermore, a shared-based key authentication example utilizing non-ITS Message Authentication codes (MAC) is also discussed. A bounded adversary is defined and considered, alongside a conditioned unbounded adversary, to check the ITS security of the final QKD keys.

The structure and strategy of this paper are as follows. Section \ref{motivation} summarizes current work to date, along with further motivation on the need for and approaches to authentication for QKD. The strategy adopted in this paper is to present formal technical descriptions of various non-ITS authentication protocols (two that leverage PQC in section \ref{protocols_sign_KEM}, and use of pre-shared key in section \ref{protocol_MAC}). This strategy must be built upon formal definitions of numerous entities, procedures, properties and quantities, which are provided in section \ref{definitions}. As the protocols leverage PQC, the required PQC is discussed in section \ref{assumptions}. Sections \ref{protocols_sign_KEM} and \ref{protocol_MAC} then present the non-ITS authentication protocols, as formal, time-ordered lists of operations, each accompanied by commentary and discussion. This formal approach provides numerous important consequences. 

Examples, that don't heavily influence this work, are that it is straightforward to evaluate the time taken (in information round trip units) and computational effort required for one round of each protocol. Furthermore, it is straightforward to convert the protocol to a flow diagram, identifying the points where an honest party would abort, having determined they are interacting with a dishonest impostor. Such a flow diagram also enables careful development of pseudo-code and thus actual computer code, to facilitate a real implementation of the protocol. Such code-development should be undertaken formally and carefully, to avoid introducing information-leakage vulnerabilities into the actual implementation of what is a fundamentally secure protocol.

For this work, however, the really important consequence from a formal presentation of the authentication protocols is security analysis. Such security analysis can be undertaken through investigation of a game, or experiment, between a challenger and two precisely specified adversaries. The outcome of the game informs on the security of the protocol. This scenario requires formal definition and establishment of a key exchange framework, which is presented in section \ref{framework}. Utilizing this framework, formal security analysis of the protocols is then presented in section \ref{security}.

Discussion of the security proof results along with a further specific example is given in section \ref{discussion}, followed by conclusions delivered in section \ref{conclusion}.

\section{Motivation, related literature and contribution}\label{motivation}

PQC signature-based, KEM-based, or even combined protocols have already been implemented to specifically authenticate a classical channel to effectively perform QKD. Benchmark parameters with standardized or variants of the National Institute of Standards and Technology (NIST) PQC algorithms~\cite{nist_pqc}, along with implemented or simulated QKD links, have been estimated in~\cite{PQCforQKD1, MoscaImplemented2024, QKDSignIplemenWang2021}. The basic idea underpinning these papers, and for this work, is to perform unauthenticated QKD and authenticate the post-processing information at some point after distilling QKD key material using PQC algorithms. An authenticated key exchange protocol with a security proof based on the ``short-term" security of PQC and ``long-term" security of QKD is also found in~\cite{MoscaAKE2013} and is strongly related to the implementation~\cite{MoscaImplemented2024}.

Care must be taken when authenticating keys with classical algorithms because impersonation or misbinding attacks may occur, as stated in the SIGMA paper~\cite{SIGMA2003}. This is why the PQC-based authentication protocols presented in this paper follow the SIGMA structure: computation of an unauthenticated ephemeral key, authentication (signature in SIGMA), and then binding of secrets with honest entities (MAC in SIGMA) using cryptographic primitives. This approach not only helps to avoid person-in-the-middle and misbinding attacks, but also provides the entity protection feature that SIGMA offers. By following the SIGMA and TLS structures~\cite{rfc8446}, non-ITS but secure authentication, confidentiality and integrity for QKD is achieved. If QKD is achieved then ITS keys are obtainable. 

A similar approach to combine PQC and QKD, proposed in the hybrid authenticated key exchange (HAKE) protocol Muckle series~\cite{Muckle, MucklePlusPlus, MucklePlus2023, MuckleKEM2024}, is to mix the different key material to obtain a final authenticated hybrid key, where its security holds as long as one of the cryptographic primitives, classical, PQC or QKD, remains secure. The goal of this work is to discuss the approach where PQC is used to authenticate QKD, whilst also being able to obtain traffic secrets, whereas in the Muckle series, the goal is to achieve redundancy by applying different cryptographic primitives that present different strengths. However, by including the public QKD post-processing information within some Muckle steps, QKD may also be authenticated by following a protocol similar to the Muckle series. In fact, the protocols to authenticate QKD within this work are strongly inspired from the Muckle series and the KEMTLS protocol~\cite{KEMTLS2020}, in addition to the SIGMA and TLS protocols.

This work addresses the separate authentication requirement that QKD presents, particularly in remote scenarios. While pre-sharing keys to authenticate is not always convenient, PQC and PKI offers a scalable solution for establishing secure QKD networks remotely. The idea is to apply the PQC approach only in the first QKD iteration, as subsequent iterations can be authenticated with QKD key material obtained in prior iterations. However, the presented PQC protocols can be reiterated for different QKD protocols, as derived QKD keys and secret state parameters can be transferred from one protocol to another securely while taking into account the failure probability (equivalently the security) of said parameters. Taking inspiration from the SIGMA protocol, the Muckle series and the KEMTLS protocol, the schemes presented in this paper enable QKD protocols to securely distill a secret key through a classical channel, authenticated due to the ``short-term" security for PQC algorithms assumption, defined in section \ref{assumptions}. Additionally, confidentiality and integrity over the classical traffic, including information of the entities, is provided. The general case of mono-authentication for QKD is covered, with specific examples similar to those in related literature. The traffic protection feature is also discussed: this feature holds against bounded adversaries, even if it is assumed that the PQC secret keys are revealed to the public after the first QKD instantiation is finalized. The security of the presented protocols is evaluated using the HAKE framework introduced and reused in the Muckle series~\cite{Muckle, MucklePlus2023, MucklePlusPlus, MuckleKEM2024}.

\section{Definitions/Preliminaries}\label{definitions}

This section provides the required definitions that are used in the security proofs of this work. The bounded and conditioned unbounded adversaries considered, Quantum Key Distribution, dual pseudorandom function, a hash function, signature, key encapsulation mechanism, and message authentication code are all defined. Some of these definitions are accompanied by game-based security models, named experiments, which highlight the security feature of interest for these primitives.

\begin{definition}\label{def1.1}\textbf{(Hybrid Polynomial-Time Adversary).} The bounded adversary considered within this work is both Probabilistic Polynomial-Time (PPT) and Quantum Polynomial-Time (QPT), making it Hybrid Polynomial-Time (HPT).

A PPT and QPT adversary represents an adversary with both classical and quantum bounded computational resources. Given an input of size n, a constant $C_{\text{PPT}}$ for the classical computations, and a constant $C_{\text{QPT}}$ for the quantum computations, the time complexity of this adversary is O(n$^{C_{\text{PPT}}}$) for classical computations and O(n$^{C_{\text{QPT}}}$) for quantum computations.

Since the considered adversary is both PPT and QPT, the adversary is able to run in parallel both QPT and PPT algorithms to solve a specific problem, if possible. If a specific problem can be solved in polynomial time by relying only on PPT or QPT algorithms independently, the time complexity of the adversary is the minimum between O(n$^{C_{\text{PPT}}}$) and O(n$^{C_{\text{QPT}}}$).

Furthermore, the adversary is able to combine PPT algorithms with QPT algorithms to perform HPT algorithms. In this case, given an input of size n, a constant $C_{T_\text{PPT}}$ for the pertinent classical computations, a constant $C_{T_\text{QPT}}$ for the pertinent quantum computations, and a constant $C_\text{parallel}$ for the classical and quantum algorithms that can run in a parallel way, the total time complexity of the adversary is the summation of the classical algorithm time complexity, O(n$^{C_{T_\text{PPT}}}$), with the quantum algorithm time complexity, O(n$^{C_{T_\text{QPT}}}$), minus the time complexity where classical and quantum algorithms can run at the same time, O(n$^{C_\text{parallel}}$). The idea behind the HPT algorithm is to solve a problem in polynomial time more efficiently, in terms of time complexity, than if an adversary is limited to only classical or quantum algorithms with their corresponding, and independent, limited resources. That can be expressed as:

\begin{equation*}
    \text{O}(\text{n}^{C_{T_\text{PPT}}}) + \text{O}(\text{n}^{C_{T_\text{QPT}}}) - \text{O}(\text{n}^{C_\text{parallel}}) < \text{min}(\text{O}(\text{n}^{C_{\text{PPT}}}), \text{O}(\text{n}^{C_{\text{QPT}}}))
\end{equation*}

Note that the HPT adversary can solve problems in polynomial time that have greater than polynomial time complexity on the classical side but not in the quantum side, by running QPT algorithms (and vice versa with PPT algorithms). A practical example could be the factorization of a composite integer to break RSA using Shor's algorithm~\cite{Shor}.

The HPT adversary is allowed to interact with the protocols presented in this paper at any time. Once the protocols are not in an active state (either because they successfully completed, or they aborted), the HPT adversary is allowed to ask an unbounded adversary to obtain the QKD keys with the condition that they were not ITS. This unbounded adversary is also limited (termed a conditioned unbounded adversary from now on), since the goal of having such a threat is to prove that the QKD keys are ITS once authentication is finalized. In other words, the conditioned unbounded adversary is hypothetical and is used as a tool in the security analysis to prove the ITS property of the final QKD keys.

\end{definition}

\begin{definition}\label{def1.2}\textbf{(Conditioned Unbounded Adversary).} An unbounded adversary is useful when defining ITS security. Unbounded adversaries have infinite resources, infinite computational power, and have access to ideal quantum and classical technologies, including ideal quantum memories.

Since the goal of this work is to prove that the protocols presented are secure against the bounded HPT adversary and to prove that the final authenticated keys are ITS, the considered conditioned unbounded adversary is allowed to have total control of the quantum channel used to perform unauthenticated QKD (Definition \ref{def2.3}), is able to read the public traffic used to distill the unauthenticated QKD keys but is not allowed to interact with anything else, i.e. with the authentication steps. Note that this conditioned unbounded adversary could be allowed to actively modify the classical traffic used to distill the QKD keys from the unauthenticated QKD step; however, that would be detected by the honest parties during the authentication step, making that degree of freedom ineffective.

The HPT adversary can interact with the conditioned unbounded adversary for two purposes. The first one is to obtain ephemeral QKD keys in the case the HPT adversary is not the one performing QKD as an honest party. This can be requested at any time, starting from the moment that the unauthenticated QKD has ended. The second purpose is to check the ITS security of QKD keys that have been successfully distilled from finalized protocols or iterations. Once the protocols finish running, the unbounded adversary is allowed to interact with the HPT adversary in a limited way. If the conditioned unbounded adversary has access to the distilled QKD keys, compromised during the unauthenticated QKD steps, then these are revealed to the HPT adversary, making the QKD keys not ITS. Otherwise, if the unbounded adversary is unable to compromise the ITS keys with the allowed interactions, then the HPT adversary receives nothing and thus learns that the QKD keys are ITS.

Furthermore, the HPT adversary is allowed to send the gathered traffic from the authentication steps to the unbounded adversary. Note that since the primitives that provide authentication, confidentiality and integrity are not ITS, the unbounded adversary can trivially break those, given the classical traffic. That is why the unbounded adversary is not allowed to transmit the compromised secrets to the HPT adversary but only the non-ITS distilled QKD keys.

Note that if authentication fails against the HPT adversary, the distilled QKD keys are not secure, thus not ITS. 

\end{definition}

\begin{definition}\label{def2.1}\textbf{(Quantum Key Distribution).} A Quantum Key Distribution (QKD) protocol uses the properties of quantum mechanics to generate a shared secret key between honest parties~\cite{GisinQC2002, QuantumCryptographyPirandola_2020, SecureQKDrealisticDevices2020}. Properties of quantum mechanics exploited to achieve secure quantum communication include the no-cloning theorem~\cite{No-Cloning-Wootters1982ASQ}, the Heisenberg uncertainty principle~\cite{heisenberg-uncert}, and quantum entanglement, depending on the QKD protocol~\cite{EntanglementQKD1991First}. QKD is divided into two parts: the preparation, sending, and measurement of qubits (quantum side) and a post-processing part for information reconciliation (classical side), typically divided into sifting, parameter estimation, error correction, and privacy amplification. The quantum side information is transmitted through an insecure quantum channel whereas the classical side information is transmitted through an authenticated classical channel. The quantum side involves QPT algorithms, while the classical side involves PPT algorithms. Since both sides are polynomial-time, the overall QKD protocol can be seen as a HPT algorithm, QKD = \{KeyGen\}, with an associated key set $\mathcal{K}$ = $\{0,1\}^*$ of arbitrary length. The KeyGen algorithm is described as:

\begin{itemize}
    \item[-] KeyGen($\kappa$, $*$) $\rightarrow$ \{($K_{\text{QKD}}$, m$_{\text{QKD}}$), $\perp$\}: The algorithm outputs the shared authenticated key $K_{\text{QKD}}$ of arbitrary size $*>0$ between honest parties, given a security parameter $\kappa$. For practical purposes, the corresponding authenticated public information m$_{\text{QKD}}$ used to distill the secret key $K_{\text{QKD}}$ is also an output. If QKD is aborted, then $\perp$ is returned.
\end{itemize}

The abortion condition strongly depends on the specific QKD protocol. For discrete-variable QKD (DV-QKD), this condition is often determined by the quantum bit error rate (QBER), which is obtained after the honest parties compare some of the assumed correlated sent and measured bits using the authenticated classical channel~\cite{GisinQC2002, SecureQKDrealisticDevices2020}. For continuous-variable QKD (CV-QKD), this condition often depends on the signal-to-noise ratio~\cite{QuantumCryptographyPirandola_2020}. The abortion condition also depends on whether the honest parties detect a tampering by an adversary within the authenticated classical channel. If an abortion occurs, the secret parameters and any revealed information from the public channel are discarded and never used again. Note that in practice, QKD may run continuously without ever aborting. However, when the keys are not secure, the final size of K$_{QKD}$ would be zero.

Although different QKD protocols employ various methods for quantum state preparation, measurement, and post-processing, they all produce an authenticated shared secret key and the corresponding authenticated public information: $K_{\text{QKD}}$ and m$_{\text{QKD}}$ are QKD protocol-agnostic.

A QKD iteration starts with honest parties exchanging and measuring quantum signals according to some agreed protocol and ends once privacy amplification is applied to their reconciled key, regardless of how many quantum signals, base sifting, parameter estimation, or other post-processing step but privacy amplification, are done in between.

\end{definition}

\begin{definition}\label{def2.2}\textbf{(Quantum Key Distribution Security).} The underlying problem behind the generation of the QKD keys relies on the bounds given by the laws of physics rather than the hardness of a mathematical problem, which makes the shared QKD secrets obtained from a QKD protocol between parties ITS from an ideal point of view. This means that the security of QKD does not depend on the computational power of the adversaries, implying that it is secure even against unbounded adversaries.

Ideally, an unbounded adversary $\mathcal{A}$ has no advantage in distinguishing between secrets distilled from a QKD protocol and random keys of the same length. The distinguishability game is defined with the experiment $\text{Exp}^{\text{ind-theory}}_{\text{QKD}, \mathcal{A}}(*)$, illustrated in Experiment \ref{exp1:QKDdistinguishability}, where the unbounded adversary $\mathcal{A}$ is able to perform a query to the QKD oracle and execute any operation within the QKD oracle's quantum and authenticated classical channels, while the QKD oracle generates the QKD key.

\begin{experiment}
\begin{enumerate}
    \item[] $\text{Exp}^{\text{ind-theory}}_{\text{QKD}, \mathcal{A}}(*)$:
    \begin{enumerate}
        \item[] $b \stackrel{\$}{\leftarrow} \{0, 1\}$
        \item[] if QKD does not abort:
        \begin{enumerate}
            \item[] ($K_{0}$, m$_{\text{QKD}}$) ${\leftarrow}$ KeyGen($*$), $K_{1} \stackrel{\$}{\leftarrow} \mathcal{K}$
        \end{enumerate}
        \item[] else:
        \begin{enumerate}
            \item[] return $\perp$ ${\leftarrow}$ KeyGen($*$)
        \end{enumerate}
        \item[] $b' \leftarrow \mathcal{A}(K_{b}, \text{m}_{\text{QKD}})$
        \item[] if $b'=b$ return 1, else return 0
    \end{enumerate}
\end{enumerate}
\caption{Distinguishability security experiment for QKD.}
\label{exp1:QKDdistinguishability}
\end{experiment}

Note that if the unbounded adversary tampers too much within the quantum channel or tampers within the authenticated classical channel, the QKD oracle will detect this and output the abort symbol. Continually tampering too much and making the QKD oracle to abort is equivalent to a denial of service (DoS) attack. Since the goal is to describe the security of QKD authenticated with PQC, DoS is not treated in this work.

In the ideal QKD scenario with ideal devices, the advantage of an unbounded adversary $\mathcal{A}$ to win the Experiment \ref{exp1:QKDdistinguishability} is zero:

\begin{equation*}
\text{Adv}^{\text{ind-theory}}_{\text{QKD}, \mathcal{A}}(*) = \left|Pr[\text{Exp}^{\text{ind-theory}}_{\text{QKD}, \mathcal{A}}(*)=1] - \frac{1}{2}\right|=0
\end{equation*}

However, the imperfections in implemented devices, the practical insecurity of the quantum channel and the information leakage and/or limitations from the usual implemented post-processing steps of QKD introduce a negligible exploitable advantage $\varepsilon_{QKD}(\kappa)$~\cite{RennerPortmann2014, PanosPirandolaEpsilonsCV}, which varies according to the considered specific QKD protocol, post-processing methods and the desired security level $\kappa$. In the same way, the authentication of the classical channel introduces a negligible exploitable advantage $\varepsilon_{\text{auth}}$. The authentication can be ITS, giving a $\varepsilon^\text{ITS}_\text{auth}$ security~\cite{ITS_auth_epsilon_Portmann2014}, or non-ITS but conditionally secure, giving a $\varepsilon^\text{PQC}_\text{auth}$ security. 

For the case where an unbounded adversary is present when ITS authentication is performed, QKD is secure if the unbounded adversary has a negligible advantage to win the experiment $\text{Exp}^{\text{ind}}_{\text{QKD}, \mathcal{A}}(\kappa, *)$, defined in the same way as the experiment \ref{exp1:QKDdistinguishability}, a result proved in~\cite{notreallybutis_composable_secure_epsilon_epsilon2_ITS_auth}:

\begin{equation*}
\text{Adv}^{\text{ind}}_{\text{QKD}, \mathcal{A}}(\kappa,*) = \left|Pr[\text{Exp}^{\text{ind}}_{\text{QKD}, \mathcal{A}}(\kappa,*)=1]-\frac{1}{2}\right|\leq\varepsilon_{QKD}(\kappa)+\varepsilon^\text{ITS}_{\text{auth}}(\kappa)
\end{equation*}

Note that the unbounded adversary wins all the times, i.e. $\text{Adv}^{\text{ind}}_{\text{QKD}, \mathcal{A}}(\kappa,*) = \frac{1}{2}$, if non-ITS authentication is performed, by realizing impersonation attacks.

In the same way, QKD is secure if all bounded HPT adversaries have a negligible advantage to win the experiment $\text{Exp}^{\text{ind}}_{\text{QKD}, \mathcal{A}}(\kappa,*)$, defined in the same way as the experiment \ref{exp1:QKDdistinguishability}:

\begin{equation*}
\text{Adv}^{\text{ind}}_{\text{QKD}, \mathcal{A}}(\kappa,*) = \left|Pr[\text{Exp}^{\text{ind}}_{\text{QKD}, \mathcal{A}}(\kappa,*)=1]-\frac{1}{2}\right|\leq\varepsilon_{QKD}(\kappa)+\varepsilon_{\text{auth}}(\kappa)
\end{equation*}

Where $\varepsilon_{\text{auth}}(\kappa) \in \{\varepsilon^\text{ITS}_{\text{auth}}(\kappa), \varepsilon^\text{PQC}_{\text{auth}}(\kappa)\}$, and the non-ITS, but post-quantum, authentication holds as discussed in section \ref{assumptions}.

\end{definition}

For the rest of this paper, $\varepsilon_{QKD}$ and $\varepsilon_{\text{auth}}$ depend implicitly on an arbitrary security parameter $\kappa$, i.e. $\varepsilon_{QKD} := \varepsilon_{QKD}(\kappa)$ and $\varepsilon_{\text{auth}} = \varepsilon_{\text{auth}}(\kappa)$. 

To give a physical interpretation to the epsilons, the probability that the practical ITS QKD keys have not been compromised by any adversary is upper bounded by $(\varepsilon_{QKD}+\varepsilon_\text{auth})$, or in other words, the probability they are not secure as intended is bounded by $(\varepsilon_{\text{QKD}} + \varepsilon_\text{auth})$. A successful QKD protocol and the derived keys are typically assumed to have a security proportional to $(\varepsilon_{QKD}+\varepsilon_\text{auth})$-secure.

\begin{definition}\label{def2.3}\textbf{(Unauthenticated Quantum Key Distribution).} Unauthenticated QKD is defined as the same as QKD from Definition \ref{def2.1} but the obtained shared key is distilled between two, not necessarily honest, parties through an insecure, unauthenticated, classical channel. This implies that the unauthenticated QKD key obtained from KeyGen, $K_{\text{QKD}}$, is not secure unless the corresponding public information m$_{\text{QKD}}$ is authenticated in further steps by the honest parties, if they are indeed both honest.

A conditioned unbounded adversary $\mathcal{A}$ is allowed to interact with the unauthenticated QKD, without targeting the authentication property of the classical channel (Definition \ref{def1.2}). Let $\text{Exp}^{\text{ind-theory}}_{\text{unauth-QKD}, \mathcal{A}}(\kappa,*)$ be the same Experiment \ref{exp1:QKDdistinguishability} but this time the unbounded adversary is not allowed to target the authentication of the classical channel from the QKD oracle. Since authentication is not targeted, the QKD oracle can be replaced by an unauthenticated QKD oracle without changing the advantage of the adversary to win the distinguishability game. The advantage of the conditioned unbounded adversary $\mathcal{A}$ to win $\text{Exp}^{\text{ind-theory}}_{\text{unauth-QKD}, \mathcal{A}}(\kappa,*)$ is the already defined $\varepsilon_{QKD}(\kappa)$.

\end{definition}

\begin{definition}\label{def2.4}\textbf{(From unauthenticated QKD to authenticated QKD).} Let two parties perform unauthenticated, and $\varepsilon_{\text{QKD}}$-secure, QKD to obtain an insecure shared key $K_{\text{QKD}}$ and the corresponding public information m$_{\text{QKD}}$ used to distill $K_{\text{QKD}}$. If the transcript m$_{\text{QKD}}$ is authenticated by honest parties in further steps using an $\varepsilon_\text{auth}$-secure composable authentication scheme, the unauthenticated QKD becomes equivalent to QKD, defined as in Definition \ref{def2.1}, making the shared key, $K_{\text{QKD}}$, ITS and $(\varepsilon_{\text{QKD}}+\varepsilon_\text{auth})$-secure~\cite{Composable_QKD_Auth}.

Assuming that the distilled and authenticated QKD key $K_{\text{QKD}}$ has length $L$, $K_{\text{QKD}}$ can be partitioned into $n$ smaller keys of arbitrary size $l_i$ with $L \geq \sum^n_{i=1} l_i$, such as $K_{\text{QKD}}=ss_{\text{QKD}_1}||\dots||ss_{\text{QKD}_n}||ss_{rest}$ where $||$ denotes concatenation (and ss is short for shared secret), the length $|ss_{\text{QKD}_i}| = l_i$ for $i = 1, ..., n$, and $|ss_{rest}| = L - \sum^n_{i=1} l_i$. Note that if any part of $K_{\text{QKD}}$, such as $ss_{\text{QKD}_i}$, is used for cryptographic purposes (e.g., renewing authentication material), that portion should not be reused for other cryptographic tasks, to avoid accidental or unexpected leakage of the secret. The remaining unused key material remains ITS and $(\varepsilon_{\text{QKD}}+\varepsilon_\text{auth})$-secure. The privacy amplification step, executed by all QKD protocols at the end, produces keys that are both uniformly random and independent with probability upper-bounded by $\varepsilon_{\text{QKD}}$, typically following the leftover hash lemma~\cite{LHL_Applications}. Privacy amplification reduces any correlation between the key bits, as well as between the key bits and the adversary's potential information, to a negligible level, which is quantified inside $\varepsilon_{{QKD}}$. The combination of privacy amplification with an epsilon composable authentication scheme makes the final keys security upper-bounded by $(\varepsilon_{{QKD}}+\varepsilon_\text{auth})$, where $\varepsilon_{QKD}(\kappa)$ includes the probability of failure given by the privacy amplification step and the security parameter $\kappa$. For example, let $K_{\text{QKD}}= ss_{\text{QKD}}||ss_{\text{rest}}$, where $ss_{\text{QKD}}$ is used for authentication with HPT secure primitives. Once the authentication is securely implemented, since the primitives used are HPT, an unbounded adversary could compromise $ss_{\text{QKD}}$, making the key non-ITS once used. However, the remaining part, $ss_{\text{rest}}$, remains ITS and $(\varepsilon_{QKD}+\varepsilon_\text{auth})$-secure, as the individual bits from the QKD protocols are indistinguishable from truly random bits with at least $1 - \varepsilon_{QKD}-\varepsilon_{\text{auth}}$ probability.

There could be instances where a single QKD iteration does not produce enough key material to run the HPT primitives within the protocols presented in sections \ref{protocols_sign_KEM} and \ref{protocol_MAC}. The concatenation of different QKD keys obtained through different QKD iterations or QKD protocols is possible. However, this implies that the security of the new concatenated key has also a new upper bound. Let $K$ be the concatenation of ``$m$" QKD keys, $K=K_1||K_2||\dots||K_m$. By Definition \ref{def2.2}, each QKD key ``$i$", $i\in\{0, 1,\dots, m\}$, is $\varepsilon_i$-secure. This implies that the concatenated key $K$ is ($1-\Pi_{i=1}^m(1-\varepsilon_i)$)-secure, which is upper bounded by $\Sigma^m_{i=1}\varepsilon_i$.

For simplicity, the epsilon security of any QKD step within the protocols presented in this paper is assumed to upper-bounded by the same value $\varepsilon_{QKD}(\kappa)$, dependent on an arbitrary security parameter $\kappa$.

\end{definition}

Next, the definition for dual pseudorandom function, signature, message authentication code, and key encapsulation mechanism and their corresponding securities are presented, similarly to within the Muckle series~\cite{Muckle, MucklePlus2023, MucklePlusPlus, MuckleKEM2024}.

\begin{definition}\label{def3}\textbf{(Dual Pseudorandom Function).} The pseudorandom functions (PRFs) used in this work are assumed to be dual PRFs. That is:

For a PPT pseudorandom function PRF: $\mathcal{K}$ $\times$ $\mathcal{M}$ $\rightarrow$ $\mathcal{R}$ there exists the dual PPT PRF$^{\text{dual}}$ defined as PRF$^{\text{dual}}$: $\mathcal{M}$ $\times$ $\mathcal{K}$ $\rightarrow$ $\mathcal{R}$ where PRF(k, m) = PRF$^{\text{dual}}$(m, k) for any bit string m $\in$ $\mathcal{M}$ and key k $\in$ $\mathcal{K}$. For a given security parameter $\kappa$, the key set $\mathcal{R}$ corresponds to the random output set $\{0, 1\}^\kappa$. The sets $\mathcal{M}$ and $\mathcal{K}$ each separately depend on $\kappa$.

The security of the PRFs and their dual counterpart are based on the distinguishing game. For a HPT adversary $\mathcal{A}$ and security parameter $\kappa$, the corresponding advantage functions are defined as:

\begin{equation*}
\text{Adv}^{\text{ind}}_{\text{PRF}, \mathcal{A}}(\kappa) = \left|Pr[\text{Exp}^{\text{ind}}_{\text{PRF}, \mathcal{A}}(\kappa)=1]-\frac{1}{2}\right|
\end{equation*}
\begin{equation*}
\text{Adv}^{\text{ind}}_{\text{PRF}^{\text{dual}}, \mathcal{A}}(\kappa) = \left|Pr[\text{Exp}^{\text{ind}}_{\text{PRF}^{\text{dual}}, \mathcal{A}}(\kappa)=1]-\frac{1}{2}\right|
\end{equation*}
where the experiments $\text{Exp}^{\text{ind}}_{\text{PRF}, \mathcal{A}}(\kappa)$ and $\text{Exp}^{\text{ind}}_{\text{PRF}^{\text{dual}}, \mathcal{A}}(\kappa)$ are defined in Experiment \ref{exp2:PRF}, such that the HPT adversary $\mathcal{A}$ is able to perform a polynomial number of queries $q$ to the PRF or PRF$^{\text{dual}}$ oracles.

\begin{experiment}
\begin{enumerate}
    \item[] $\text{Exp}^{\text{ind}}_{\text{PRF}^*, \mathcal{A}}(\kappa)$:
    \begin{enumerate}
        \item[] $b \stackrel{\$}{\leftarrow} \{0, 1\}$
        \item[] k $\stackrel{\$}{\leftarrow}$ $\mathcal{K}$
        \item[] for i $\leftarrow$ 1 to q:
        \begin{enumerate}
            \item[] m$_i$ $\leftarrow \mathcal{A}(r_1,\dots,r_{i-1}) \in \mathcal{M}$
            \item[] if * = dual r$_{0_i}$ ${\leftarrow}$ PRF$^\text{dual}$(m$_i$, k), else r$_{0_i}$ ${\leftarrow}$ PRF(k, m$_i$) 
            \item[] $r_{1_i} \stackrel{\$}{\leftarrow} \mathcal{R} \in \{0,1\}^\kappa$
        \end{enumerate}
        \item[] $b' \leftarrow \mathcal{A}(r_{b_1},\dots,r_{b_q})$
        \item[] if $b'=b$ return 1, else return 0
    \end{enumerate}
\end{enumerate}
\caption{Distinguishability security experiment for PRF or PRF$^\text{dual}$.}
\label{exp2:PRF}
\end{experiment}

The pseudorandom functions PRF and PRF$^\text{dual}$ are PPT, QPT or HPT secure if for all respective PPT, QPT or HPT adversaries $\mathcal{A}$, the advantages $\text{Adv}^{\text{ind}}_{\text{PRF}, \mathcal{A}}(\kappa)$ and $\text{Adv}^{\text{ind}}_{\text{PRF}^{\text{dual}}, \mathcal{A}}(\kappa)$ are negligible in the security parameter $\kappa$. That is, $\text{Adv}^{\text{ind}}_{\text{PRF}, \mathcal{A}}(\kappa)$ $\leq$ negl$_{\text{PRF}}(\kappa)$ and $\text{Adv}^{\text{ind}}_{\text{PRF}^{\text{dual}}, \mathcal{A}}(\kappa)$ $\leq$ negl$_{\text{PRF}^{\text{dual}}(\kappa)}$, where negl($\kappa$) is a function that takes as input the security parameter $\kappa$ and defined as: there exists $\kappa$ $>$ $\kappa_0$ such that negl($\kappa_0$) $<$ $\frac{1}{p(\kappa_0)}$ for every positive polynomial $p$.

\end{definition}

\begin{definition}\label{def4}\textbf{(Hash function).} The hash function H of this paper is assumed to be at least weakly collision resistant. That is:

Given a known bit string $m\in\mathcal{M}$, the PPT hash function H: $\mathcal{M} \rightarrow \mathcal{D}$, and the corresponding digest H$(m)\in\mathcal{D}$, an HPT adversary has negligible advantage Adv$_{\text{H},A}(\kappa)$ $\leq$ negl$_H(\kappa)$ of finding $m'\in\mathcal{M}$ such that H$(m)=\text{H}(m')$, given a security parameter $\kappa$.

\end{definition}

\begin{definition}\label{def5.1}\textbf{(Signature - $\Sigma$).} Signatures are composed by three PPT algorithms $\Sigma$ = \{KeyGen, Sign, Ver\}, with an associated bit string set $\mathcal{M}$, defined as:

\begin{enumerate}
    \item[-] KeyGen($\kappa$) $\rightarrow$ (sk, pk): The algorithm takes as input a security parameter $\kappa$ and outputs a secret key sk with the corresponding public key pk.
    \item[-] Sign(sk, m) $\rightarrow \sigma$: The algorithm takes as input a secret key sk and a bit string m $\in \mathcal{M}$ and outputs a signature tag $\sigma$.
    \item[-] Ver(pk, m, $\sigma$) $\rightarrow \{0, 1\}$: The algorithm takes as input a public key pk, a bit string m $\in \mathcal{M}$ and a signature tag $\sigma$ and outputs a bit $\in \{0,1\}$.
\end{enumerate}

$\Sigma$ is correct if for all (sk, pk) $\leftarrow$ KeyGen($\kappa$) and $\sigma$ $\leftarrow$ Sign(sk, m) given any security parameter $\kappa \in \mathbb{N}$ and bit string m $\in \mathcal{M}$, then Ver(pk, m, $\sigma$) = 1:

\begin{equation*}
    \text{Pr}\left[\text{Ver}(\text{pk}, \text{m}, \text{Sign(sk,m)})=1\right] = 1
\end{equation*}

\end{definition}

\begin{definition}\label{def5.2}\textbf{(EUF-CMA security of $\Sigma$).} For a HPT adversary $\mathcal{A}$, the advantage function for existential unforgeability under chosen message attacks (EUF-CMA) of $\Sigma$ is:

\begin{equation*}
    \text{Adv}^{\text{EUF-CMA}}_{\Sigma, \mathcal{A}}(\kappa)=\text{Pr}\left[\text{Exp}^{\text{EUF-CMA}}_{\Sigma, \mathcal{A}}\left(\kappa\right)=1\right]
\end{equation*}

Where the experiment $\text{Exp}^{\text{EUF-CMA}}_{\Sigma, \mathcal{A}}(\kappa)$ is defined in Experiment \ref{exp3:EUFCMASignature}.

\begin{experiment}
\begin{enumerate}
    \item[] $\text{Exp}^{\text{EUF-CMA}}_{\Sigma, \mathcal{A}}(\kappa)$:
    \begin{enumerate}
        \item[] (sk, pk) $\leftarrow$ KeyGen($\kappa$), $\mathcal{Q}$ $\leftarrow$ $\emptyset$
        \item[] for i $\leftarrow$ 1 to q:
        \begin{enumerate}
            \item[] m$_i$ $\leftarrow \mathcal{A}(\text{pk}, \sigma_1,\dots,\sigma_{i-1}) \in \mathcal{M}$
            \item[] $\sigma_i$ ${\leftarrow}$ Sign(sk, m$_i$), $\mathcal{Q} \leftarrow \mathcal{Q}\cup\{\text{m}_i\}$
        \end{enumerate}
        (m*, $\sigma$*) $\leftarrow \mathcal{A}(\text{pk}, \sigma_1,\dots,\sigma_q)$ 
        \item[] if Ver(pk, m*, $\sigma$*) $=1$ AND m* $\notin \mathcal{Q}$ return 1, else return 0
    \end{enumerate}
\end{enumerate}
\caption{EUF-CMA security experiment for signature $\Sigma$.}
\label{exp3:EUFCMASignature}
\end{experiment}

vThe signature scheme $\Sigma$ is PPT, QPT or HPT EUF-CMA secure if for all respective PPT, QPT or HPT adversaries $\mathcal{A}$, the advantage $\text{Adv}^{\text{EUF-CMA}}_{\Sigma, \mathcal{A}}(\kappa)$ is negligible in the security parameter $\kappa$. As a formal inequality, $\text{Adv}^{\text{EUF-CMA}}_{\Sigma, \mathcal{A}}(\kappa)$ $\leq$ negl$_{\Sigma,\text{EUF-CMA}}(\kappa)$.

\end{definition}

\begin{definition}\label{def6.1}\textbf{(Key Encapsulation Mechanism - KEM).} Key Encapsulation Mechanisms (KEM) are composed by three PPT algorithms KEM = \{KeyGen, Enc, Dec\}, with an associated key set $\mathcal{K}$, defined as:

\begin{enumerate}
    \item[-] KeyGen($\kappa$) $\rightarrow$ (sk, pk): The algorithm takes as input a security parameter $\kappa$ and outputs a secret key sk with the corresponding public key pk.
    \item[-] Enc(pk) $\rightarrow$ (k, c): The algorithm takes as input a public key pk and outputs a key k $\in \mathcal{K}$ and the corresponding encapsulation ciphertext c.
    \item[-] Dec(sk, c) $\rightarrow$ \{k, $\perp$\}: The algorithm takes as input a secret key sk and a ciphertext c and outputs the decapsulated k $\in \mathcal{K}$ or $\perp$ in case of failure.
\end{enumerate}

KEM is correct if for all (sk, pk) $\leftarrow$ KenGen($\kappa$) and (k, c) $\leftarrow$ Enc(pk) given any security parameter $\kappa \in \mathbb{N}$, then k $\leftarrow$ Dec(sk, c):

\begin{equation*}
    \text{Pr[Dec(sk, c)} = \text{k]} = 1
\end{equation*}

\end{definition}

\begin{definition}\label{def6.2}\textbf{(IND-CPA and IND-CCA security of KEM).} For a HPT adversary $\mathcal{A}$, the advantage functions for indistinguishability under chosen-ciphertext attacks (IND-CCA) and for indistinguishability under chosen-plaintext attacks (IND-CPA) of KEM are:

\begin{equation*}
    \text{Adv}^{\text{IND-CPA}}_{\text{KEM}, \mathcal{A}}(\kappa)=\left|\text{Pr}\left[\text{Exp}^{\text{IND-CPA}}_{\text{KEM}, \mathcal{A}}\left(\kappa\right)=1\right] - \frac{1}{2}\right|
\end{equation*}
\begin{equation*}
    \text{Adv}^{\text{IND-CCA}}_{\text{KEM}, \mathcal{A}}(\kappa)=\left|\text{Pr}\left[\text{Exp}^{\text{IND-CCA}}_{\text{KEM}, \mathcal{A}}\left(\kappa\right)=1\right] - \frac{1}{2}\right|
\end{equation*}
where the experiments $\text{Exp}^{\text{IND-CPA}}_{\text{KEM}, \mathcal{A}}(\kappa)$ and $\text{Exp}^{\text{IND-CCA}}_{\text{KEM}, \mathcal{A}}(\kappa)$ are defined in Experiment \ref{exp4:IND-KEM}.

\begin{experiment}
\begin{enumerate}
    \item[] $\text{Exp}^{\text{IND-*}}_{\text{KEM}, \mathcal{A}}(\kappa)$:
    \begin{enumerate}
        \item[] (sk, pk) $\leftarrow$ KeyGen($\kappa$), $\mathcal{Q}$ $\leftarrow$ $\emptyset$, $b \stackrel{\$}{\leftarrow} \{0, 1\}$
        \item[] for i $\leftarrow$ 1 to q:
        \begin{enumerate}
            \item[] k$_i$ $\leftarrow \mathcal{A}(\text{pk}, \text{c}_1,\dots,\text{c}_{i-1}) \in \mathcal{K} := \{0,1\}^\kappa$
            \item[] $\text{c}_i$ ${\leftarrow}$ Enc(pk, k$_i$), $\mathcal{Q} \leftarrow \mathcal{Q}\cup\{\text{c}_i\}$
        \end{enumerate}
        k$^*_0$, k$^*_1$ $\leftarrow \mathcal{A}(\text{c}_1,\dots,\text{c}_q) \in \mathcal{K}$ 
        \item[] c$^*$ $\leftarrow$ Enc(pk, k$_b$)
        \item[] if * = CCA:
        \begin{enumerate}
            \item[] for i $\leftarrow$ 1 to q$'$:
            \begin{enumerate}
                \item[] c$'_i \leftarrow \mathcal{A}(\text{pk}, \text{c}_1,\dots,\text{c}_{q}, \text{c}^*, \text{k}'_1,\dots, \text{k}'_{i-1})$
                \item[] k$'_i \leftarrow$ Dec(sk, c$'_i$), $\mathcal{Q} \leftarrow \mathcal{Q}\cup\{\text{c}'_i\}$
            \end{enumerate}
            \item[] $b' \leftarrow \mathcal{A}(\text{pk}, \text{c}_1,\dots,\text{c}_{q}, \text{c}^*, \text{k}'_1,\dots, \text{k}'_{\text{q}'})$
        \end{enumerate}
        \item[] else:
        \begin{enumerate}
            \item[] $b' \leftarrow \mathcal{A}(\text{pk}, \text{c}_1,\dots,\text{c}_{q}, \text{c}^*)$
        \end{enumerate} 
        \item[] if $b'=b$ AND c$^*$ $\notin \mathcal{Q}$ return 1, else return 0
    \end{enumerate}
\end{enumerate}
\caption{IND-CPA and IND-CCA security experiments for KEM.}
\label{exp4:IND-KEM}
\end{experiment}

The KEM scheme is PPT, QPT or HPT IND-CPA and/or IND-CCA secure if for all respective PPT, QPT or HPT adversaries $\mathcal{A}$, the advantages $\text{Adv}^{\text{IND-CPA}}_{\text{KEM}, \mathcal{A}}(\kappa)$ and/or $\text{Adv}^{\text{IND-CCA}}_{\text{KEM}, \mathcal{A}}(\kappa)$ are negligible in the security parameter $\kappa$. As a formal inequality, $\text{Adv}^{\text{IND-CPA}}_{\text{KEM}, \mathcal{A}}(\kappa)$ $\leq$ negl$_{\text{KEM},\text{IND-CPA}}(\kappa)$ and/or $\text{Adv}^{\text{IND-CCA}}_{\text{KEM}, \mathcal{A}}(\kappa)$ $\leq$ negl$_{\text{KEM},\text{IND-CCA}}(\kappa)$.

The KEM CPA security is not used for the rest of this paper and only the CCA security is relevant. However, the experiment is presented because Definition \ref{def8.2} makes use of it. 

\end{definition}

\begin{definition}\label{def7.1}\textbf{(Message Authentication Codes - MACs).} MACs are composed by three PPT algorithms MAC = \{KeyGen, Auth, Ver\}, with an associated key set $\mathcal{K}$ and bit string set $\mathcal{M}$, defined as:

\begin{enumerate}
    \item[-] KeyGen($\kappa$) $\rightarrow$ sk: The algorithm takes as input a security parameter $\kappa$ and outputs a secret key sk $\in \mathcal{K} := \{0,1\}^\kappa$.
    \item[-] Auth(sk, m) $\rightarrow \tau$:  The algorithm takes as input a secret key sk $\in \mathcal{K}$ and a bit string m $\in \mathcal{M}$ and outputs an authentication tag $\tau$.
    \item[-] Ver(sk, m, $\tau$) $\rightarrow \{0,1\}$: The algorithm takes as input a secret key sk $\in \mathcal{K}$, a bit string m $\in \mathcal{M}$ and an authentication tag $\tau$ and outputs a bit $\in \{0,1\}$.
\end{enumerate}

MAC is correct if for all (sk, pk) $\leftarrow$ KeyGen($\kappa$) and $\tau \leftarrow$ Auth(sk, m) given any security parameter $\kappa \in \mathbb{N}$ and bit string m $\in \mathcal{M}$, then Ver(sk, m $\tau$) $=1$:

\begin{equation*}
    \text{Pr}\left[\text{Ver(sk, m, Auth(sk,m)) = 1}\right] =1
\end{equation*}

\end{definition}

\begin{definition}\label{def7.2}\textbf{(EUF-CMA security of MAC).} For a HPT adversary $\mathcal{A}$, the advantage function for existential unforgeability under chosen message attacks (EUF-CMA) of MAC is:

\begin{equation*}
    \text{Adv}^{\text{EUF-CMA}}_{\text{MAC}, \mathcal{A}}(\kappa)=\text{Pr}\left[\text{Exp}^{\text{EUF-CMA}}_{\text{MAC}, \mathcal{A}}\left(\kappa\right)=1\right]
\end{equation*}

Where the experiment $\text{Exp}^{\text{EUF-CMA}}_{\text{MAC}, \mathcal{A}}(\kappa)$ is defined in Experiment \ref{exp5:EUFCMA-MAC}.

\begin{experiment}
\begin{enumerate}
    \item[] $\text{Exp}^{\text{EUF-CMA}}_{\text{MAC}, \mathcal{A}}(\kappa)$:
    \begin{enumerate}
        \item[] sk $\leftarrow$ KeyGen($\kappa$), $\mathcal{Q}$ $\leftarrow$ $\emptyset$
        \item[] for i $\leftarrow$ 1 to q:
        \begin{enumerate}
            \item[] m$_i$ $\leftarrow \mathcal{A}(\tau_1,\dots,\tau_{i-1}) \in \mathcal{M}$
            \item[] $\tau_i$ ${\leftarrow}$ Auth(sk, m$_i$), $\mathcal{Q} \leftarrow \mathcal{Q}\cup\{\text{m}_i\}$
        \end{enumerate}
        (m*, $\tau$*) $\leftarrow \mathcal{A}(\tau_1,\dots,\tau_q)$ 
        \item[] if Ver(sk, m*, $\tau$*) $=1$ AND m* $\notin \mathcal{Q}$ return 1, else return 0
    \end{enumerate}
\end{enumerate}
\caption{EUF-CMA security experiment for MAC.}
\label{exp5:EUFCMA-MAC}
\end{experiment}

The MAC scheme is PPT, QPT or HPT EUF-CMA secure if for all respective PPT, QPT or HPT adversaries $\mathcal{A}$, the advantage $\text{Adv}^{\text{EUF-CMA}}_{\text{MAC}, \mathcal{A}}(\kappa)$ is negligible in the security parameter $\kappa$. As a formal inequality, $\text{Adv}^{\text{EUF-CMA}}_{\text{MAC}, \mathcal{A}}(\kappa)$ $\leq$ negl$_{\text{MAC},\text{EUF-CMA}}(\kappa)$.

An ITS MAC is a MAC that is secure for all unbounded adversaries $\mathcal{A}$: the advantage $\text{Adv}^{\text{EUF-CMA}}_{\text{MAC}, \mathcal{A}}(\kappa)$ is negligible in the security parameter $\kappa$.

\end{definition}

\begin{definition}\label{def8.1}\textbf{(Authenticated Encryption with Associated Data - AEAD).} An AEAD scheme is composed of three PPT algorithms AEAD = \{KeyGen, Enc, Dec\}, with an associated key space $\mathcal{K}$, message space $\mathcal{M}$, associated data space $\mathcal{AD}$, and ciphertext space $\mathcal{C}$, defined as follows:

\begin{enumerate}
    \item[-] KeyGen($\kappa$) $\rightarrow$ k: The algorithm takes as input a security parameter $\kappa$ and outputs a key k $\in$ $\mathcal{K}$ chosen uniformly at random.
    \item[-] Enc(k, A, m) $\rightarrow$ c: The encryption algorithm takes as input a key k $\in$ $\mathcal{K}$, associated data A $\in$ $\mathcal{AD}$, and a message m $\in$ $\mathcal{M}$, and outputs a ciphertext c $\in$ $\mathcal{C}$.
    \item[-] Dec(k, A, c) $\rightarrow$ \{m, $\perp$\}: The decryption algorithm takes as input a key k $\in$ $\mathcal{K}$, associated data A $\in$ $\mathcal{AD}$, and a ciphertext c $\in$ $\mathcal{C}$, and outputs the corresponding message m $\in$ $\mathcal{M}$ if c is valid, $\perp$ otherwise.
\end{enumerate}

The AEAD scheme is said to be correct if for all keys 
k output by KeyGen($\kappa$), for all m $\in$ $\mathcal{M}$ and A $\in$ $\mathcal{AD}$, it holds that:

\begin{equation*}
    \text{Dec(k, A, Enc(k, A, m))} = \text{m}
\end{equation*}

\end{definition}

\begin{definition}\label{def8.2}\textbf{(IND-CPA and INT-CTXT Security of AEAD).} For any HPT adversary $\mathcal{A}$, the advantage functions for indistinguishability under chosen plaintext attacks (IND-CPA) and for ciphertext integrity (INT-CTXT) of an AEAD scheme are defined as:

\begin{equation*}
    \text{Adv}_{\text{AEAD},\mathcal{A}}^\text{IND-CPA}(\kappa) = \left| \text{Pr} \left[ \text{Exp}_{\text{AEAD},\mathcal{A}}^\text{IND-CPA}(\kappa) = 1 \right] - \frac{1}{2} \right|
\end{equation*}
\begin{equation*}
    \text{Adv}_{\text{AEAD},\mathcal{A}}^\text{INT-CTXT}(\kappa) = \text{Pr} \left[ \text{Exp}_{\text{AEAD},\mathcal{A}}^\text{INT-CTXT}(\kappa) = 1 \right]
\end{equation*}

The experiment $\text{Exp}_{\text{AEAD},\mathcal{A}}^\text{IND-CPA}(\kappa)$ is the same as Experiment \ref{exp4:IND-KEM} but there is no public key involved. The $\text{Exp}_{\text{AEAD},\mathcal{A}}^\text{INT-CTXT}(\kappa)$ Experiment is defined in Experiment \ref{exp6:INT-CTXT-AEAD}.

\begin{experiment}
\begin{enumerate}
    \item[] $\text{Exp}_{\text{AEAD},\mathcal{A}}^\text{INT-CTXT}(\kappa):$
    \begin{enumerate}
        \item[] k $\leftarrow$ KeyGen$(\kappa)$, $\mathcal{Q} \leftarrow \emptyset$
        \item[] for $i\leftarrow 1$ to q:
        \begin{enumerate}
            \item[] (A$_i$, m$_i$) $\leftarrow$ $\mathcal{A}(\text{A}_1, \text{m}_1,\text{c}_1,\dots,\text{A}_{i-1}, \text{m}_{i-1},\text{c}_{i-1})$
            \item[] c$_i \leftarrow$ Enc(k, A$_i$, m$_i$), $\mathcal{Q} \leftarrow \mathcal{Q}\cup \{(\text{A}_i,\text{c}_i)\}$
        \end{enumerate}
        \item[] (A$^*$, c$^*$) $\leftarrow$ $\mathcal{A}(\text{A}_1, \text{m}_1,\text{c}_1,\dots,\text{A}_q,\text{m}_q,\text{c}_q)$ $\in$ $\mathcal{AD} \times \mathcal{C}$
        \item[] m$^*$ $\leftarrow$ Dec(k, A$^*$, c$^*$)
        \item[] if m$^*\neq \perp$ and (A$^*$, c$^*$) $\notin \mathcal{Q}$ return 1, else return 0.
    \end{enumerate}
\end{enumerate}
\caption{INT-CTXT security experiment for AEAD.}
\label{exp6:INT-CTXT-AEAD}
\end{experiment}

An AEAD scheme is PPT, QPT or HPT IND-CPA secure, and INT-CTXT secure, if for all respective PPT, QPT or HPT adversaries $\mathcal{A}$, the advantage $\text{Adv}_{\text{AEAD},\mathcal{A}}^\text{IND-CPA}(\kappa)$, and the advantage $\text{Adv}_{\text{AEAD},\mathcal{A}}^\text{INT-CTXT}(\kappa)$, are negligible in the security parameter $\kappa$. As a formal inequality, $\text{Adv}_{\text{AEAD},\mathcal{A}}^\text{IND-CPA}(\kappa)$ $\leq$ negl$_{\text{AEAD},\text{IND-CPA}}(\kappa)$ and $\text{Adv}_{\text{AEAD},\mathcal{A}}^\text{INT-CTXT}(\kappa)$ $\leq$ negl$_{\text{AEAD},\text{INT-CTXT}}(\kappa)$.

Note that if an AEAD algorithm is both IND-CPA secure and INT-CTXT secure then the AEAD algorithm is also IND-CCA secure. Note that IND-CCA security implies IND-CPA security but not INT-CTXT security~\cite{IND-CPA-INT-CTXT-IND-CCA-Relation}. 

\end{definition}

In the next section, useful assumptions on the PQC algorithms used to authenticate QKD are presented.

\section{Post-quantum cryptography and assumptions made in this paper}\label{assumptions}

\subsection{Post-quantum security}

Post-quantum cryptographic algorithms are designed to be secure against HPT adversaries and are believed to withstand attacks from any classical, quantum or hybrid algorithm that can run in polynomial time. Practically, this means there is no known classical or quantum algorithm that can break a PQC algorithm within a feasible time frame, typically considered to be at least 30 years.

Post-quantum secure signatures, KEMs and other asymmetric cryptographic algorithms, often referred to as PQC algorithms, are relatively recent technologies undergoing standardization by relevant organizations such as NIST~\cite{nist_pqc}. As mentioned earlier, they are believed to be HPT secure. PQC can rely on public key infrastructure (PKI) to manage keys and certificates delivered by a certificate authority (CA), specially for authentication purposes. In the schemes presented in this paper, PQC is used to authenticate honest parties, relying on PKI and post-quantum CAs. The topic of creating a post-quantum secure CA is out of scope.

PRFs, MACs and other symmetric cryptographic PPT algorithms are considered to be quantum resistant. However, they are not typically included when discussing PQC. The primary quantum algorithm found so far that decreases the security of these algorithms, in this case quadratically, is Grover's algorithm~\cite{Grover}. To maintain the security of the symmetric cryptographic PPT algorithms against Grover's algorithm, the size of the secret material must be at least doubled. For the rest of this paper, symmetric cryptographic algorithms are assumed to be HPT secure.

\subsection{Practical security}

Before discussing why PQC is suitable for authenticating QKD users, it is important to note that PQC and QKD encompass a wide variety of protocols that offer different methods for establishing secure communications between honest parties. Both technologies should complement each other, rather than displace one another. Furthermore, classical cryptography, known to be not post-quantum, can be integrated with PQC and QKD to provide redundant security. Although this topic is out of scope of this paper, examples of hybridization involving QKD, PQC and classical cryptography can be found in the Muckle series~\cite{Muckle, MucklePlus2023, MucklePlusPlus, MuckleKEM2024}.

As the security of PQC algorithms relies on the belief that the hardness of certain mathematical problems holds against both classical and quantum adversaries, there is no proof that this will remain true for the future. There is always uncertainty about when a classical, quantum, or hybrid algorithm might be developed that could break the hardness of the mathematical problems underlying PQC algorithms. An example of this happening in the past is Shor's algorithm, which, along with advancements in quantum computing technology, motivated the standardization of PQC algorithms. 

However, since PQC is built on the assumed hardness of certain mathematical problems, there is a potential gap that could be exploited by honest parties to perform cryptographic operations, such as using PQC for authentication in QKD. The argument relies on the premise that if honest parties can perform their authentication operations and exchanges faster than an HPT adversary can compromise the secret material used in the PQC algorithms by targeting and processing the public material, such as public keys and ciphers, then the honest parties can be assured that no malicious activity targeting authentication has been effectuated by any HPT adversary. The secret keys obtained through a QKD protocol, where the authentication of the classical channel is secured through a PQC protocol, are ITS and $(\varepsilon_{QKD}+\varepsilon^\text{PQC}_\text{auth})$-secure if honest parties are sure that no successful efficient attack by any HPT adversary has been possible within the time required to authenticate the classical information needed to distill the QKD keys.

A general and conservative runtime condition for the honest parties to be sure that no attack targeting authentication obtained through a PQC protocol has been possible is presented in equation \ref{eq1:runtime}.

\begin{equation}
    \text{T}_A + \text{T}_B + \text{T}_T < \text{T}_{HPT}
    \label{eq1:runtime}
\end{equation}

Where $\text{T}_A$ and $\text{T}_B$ are the runtimes required for honest parties A and B to perform the computational operations involving the authentication PQC protocol, $\text{T}_T$ is the time accounting for transmission, network delays and other non-trivial times, and $\text{T}_{HPT}$ is the time required for an HPT adversary to break the PQC algorithm in question. Note that fractions of T$_A$, T$_B$ and/or T$_T$ may overlap and such overlap is not taken in account. $\text{T}_{HPT}$ is the time required for the adversary to obtain a non-negligible advantage function, defined by the corresponding security experiments/games, which accounts for the fact that the adversaries gain some (negligible or non-negligible) information about the secret material after every query. The adversary time $\text{T}_{HPT}$ starts counting once any kind of public information generated by the PQC algorithms from the honest parties is obtained.

While the runtimes from the left side of equation \ref{eq1:runtime}, T$_A$, T$_B$ and T$_T$, are more or less feasible to estimate via benchmarking, time complexity theory and risk assessments, the adversary runtime T$_{HPT}$ is far from being trivial for several reasons. One reason is that honest parties have no clue when an adversary has taken possession of any public information computed through PQC algorithms by them. Another reason is that the runtime T$_{HPT}$ may be overestimated, as the more public information the HPT adversary is capable of gathering, the more resources it has to break the PQC algorithms faster than predicted. Furthermore, it is hard to know exactly what the computational capacities of the adversary are. Unpredictable advancements in classical and quantum computation fields, as well as unknown attacks, further complicate the task of characterizing the adversary runtime. After characterization of the HPT adversary runtime T$_{HPT}$, this could be expressed in T$_{HPT}$-term CAs that would revoke provided certificates once their lifetime surpasses a bound, dependent on an estimation of the HPT adversary runtime T$_{HPT}$.

However, since PQC (or rather hybrid classical and PQC schemes) are becoming the norm of modern PPT cryptography, it is with the same legitimacy that the assumption that equation \ref{eq1:runtime} holds for some justified HPT runtime T$_{HTP}$, a time that could range from seconds to days. After that time has passed, a new pair of secret and public keys would need to be generated along with a T$_{HPT}$-term certificate to validate the identity of the honest parties behind them.

Thus, the PQC algorithms used in the following sections are considered to be T$_{HPT}$-term HPT secure, according to equation \ref{eq1:runtime} and Definition \ref{def9} given below.

\begin{definition}\label{def9}\textbf{(T$_{HPT}$-term HPT security).} An algorithm is T$_{HPT}$-term HPT secure (or T$_{HPT}$-term secure) if the advantage function, as defined by the corresponding experiments/games, remains negligible in the relevant security parameter against a HPT adversary for a specified time T, T $<$ T$_{HPT}$. Said advantage function becomes no longer negligible once T $\approx$ T$_{HPT}$.

\end{definition}

Additionally, the symmetric algorithms and the dual PRF used in the following sections are considered to be long-term HPT secure according to the Definition \ref{def10}.

\begin{definition}\label{def10}\textbf{(Long-term HPT security).} An algorithm is long-term HPT secure if the advantage function, as defined by the corresponding experiments/games, remains negligible in the relevant security parameter against a HPT adversary for an arbitrarily long time, typically ranging from 10 to 30 years.
Note that if T$_{HPT}$ is estimated to be large enough, Definition \ref{def9} and Definition \ref{def10} are equivalent.

\end{definition}

To prevent the threat from Grover's algorithm~\cite{Grover}, the secret keys used inside the long-term HPT secure symmetric primitives, namely dual PRF, AEAD and MAC, have to be at least doubled compared to the usual sizes that resist a non-quantum search threat. For example, to have a bit security close to 256 bits, the secret input length of the PRF has to be 512 bits or more.

\section{PQC authentication for QKD}\label{protocols_sign_KEM}

A crucial component that ensures QKD is ITS and $(\varepsilon_\text{QKD}+\varepsilon_\text{auth})$-secure is the presence of authenticated channels. Specifically, it is essential that honest parties can authenticate the classical traffic used to distill QKD keys.

As shown in the security proof, in section \ref{security}, if the authentication algorithms are T$_{HPT}$-term HPT secure, following Definition \ref{def9}, honest parties have a time window, strictly smaller than T$_{HPT}$, during which they are confident that no adversary can execute any attack targeting authentication and integrity. Consequently, as shown in section \ref{security}, the following PQC authentication protocols for QKD would remain secure as long as Definition \ref{def9} holds, ensuring that the distilled QKD keys are ITS, $(\varepsilon_\text{QKD} + \varepsilon^\text{PQC}_\text{auth})$-secure, and shared by honest parties.

Stronger security for authentication can be adopted once the PQC authentication for QKD protocols have been executed at least once. Given that honest parties will share a secret at that point, ITS authentication can then be achieved using the commonly employed Carter-Wegman MAC construction, which is based on strongly universal hash functions~\cite{ITS-MAC-WEGMAN1981265, ITS-MAC-secret-key-rate, quantumsecureMACs, ITS-MAC-General-Stinson1994}. A practical, lightweight implementation can be constructed following~\cite{ITS-MAC-secret-key-rate}, where the type of MAC is defined and optimized to require a minimal amount of secret key material for authentication in each QKD iteration, given an arbitrary authentication security $\varepsilon^\text{ITS}_\text{auth}$. ITS authentication is not the scope of this work and the topic is mostly omitted in the following sections.

This section discusses the general case where PQC primitives, alongside symmetric primitives, are used to provide authentication, confidentiality and integrity to QKD. Some examples of PQC authentication for QKD are given. The security of the protocols presented is discussed in the subsequent section.

\subsection{General case}\label{general-case}

\begin{figure*}[!ht]
\centering
\includegraphics[width=\textwidth]{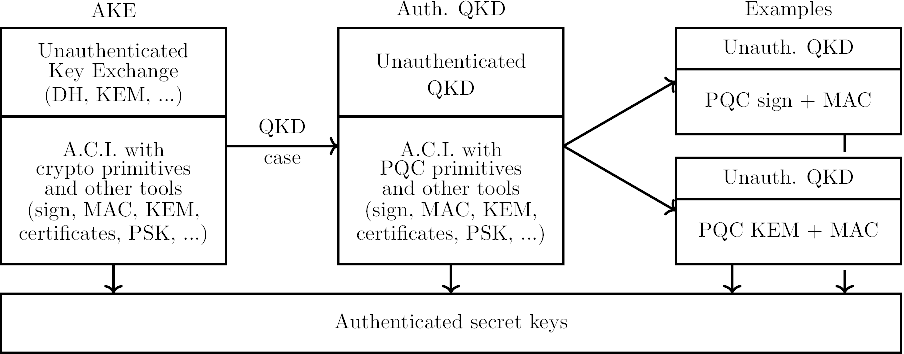}
\caption{Authenticated key exchange general steps (left); Post-quantum AKE applied to QKD (center); Examples of AKE based on QKD (right). A.C.I. = Authentication, Confidentiality and Integrity.}
\label{fig:generalAKE}
\end{figure*}

One of the most used authentication mechanisms for key exchange at the time of writing this paper is the Transport Layer Security (TLS) protocol. The latest version, TLS 1.3~\cite{rfc8446}, implements Diffie-Hellman (DH) exchanges, with authentication inspired from the SIGMA protocol, using digital certificates. The authenticated key exchange (AKE) steps from TLS 1.3 and SIGMA may be generalized as the creation of an authenticated key and the authentication step as shown in figure \ref{fig:generalAKE} (left).

The first step of an AKE corresponds to the creation of an unauthenticated key between two parties, often referred as an ephemeral key since its generation and usage is typically attributed to a specific task making them have a short lifespan. The ephemeral key may be computed through different methods such as a DH exchange, elliptic-curve DH protocol, KEM exchange, among other protocols. The second step of the AKE is to verify the entities behind the key exchange while keeping confidentiality and the integrity of the traded messages. This is often achieved by relying on cryptographic primitives such as signatures, MACs and/or KEMs, along with related tools like pre-shared key (PSK) material or certificates. Specifically, if honest parties have PSK material, authentication typically relies on symmetric primitives such as MACs, which can be designed to be ITS if desired~\cite{ITS-MAC-WEGMAN1981265, ITS_auth_epsilon_Portmann2014, ITS-MAC-General-Stinson1994}. Otherwise, parties have to rely on asymmetric cryptography, such as signatures and KEMs, and trust-based mechanisms, such as public key infrastructure (PKI) along with certificate authorities (CA), which bind public keys to entities. 

This AKE architecture can be used to authenticate classical information to make a QKD protocol and the corresponding distilled keys secure, by replacing the traditional unauthenticated key exchange by an unauthenticated QKD key obtained through an unauthenticated QKD protocol following Definition \ref{def2.3} (Figure \ref{fig:generalAKE}, center). A QKD protocol can be executed by two parties without really needing to immediately authenticate the public information required to distill a secret QKD key, the post-processing information. Certainly, this would open the person-in-the-middle vulnerability if no authentication process occurs after an unauthenticated QKD protocol is finalized. The idea behind the authentication is to rely on PKI, along with CA and certificates, and T$_{HPT}$-term HPT secure primitives (equivalent to T$_{HPT}$-term secure PQC), such as signatures and KEMs. The information utilized by the honest parties to distill a QKD key through an unauthenticated QKD protocol, such as the post-processing information, is authenticated during the authentication step with an $\varepsilon^\text{PQC}_\text{auth}$ scheme and by Definition \ref{def2.4}, the QKD keys become $(\varepsilon_\text{QKD}+\varepsilon^\text{PQC}_\text{auth})$-secure. The process of authenticating the required information at the end of a scheme is known as ``mono-authentication", and is also found in~\cite{ITS-MAC-secret-key-rate} and~\cite{MoscaImplemented2024}, for the QKD case.

The other properties of interest are confidentiality and integrity. By using the ephemeral keys, along with the derived classical keys through the AKE scheme, honest parties derive key material to encrypt and decrypt the messages exchanged between them, often relying on symmetric algorithms such as the Authenticated Encryption with Associated Data (AEAD) primitive~\cite{AEADreference}, on deriving keys to compute MAC tags to keep in check the integrity of the traded messages. Confidentiality over the entities' information is often called entity protection. The goal of entity protection is to encrypt and conceal the exchanged information between honest parties, such as certificates and ciphers, so that adversaries or eavesdroppers cannot determine the identities behind the traded messages. The entity protection feature within the AKE schemes can be optional by giving up on confidentiality.

The entity protection feature usually provides security against passive adversaries for both parties, and against active adversaries for one of the parties. A passive adversary is limited to eavesdropping on the communications between honest parties, whereas an active adversary can interact with the traded data and actively engage in communications by sending and receiving information. Once an AKE protocol with the entity protection feature is successfully completed between honest parties, it is safe to assume that the concerned entities are protected with the same security as the underlying symmetric protocol used to hide the entities (and/or tags), regardless of whether the adversaries are passive or active.

In the context of a new user wanting to join an existing quantum network, one could argue that it is best to provide entity protection against active adversaries to the existing network user, rather than to the new user. In this way, active adversaries are unable to identify the entities within the network. However, if an adversary knows that a new user is attempting to join the network, the adversary could potentially identify the entity behind the new user and do so for every user joining the network, thereby revealing the entities within the network as it grows. A potential solution for this issue is to provide entity protection against active adversaries to the new user and use a generic entity name for the users within the network. Once the new user is within the network, the real identities can be exchanged by following subsequent procedures that are not commented in this work.

The presented protocols in this work provide entity protection in the same way that SIGMA or TLS does~\cite{SIGMA2003, rfc8446}. This relates to the known traffic secrets that TLS and similar protocols such the Muckle series present. The symmetric encryption of ``message information" using primitives such as AEAD with a key ``key", mainly used to provide entity protection (and tag protection), is noted as $\{$message information$\}_{key}$ within the protocols. Note that the messages ``m$_i$" within the digest H(m$_i$) correspond to the message before encryption if ``$\{$message information$\}_{key}$" is sent, or decrypted if $\{$message information$\}_{key}$ is received, ``message information" rather than the entire encrypted bit string $\{$message information$\}_{key}$. For example, given an exchanged message m$_i$: $\{$message information$\}_{key}$ between parties, H(m) = H(message information) is used, rather than H($\{$message information$\}_{key}$). Note that ``message information" may contain optional traffic other than that presented in the protocol, but the security of the protocol is not affected by this optional traffic.

Examples of authentication methods for QKD, which are described in more detail below, are T$_{HPT}$-term HPT secure PQC signatures and MAC, or T$_{HPT}$-term HPT secure KEM and MAC (figure \ref{fig:generalAKE}, right). The signature then MAC is similar to the SIGMA protocol (and closely related to TLS 1.3) but using an ephemeral key obtained through unauthenticated QKD, while the KEM then MAC process is similar to the KEMTLS protocol~\cite{KEMTLS2020} but with the same ephemeral unauthenticated QKD key.

The presented schemes assume that honest parties start interacting with each other without any pre-shared keys. To secure the first iteration of QKD, PQC is used for authentication at least for the initial QKD iteration. Once sufficient QKD key material has been obtained and authenticated using PQC, the authentication for QKD can rely on HPT-secure MAC or even ITS-MAC, as mentioned before. However, the presented PQC authentication for QKD examples can be used successively by keeping continuity of the secret state parameter, denoted $SecState$ in the protocols.

\subsection{QKD authenticated with signature then MAC}\label{sign}

Despite the wide variety of QKD protocols~\cite{QuantumCryptographyPirandola_2020, SecureQKDrealisticDevices2020}, all can rely on PQC for authentication for the first instantiation. An unauthenticated QKD may be seen as a black box that outputs unauthenticated ephemeral keys that have been obtained through some public information that must be authenticated at some point to make the whole QKD scheme secure, following Definitions \ref{def2.1}, \ref{def2.3} and \ref{def2.4}. Regarding the QKD black box, special care has to be taken in terms of the order of exchange of the encoding and decoding bases to sift the raw QKD keys. The initiator sends its encoding bases only when the already used decoding bases from the responder have been received: this is to avoid attacks where quantum memories are involved. The order of the exchanged public information needed to distill a QKD key must follow the steps as if the classical channel was authenticated. The QKD protocol used and how the unauthenticated QKD key is distilled are not discussed in this paper, as the authentication (PQC), confidentiality, (PRF and AEAD) and integrity (PQC and/or MAC) approaches are agnostic regarding this matter.

For the rest of this paper, the QKD protocols are seen as black boxes that provide an unauthenticated shared secret $K_{QKD} =\text{ss}_{QKD_1}||\dots||ss_{QKD_n}||\text{ss}_{rest}$ and the traded public information to be authenticated m$_{QKD}$, as specified in Definitions \ref{def2.1} and \ref{def2.3}. If the bit length of the key $K_{QKD}$ is not sufficient to run the HPT primitives given by the protocols, honest parties can perform QKD again to produce $K'_{QKD}$ and the corresponding classical traffic $m'_{QKD}$. The authentication algorithms will authenticate both transcripts, $m_{QKD}$ and $m'_{QKD}$, to make the individual keys, $K_{QKD}$ and $K'_{QKD}$. Since this would make the epsilon security of the concatenated key more complex, let us assume that the QKD protocols always produce keys long enough to run the HPT primitives. This assumption is realistic because honest parties have a prior estimate of the minimum key length required to be extracted from QKD in order to authenticate, and even to store for further applications. In this case, the QKD keys are always $\varepsilon_{QKD}(\kappa)$. The intention of honest parties is to derive more key material than what is used in the authentication process, hence they will perform authentication only after distilling enough unauthenticated QKD keys/bits, which is feasible as QKD typically post-processes blocks of $10^5$ bits or more.

\begin{figure*}[!ht]
\centering
\includegraphics[width=\textwidth]{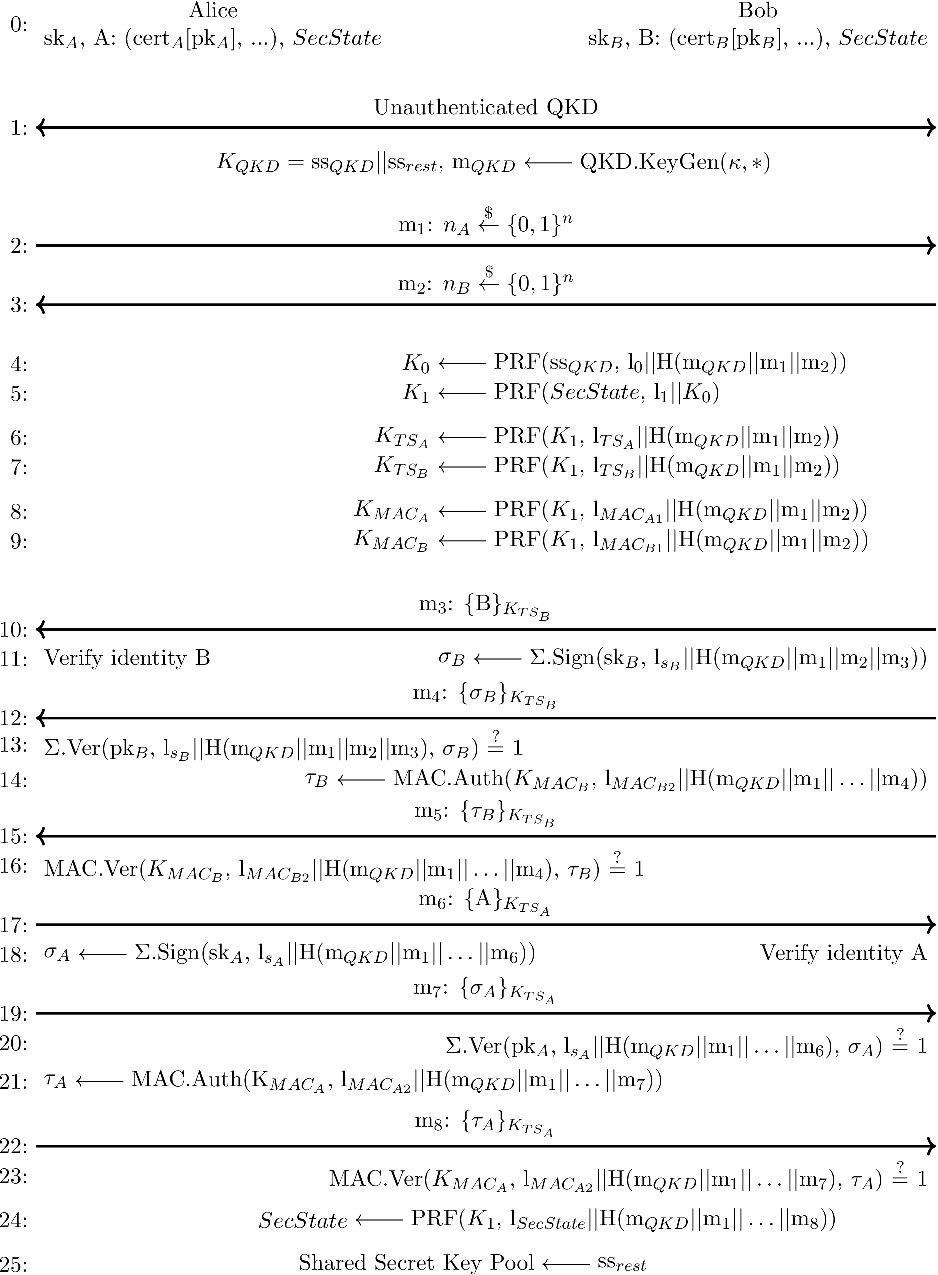}
\caption{Example of QKD mono-authenticated with PQC signature then MAC.}
\label{signHAKEauthFurther}
\end{figure*}

Within the figures below, the QKD exchange is illustrated by a black double-headed arrow that connects the QKD initiator Alice and the QKD responder Bob. For the rest of the paper, Alice is set as the initiator, whereas Bob is the responder. The authentication provided by the signature then MAC scheme is illustrated in figure \ref{signHAKEauthFurther}. As typical in the current state of the art, the single headed arrows correspond to the transmission of classical information through a public and non-authenticated channel.

\paragraph*{Step by step.} The steps for the signature then MAC protocol are presented below. The security analysis is presented in section \ref{security}.

0: Before starting the protocol, the initiator Alice and the responder Bob have known parameters. Alice and Bob have their corresponding T$_{HPT}$-term HPT secret key, sk$_A$ or sk$_B$, from the PQC signature algorithm and the corresponding public key, pk$_A$ or pk$_B$, within a digital certificate, cert$_A$ or cert$_B$, provided by a legitimate CA. Entity parameters such as certificates are placed inside the parameters A and B for Alice and Bob, respectively. Both parties also share the secret state, $SecState$, that is updated at the end of each execution of the protocol and carry on from one application of the protocol to another. This secret parameter, $SecState$, is initialized to an arbitrary value: it could be initialized at 0 or even take the value given by PRF($ss_{QKD}$, label$||$H()) with some bit string that composes the label, for example. Note that the secret keys, sk$_A$ and sk$_B$, are revealed to no one, including honest parties. The entity parameters, A and B, may be public in the case that entity concealment is not required, otherwise they are secret and only traded following the presented protocol. As discussed in section \ref{general-case}, the entity protection feature is secure against passive adversaries for both parties, and secure against active adversaries for one of the parties.

1: The protocol starts with the initiator Alice and the responder Bob performing an unauthenticated QKD protocol following Definition \ref{def2.3}. Both parties obtain the QKD traffic to authenticate, m$_{QKD}$, which is public information, and the corresponding unauthenticated QKD bit string $K_{QKD} = \text{ss}_{QKD}||\text{ss}_{rest}$, which is secret data. In this example of a protocol, the first key, ss$_{QKD}$, acts as an ephemeral key.

If the unauthenticated QKD is badly performed between honest parties, the condition unbounded adversary is able to obtain a non-negligible amount of information of the unauthenticated QKD bit string, $K_{QKD} = \text{ss}_{QKD}||\text{ss}_{rest}$, out of the public QKD traffic m$_{QKD}$. By consequence, the final QKD keys do not hold the promised $\varepsilon_{QKD}$ security, even if authentication is successful, and thus are not secure.

2, 3: Parties exchange random and unique nonces, $n_A$ and $n_B$. The nonces are required to prevent replay attacks, misbinding attacks, and to avoid derived vulnerabilities~\cite{SIGMA2003}.

4: Parties derive key material from the QKD material ss$_{QKD}$ thanks to a dual PRF. The derived key $K_0$ binds the unauthenticated QKD key ss$_{QKD}$ to the public information m$_{QKD}$ and the exchanged nonces. A hash function H is applied to the classical information to compress large bit strings into sizes that fit the dual PRF's input requirements. This adds another potential layer of vulnerability, as an adversary/Eve could tamper with the classical messages such as to arrange that the public information that Alice has, m$_{QKD_A}$, and the public information that Bob has, m$_{QKD_B}$, do not match, but their digests do, for example, H(m$_{QKD_A}||\text{m}_1||\text{m}_2)=$ H(m$_{QKD_B}||\text{m}_1||\text{m}_2$). The tampering would allow the HPT adversary to gain some information about intermediate keys of the protocols, the output of the protocol or even the output $K_{QKD}$ of the QKD protocol. This potential vulnerability is discussed further in the security analysis, section \ref{security}. The considered hash function H follows Definition \ref{def4} and does not require a secret key to work. On the same note, if the HPT adversary is passive and does not obtain the shared secret ss$_{QKD}$, a guess of the PRF output or a tentative forgery is the only way to obtain the intermediate key $K_0$, since the secret ss$_{QKD}$ is not known, as also discussed in the security proof.

The labels l$_i$, $i \in \{0,1, TS_A, TS_B, MAC_{A1}, MAC_{B1}, s_A, s_B, MAC_{A2}, MAC_{B2}, SecState\}$, presented inside the cryptographic primitives, correspond to unique (from each other) bit strings that are public information. The labels allow for domain separation for each intermediate key, provide security against reuse attacks and bind context to the derived keys, as commented in this NIST recommendation strongly related to PRFs~\cite{KeyDerviPRF_NIST:SP800-108r1-upd1}.

5: The obtained key $K_0$ is combined with the secret state $SecState$ using dual PRFs to derive the intermediate key $K_1$. The adversary would require to know both of the parameters, $SecState$ and $K_0$, to be able to derive $K_1$, otherwise a forgery or a guess is the only way to obtain it, for which the success probability is low since PRFs are assumed to be long-term HPT secure (Definition \ref{def10}). 

The parameter $SecState$ is known at the first iteration of authentication and gets updated at the very end of the protocol, once the authentication gets accepted by both parties. The inclusion and update of the secret state parameter $SecState$ provides a weak form of post-compromise security, in the sense that if an active adversary learns all secrets of a session, security can be recovered only if honest parties complete at least one session in which the adversary has become passive. 

6, 7: If entity protection or traffic protection is enabled, the traffic secrets, $K_{TS_A}$ and $K_{TS_B}$, are derived from $K_1$. The HPT adversary requires to know $K_1$ in order to derive the traffic secrets.

Traffic protection, which includes entity protection, against passive attackers is achieved for the initiator and responder at all times. If the secret state $SecState$ is a known value to an eavesdropper, either at the initialization step or because it was leaked, and depending on how messages are traded between the initiator and the responder, entity protection against active attacks is achieved for either the initiator or the responder, but not both. In this case, traffic protection against active adversaries for both honest parties is not possible to achieve, as the first traffic protection encryption is always performed with an unauthenticated key, and the parameter $SecState$ is known by the adversary: anyone impersonating can compute $K_1$ if the shared parameter $SecState$ is not secret.

The presented protocol \ref{signHAKEauthFurther} achieves active entity protection only for the initiator, but this can be changed to the responder if the order of who reveals their identity first is reversed. An eavesdropper/adversary Eve would be required to have QKD equipment in order to realize an active attack and learn the identity of one of the parties: her presence would be noticed afterwards, when parties require the PQC secret keys to execute PQC algorithms and the protocol would be aborted by the honest party. In the case that the shared parameter $SecState$ is secret, then entity protection against active adversaries is achieved by both honest parties.

8, 9: The MAC keys, $K_{MAC_A}$ and $K_{MAC_B}$, are derived from $K_1$. These keys are used to create and verify MAC tags. The HPT adversary requires to know $K_1$ in order to derive the MAC keys.

10: If entity protection is enabled, the responder Bob AEAD encrypts the entity information B with the key $K_{TS_B}$ and the output, \{B\}$_{K_{TS_B}}$, is sent to the initiator Alice. Otherwise, the entity information B is public knowledge.

11: The initiator Alice verifies the identity B thanks to public information, such as the public key from the responder Bob's certificate. The responder Bob computes the signature tag $\sigma_B$ by signing the public information l$_{s_B}||$H(m$_{QKD}||$m$_1||$m$_2||$m$_3$) with the secret key sk$_B$. Note that the HPT adversary would have a negligible chance to create such tag, and thus be detected by Alice in the verification step.

12: If traffic protection is enabled, the responder Bob AEAD encrypts $\sigma_B$ with the key $K_{TS_B}$ and sends the output, \{$\sigma_B$\}$_{K_{TS_B}}$, to the initiator Alice. Otherwise, $\sigma_B$ is sent directly.

13: The initiator Alice (decrypts the received output with the corresponding traffic secret and) verifies the received signature tag with public information l$_{s_B}||$ H(m$_{QKD}||$m$_1||$m$_2||$m$_3$) and the responder Bob's public key pk$_B$. Here is where Alice detects HPT adversaries that have tried to forge a signature tag pretending to be Bob. Furthermore, this verification step is what provided entity protection against active adversaries since Alice will abort if the verification is not successful, thus never sending her identity.

14: The responder Bob computes the authentication tag $\tau_B$ of the public information l$_{MAC_{B2}}||$H(m$_{QKD}||\text{m}_1||\dots||$m$_4$) with the derived secret key $K_{MAC_B}$.

15: If traffic protection is enabled, the responder Bob AEAD encrypts $\tau_B$ with the traffic secret $K_{TS_B}$ and sends the output, \{$\tau_B$\}$_{K_{TS_B}}$, to the initiator Alice. Otherwise, $\tau_B$ is sent directly. 

16: The initiator Alice (decrypts the received output with the corresponding traffic secret and) verifies the received authentication tag, $\tau_B$, with the public information l$_{MAC_{B2}}$ $||$H(m$_{QKD}||\text{m}_1||\dots||$m$_4$) and the derived secret key $K_{MAC_B}$. If the adversary does not know $K_{MAC_B}$, then any forgery attempt of $\tau_B$ is detected here with high chance, as taken in account within the security proof. 

17: If the verification of the responder Bob's identity, the received signature and the received authentication tag are successful, then the initiator Alice AEAD encrypts the entity information A with the key $K_{TS_A}$, in the case that entity protection is enabled. The output \{A\}$_{K_{TS_A}}$ is sent to the responder Bob. If entity protection is not enabled, the entity information A is sent directly.

18: The responder Bob (decrypts the received output with the corresponding traffic secret and) verifies the identity A, utilizing public information such as the public key from the initiator Alice's certificate. The initiator Alice computes the signature tag $\sigma_A$ by signing the public information l$_{s_A}||$H(m$_{QKD}||$m$_1||\dots||$m$_6$) with the secret key sk$_A$. Note that the HPT adversary would have a negligible chance to create such tag, and thus be detected by Bob in the verification step.

19: If traffic protection is enabled, the initiator Alice AEAD encrypts $\sigma_A$ with the key $K_{TS_A}$ and sends the output, \{$\sigma_A$\}$_{K_{TS_A}}$, to the responder Bob. Otherwise, $\sigma_A$ is sent directly.

20: The responder Bob (decrypts the received output with the corresponding traffic secret and) verifies the received signature tag with public information l$_{s_A}$ $ ||$H(m$_{QKD}$$||$m$_1||\dots$$||$m$_6$) and the initiator Alice's public key pk$_A$. Here is where Bob detects HPT adversaries that have tried to forge a signature tag pretending to be Alice.

21: The initiator Alice computes the authentication tag $\tau_A$ of the public information l$_{MAC_{A2}}||$H(m$_{QKD}||\text{m}_1||\dots||\text{m}_7$) with the derived secret key $K_{MAC_A}$.

22: If traffic protection is enabled, the initiator Alice AEAD encrypts $\tau_A$ with the key $K_{TS_A}$ and sends the output ,\{$\tau_A$\}$_{K_{TS_A}}$, to the responder Bob. Otherwise, $\tau_A$ is sent directly.

23: The responder Bob (decrypts the received output with the corresponding traffic secret and) verifies the received authentication tag with the public information l$_{MAC_{A2}}$  $||$ H(m$_{QKD}||\text{m}_1||\dots||\text{m}_7$) and the derived secret key $K_{MAC_A}$. If the adversary does not know $K_{MAC_A}$, then any forgery attempt of $\tau_A$ is detected here with high chance, as taken in account within the security proof of section \ref{security}.

Once both authentication tags have been verified (step 23), the PQC secrets, sk$_A$ and sk$_B$, can be revealed to any HPT adversary without compromising the security of the now authenticated QKD material ss$_{rest}$, ss$_{QKD}$, or the concealment of traffic, including the information of entities.

Alice does not have feedback on Bob's final MAC verification as the protocol is. Thus, denial of service of the final interaction would force Alice to update the final keys whereas Bob would have the precedent values. This could be prevented by adding an authenticated acknowledgment message along with a timeout mechanism. The guarantee of liveness (i.e. the protocol finalizes with honest parties having the same values, despite an adversary interference) will not be discussed further in this work.

After this step 23 is successful, parties can derive session secrets out of the intermediate $K_1$ to be used in other cryptographic purposes, similar to in TLS~\cite{rfc8446} or Muckle~\cite{Muckle, MucklePlus2023, MuckleKEM2024}, following the same approach the secret state $SecState$ is updated, by using more unique labels. This is not explicitly written down in the protocol, since the goal is to provide authentication, confidentiality, and integrity to the classical traffic of QKD protocols in order to obtain keys given by QKD. These final keys are ITS with an epsilon security to be calculated according to the QKD and authentication security levels. The final security of the keys is given within the security proof of section \ref{security}.

24: Both parties update the secret state $SecState$ by deriving it from the intermediate key $K_1$ and the classical traffic traded so far, using a PRF.

25: Honest parties keep the unused QKD bits, ss$_{rest}$, in a pre-agreed shared secret key pool.

Once the protocol is finalized, the authenticated key ss$_{rest}$ is ready to be used in other cryptographic applications or can even be used to derive additional secret key material.

If authentication is secure, the quantum nature of QKD provides naturally forward security, i.e. compromising all the secrets of a session, including the T$_{HPT}$-term PQC secrets, sk$_A$ and sk$_B$, the secret state, $SecState$, and the kept shared bits, ss$_{rest}$, does not compromise the rest of the keys that are stored inside the shared secret key pool in the previous iterations of the protocol.

As shown in the security analysis, in section \ref{security}, the secret material returned by the PQC authentication protocol \ref{signHAKEauthFurther}, namely $SecState$ and ss$_{rest}$, are known only to the authenticated parties (with high probability, 1 - a negligible quantity determined in the security proof), the initiator Alice and the responder Bob, if the T$_{HPT}$-term HPT security assumption (Definition \ref{def9}) applied to the PQC algorithm holds while the authentication protocol is in an active state against an HPT adversary. In this case, Definition \ref{def2.4} makes the initial QKD key $K_{QKD}$ and its segments ITS and $(\varepsilon_\text{QKD}+\varepsilon^\text{PQC}_\text{auth})$-secure. Because the assumption that AEAD is long-term HPT (Definition \ref{def10}) secure holds, the traffic protection tags are long-term HPT secure as well. This implies that an unbounded adversary would be able to compromise the traffic secrets, $K_{TS_A}$ and $K_{TS_B}$, and trace back until compromising the old, already used, secret state $SecState$ (step 0) and ss$_{QKD}$ from $K_1$ and $K_0$, respectively. Additionally, this compromises the updated secret state $SecState$ as well (step 24). In other words, after the protocol is finalized, the key material $ss_{QKD}$, $K_0$, $K_1$ and all derived keys, including the secret state $SecState$, are not ITS but rather long-term HPT-secure. Note that this does not affect the ITS security of the final keys derived in the present and past sessions, stored inside the shared secret key pool (step 25).

The PQC signature authentication process shown in figure \ref{signHAKEauthFurther} requires 1.5 round-trip times (RTT) and could be initiated within the last interactions of the post-processing procedure, within the unauthenticated QKD step, depending on which party has to send the certificate first, which is pre-agreed according to who needs the most entity protection against active adversaries.

Additionally, the presented protocol could be modified to cover the case where both parties have knowledge of the certificates, or when only one party needs to authenticate, as is the case for some client-server interactions.

Since the protocol takes as input the ephemeral key ss$_{QKD}$, the minimum QKD key length required to execute \ref{signHAKEauthFurther} is exactly the size of the PRF input. On the other hand, if ITS MAC authentication is desired, some shared secret key pool of considerable size should be anticipated and kept, as ITS MAC consumes key material every time an authentication tag has to be sent, on top of having enough keys to not exhaust the key pool in case of denial of service attack.

Additional redundant protection may be added, for example by initiating the protocol with an ephemeral PQC key obtained through a PQC KEM exchange. The ephemeral PQC key can be used to provide an additional layer of confidentiality to the traded QKD public information, in the same way traffic protection works within the authentication protocol from figure \ref{signHAKEauthFurther}. If PQC redundancy is added, an eavesdropper would need to execute PQC algorithms, possess QKD equipment, and know the secret state parameter, $SecState$, to carry out an active attack on the entity protection feature and identify the vulnerable party. Eventually, the active attacker would be detected with high probability by the honest party, once the PQC secret material is required to perform authentication.

The next subsection provides an example of authentication that does not rely on signatures.

\subsection{QKD authenticated with KEM then MAC}\label{KEM}

Previous implementations have found that PQC HAKE protocols based on certain KEM algorithms run with fewer computational cycles and with lower memory requirements, compared to the signature-based protocols, but with the additional cost of an extra RTT~\cite{MuckleKEM2024, KEMTLS2020}. Following the same motivation, this subsection presents an alternative approach to mono-authenticate QKD with PQC KEM. Note that KEM and signature protocols could be combined to achieve redundant protection where both of the PQC algorithms used to authenticate are built on different mathematical problems~\cite{nist_pqc}.

\begin{figure*}[!ht]
\centering
\includegraphics[width=\textwidth]{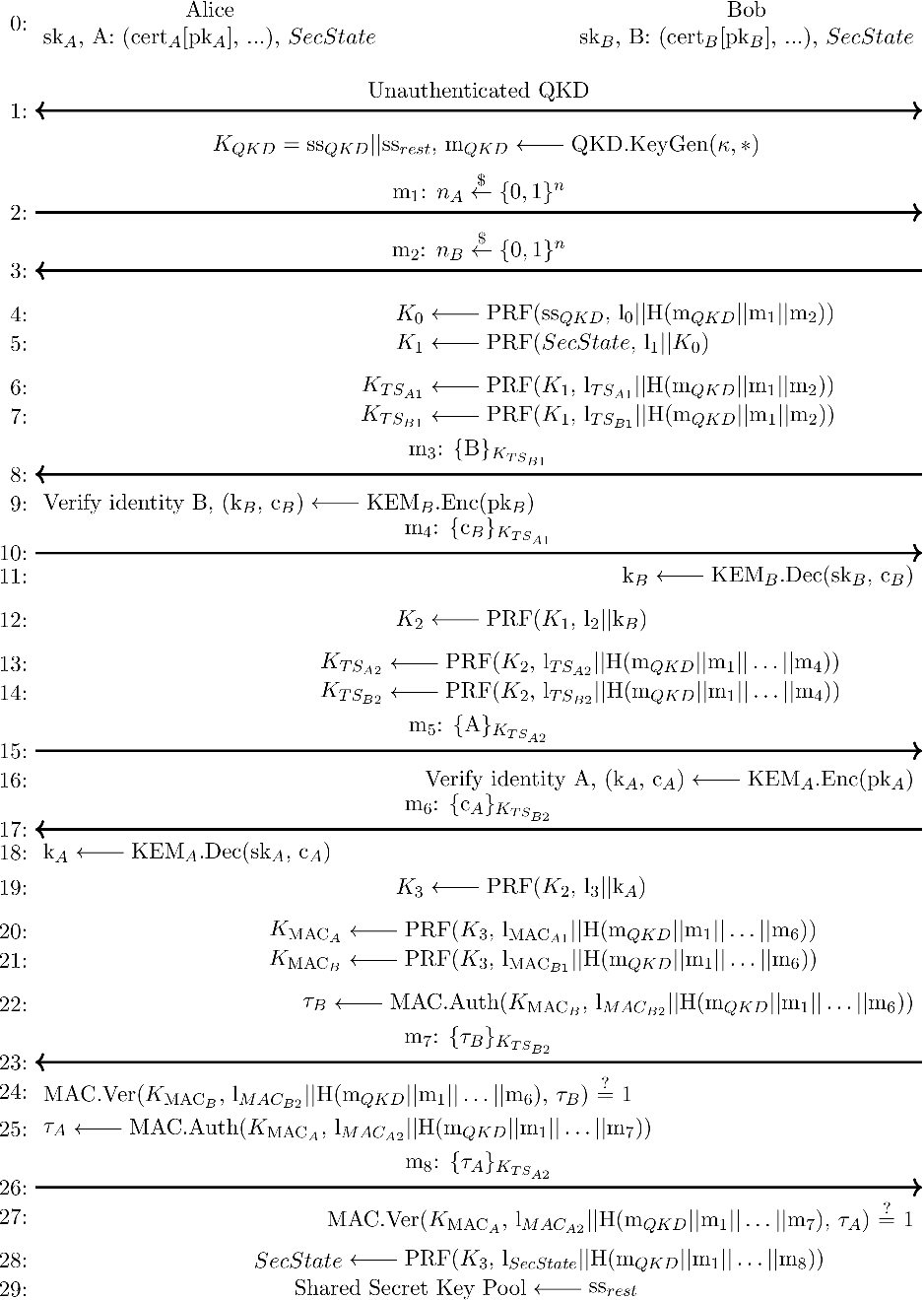}
\caption{Example of QKD mono-authentication with PQC KEM then MAC.}
\label{KEMHAKEauthFurther}
\end{figure*}

The PQC KEM-based authentication protocol follows a similar structure to the sign then MAC protocol from figure \ref{signHAKEauthFurther}. The authentication mechanism relies on T$_{HPT}$-term HPT secure KEM algorithms, rather than signatures. The protocol is presented in figure \ref{KEMHAKEauthFurther}. 

\paragraph*{Step by step.} The steps for the KEM then MAC protocol are presented below. The security analysis is presented in section \ref{security}.

0, 1, 2, 3, 4, 5, 6, 7, 8: Same as steps 0, 1, 2, 3, 4, 5, 6, 7, 10, respectively, from the signature then MAC case (figure \ref{signHAKEauthFurther}). In this case, the T$_{HPT}$-term HPT secret keys, sk$_A$ and sk$_B$, are associated with a PQC KEM algorithm. Alice and Bob can have two different KEM algorithms, denoted as KEM$_A$ and KEM$_B$ in figure \ref{KEMHAKEauthFurther}. The choice of the KEMs is arbitrary and depends on the needs of the honest parties, including security levels, diversification of PQC algorithms, performance efficiency or memory requirements.

The labels l$_i$, $i \in \{0,$ $1,$ $TS_{A1},$ $TS_{B1},$ $2,$ $TS_{A2},$ $TS_{B2},$ $3,$ $MAC_{A1},$ $MAC_{B1},$ $MAC_{A2},$ $MAC_{B2}\}$ presented inside the PRFs, correspond to unique bit strings that are public information. The labels allow for domain separation for each intermediate key, provides security against reuse attacks and binds context to the derived keys, as commented in this NIST recommendation strongly related to PRFs~\cite{KeyDerviPRF_NIST:SP800-108r1-upd1}.

9: The initiator Alice verifies the identity B using public information, such as the public key from the responder Bob's certificate. If the verification is successful, the initiator Alice computes the encapsulation c$_B$ of a random key k$_B$ by using the encapsulation algorithm KEM$_B$.Enc along with the responder Bob's public key pk$_B$.

10: If traffic protection is enabled, the initiator Alice AEAD encrypts c$_B$ with the key $K_{TS_{A1}}$ and sends the output, \{c$_B$\}$_{K_{TS_{A1}}}$, to the responder Bob. Otherwise, c$_B$ is sent directly.

11: The responder Bob (decrypts the received output with the corresponding traffic secret and) decapsulates the random secret k$_B$ from the encapsulation cipher c$_B$ using the decapsulation algorithm KEM$_B$.Dec along with the secret key sk$_B$. Without the knowledge of the secret key sk$_B$, an HPT adversary is not able to decapsulate the random secret k$_B$ with high probability as taken into account within the security proof.

12: The shared key k$_B$ is combined with the intermediate key $K_1$ using a dual PRF to obtain the intermediate key $K_2$. An adversary without k$_B$ has negligible probability of forging $K_2$.

13, 14: If the entity protection feature is enabled, the entity protection keys, $K_{TS_{A1}}$ and $K_{TS_{B1}}$, are derived from $K_2$. An adversary without $K_2$ has negligible probability of forging both.

15: If entity protection is enabled, the initiator Alice AEAD encrypts the entity information A with the key $K_{TS_{A2}}$ and sends the output, \{A\}$_{K_{TS_{A2}}}$, to the responder Bob. Otherwise, the entity information A is public knowledge.

16: The responder Bob (decrypts the received output with the corresponding traffic secret and) verifies the identity A using public information, such as the public key from the initiator Alice's certificate. If the verification is successful, the responder Bob computes the encapsulation c$_A$ of a random key k$_A$ using the encapsulation algorithm KEM$_A$.Enc along with the initiator Alice's public key pk$_A$. Note that in the case Eve/adversary is impersonating Bob, since $K_2$ has negligible chance to be forged, so is the traffic secret $K_{TS_{A2}}$, giving the entity protection against active adversaries, discussed in section \ref{general-case}.

17: If traffic protection is enabled, the responder Bob AEAD encrypts c$_A$ with the key $K_{TS_{A2}}$, and sends the output, \{c$_A$\}$_{K_{TS_{A2}}}$, to the initiator Alice. Otherwise, c$_A$ is sent directly.

18: The initiator Alice (decrypts the received output with the corresponding traffic secret and) decapsulates the random secret k$_A$ from the encapsulation cipher c$_A$ using the decapsulation algorithm KEM$_A$.Dec along with the secret key sk$_A$. Without the knowledge of the secret key sk$_A$, an HPT adversary is not able to decapsulate the random secret k$_A$ with high probability, as taken into account within the security proof.

19: The shared key k$_A$ is combined with the intermediate key $K_2$ using a dual PRF to obtain the intermediate key $K_3$. An adversary without k$_A$ has negligible probability of forging $K_3$.

20, 21: The MAC keys, $K_{MAC_A}$ and $K_{MAC_B}$, are derived from $K_3$. An adversary without $K_3$ has negligible probability of forging both.

22: The responder Bob computes the authentication tag $\tau_B$ of the public information l$_{MAC_{B2}}||$H(m$_{QKD}||\text{m}_1||\dots||\text{m}_6$) with the derived secret key $K_{MAC_{B}}$. 

23: If traffic protection is enabled, the responder Bob AEAD encrypts $\tau_B$ with the key $K_{TS_{B2}}$ and sends the output, \{$\tau_B$\}$_{K_{TS_{B2}}}$, to the initiator Alice. Otherwise, $\tau_B$ is sent directly.

24: The initiator Alice (decrypts the received output with the corresponding traffic secret and) verifies the received authentication tag with the public information l$_{MAC_{B2}}$ $||$ H(m$_{QKD}||\text{m}_1||\dots||\text{m}_6$) and the derived secret key $K_{MAC_B}$. If an HPT adversary tried to forge the MAC tag, Alice would detect this within the verification step with a high probability, as taken into account within the security proof.

25: The initiator Alice computes the authentication tag $\tau_A$ of the public information l$_{MAC_{A2}}$$||$H(m$_{QKD}$$||\text{m}_1||\dots||\text{m}_7$) with the derived secret key $K_{MAC_A}$.

26: If traffic protection is enabled, the initiator Alice AEAD encrypts $\tau_A$ with the key $K_{TS_{A2}}$ and sends the output, \{$\tau_A$\}$_{K_{TS_{A2}}}$, to the responder Bob. Otherwise, $\tau_A$ is sent directly.

27: The responder Bob (decrypts the received output with the corresponding traffic secret and) verifies the received authentication tag with the public information l$_{MAC_{A2}}||$H(m$_{QKD}$ $||\text{m}_1||\dots||\text{m}_7$) and the derived secret key $K_{MAC_A}$. If an HPT adversary tried to forge the MAC tag, Bob would detect this within the verification step with a high probability, as taken into account within the security proof.

28: Both parties update the secret state $SecState$ by deriving it from the intermediate key $K_3$ and the classical traffic traded so far.

29: Honest parties keep the unused QKD bits, ss$_{rest}$, in a pre-agreed shared secret key pool.

Entity protection is provided in the same way as in the signature-based protocol. The comments specified inside the signature then MAC version also apply here, since they are agnostic of which PQC algorithm is used to authenticate.

As shown in the security analysis, in section \ref{security}, the secret material returned by the PQC authentication protocol \ref{signHAKEauthFurther}, namely $SecState$, and ss$_{rest}$, are known only to the authenticated parties (with high probability), the initiator Alice and the responder Bob, if the T$_{HPT}$-term HPT security assumption (Definition \ref{def9}) applied to the PQC algorithm holds while the authentication protocol is in an active state against HPT adversaries, as shown in the security proof, section \ref{security}. In this case, Definition \ref{def2.4} makes the initial QKD key and its segments ITS and $(\varepsilon_\text{QKD}+\varepsilon^\text{PQC}_\text{auth})$-secure. Given the assumption that AEAD is long-term HPT (Definition \ref{def10}) secure holds, the entity protection tags are long-term HPT secure. This implies that an unbounded adversary would be able to compromise the entity protection keys, $K_{TS_{A1}}$ and $K_{TS_{B1}}$, and trace back until compromising the old, already used, secret state $SecState$ (step 0) and ss$_{QKD}$ from $K_1$ and $K_0$, respectively. In other words, after the protocol is finalized, the QKD key ss$_{QKD}$, and the secret state $SecState$ (step 0), are not ITS but rather long-term HPT-secure. The remaining intermediate keys, $K_2$ and $K_3$, are long-term HPT secure whereas the PQC secrets k$_A$ and k$_B$ are T$_{HPT}$-term HPT secure. Since the intermediate key $K_3$ is long-term secure, the derived secret state $SecState$ (step 28) is long-term secure too. Note that this does not affect the ITS security of the final keys derived in the present and past sessions, stored inside the shared secret key pool (step 29).

Similarly with the signature protocol, if ITS MAC authentication is desired for the next QKD iterations, the shared secret key pool has to be filled according to some agreed size between the honest parties. Indeed, once the shared secret key pool is filled sufficiently, (ITS or non-ITS) authentication without relying on PQC in posterior QKD iterations may be undertaken.

Since this work focuses rather on non-ITS primitives, the next section provides an example on how to achieve authentication using a non-ITS MAC algorithm by taking a strong inspiration from the original Muckle protocol~\cite{Muckle}.

\section{Further QKD iterations: shared key scenario}\label{protocol_MAC}

Once parties have performed PQC authentication for QKD in the first iteration successfully, authentication can stop relying on asymmetric cryptography since the shared secret key pool contains some ITS and $(\varepsilon_\text{QKD}+\varepsilon_\text{auth})$-secure keys stored inside. Authentication relying on symmetric but non-ITS, long-term HPT-secure, primitives such as most MAC functions is possible. In the same way, authentication relying on ITS primitives such as ITS-MAC algorithms is achievable. This section discusses an example for long-term HPT-secure authentication (non-ITS).

Theorem \ref{theorem1}, obtained after performing the security analysis, in section \ref{security}, shows that the security of QKD keys, that are stored within the shared secret key pool, decreases in a polynomial manner with the number of parties, sessions and iterations. This implies that the keys generated in previous QKD iterations have an advantage to be used for further authentication steps regarding the keys generated in later QKD iterations.

\begin{figure*}[!ht]
\centering
\includegraphics[width=\textwidth]{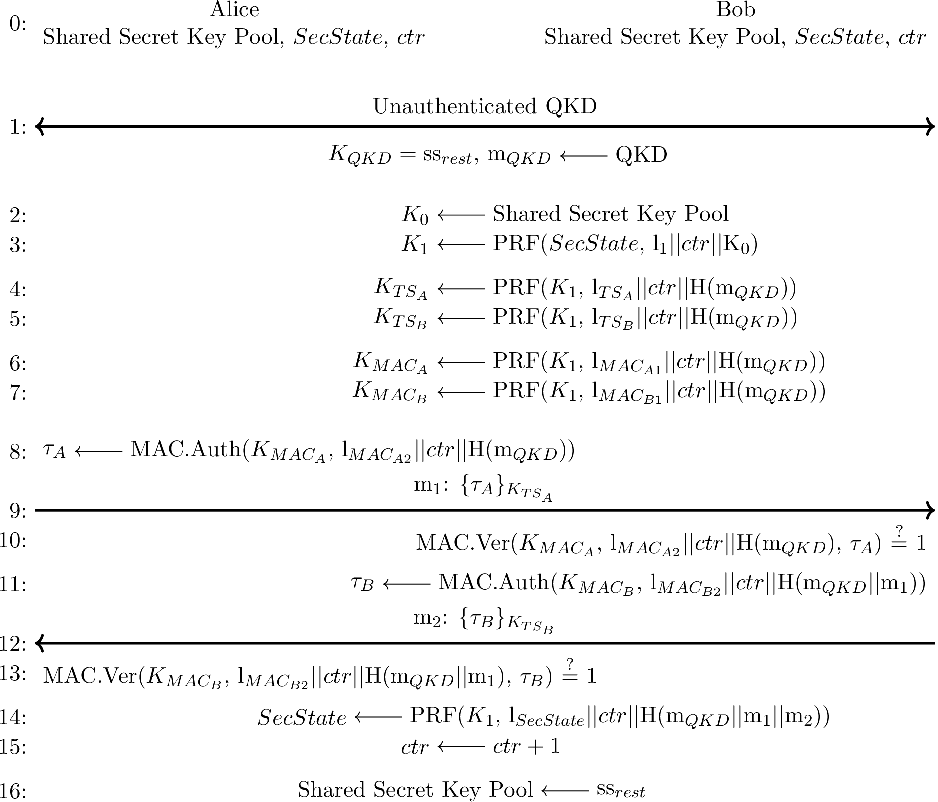}
\caption{Example of shared key-based long-term HPT-secure mono-authentication for QKD.}
\label{renewalQKDonlyAuth}
\end{figure*}

Further QKD iterations can be mono-authenticated securely thanks to the now non-empty shared secret key pool and the long-term HPT-secure MAC algorithms. An example protocol is presented in figure \ref{renewalQKDonlyAuth}.

\paragraph*{Step by step.} The steps for the shared key-based mono-authentication for QKD using long-term HPT-secure MAC are presented below. The security analysis is presented in section \ref{security}.

0: The initiator Alice and the responder Bob hold a non-empty $\varepsilon$-secure ITS shared secret key pool, where $\varepsilon$ is estimated in previous steps, and the secret state parameter $SecState$, given by previous iterations, or initialized at a known, and public, value with the condition that the shared secret key pool has enough key material to run the presented MAC protocol.
They also share a public counter $ctr$, initialized at 0. In this occasion public key infrastructure is not required since honest parties share secrets that no adversary has (with negligible probability), hence there is no need to have certificates. 

1: The protocol starts with the initiator Alice and the responder Bob performing an unauthenticated QKD protocol following Definition \ref{def2.3}. Both parties obtain the public information m$_{QKD}$ and the corresponding unauthenticated QKD bit string $K_{QKD} =\text{ss}_{rest}$.

2: Honest parties agree on a shared secret $K_0$, taken from the shared secret key pool. Each key has a security level associated and most likely an unique ID to differentiate them easily. Earlier QKD keys present a better security than the ones distilled in later QKD iterations: this could be a selection criteria.

3: The obtained key $K_0$ is combined with the secret state $SecState$ using dual PRFs to derive the intermediate keys $K_1$. An HPT adversary without the key $K_0$ has negligible probability to forge the intermediate key $K_1$.

The labels l$_i$, $i \in \{1, TS_A, TS_B, MAC_{A1}, MAC_{B1}, MAC_{A2}, MAC_{B2}, SecState\}$, presented inside the cryptographic primitives, correspond to unique (with each other) bit strings that are public information. The labels, combined with the counter $ctr$, allow for domain separation for each intermediate key, provide security against reuse attacks and bind context to the derived keys, as commented in this NIST recommendation strongly related to PRFs~\cite{KeyDerviPRF_NIST:SP800-108r1-upd1}.

4, 5: If traffic protection is enabled, the traffic secrets, $K_{TS_A}$ and $K_{TS_B}$, are derived from $K_1$. An HPT adversary without the intermediate key $K_1$ has negligible probability to forge both traffic secrets.

Traffic secrets are needed when honest parties want confidentiality regarding the message structure or metadata. Otherwise, the derivation of traffic secrets is optional as the MAC tags, $\tau_A$ and $\tau_B$, along with the shared keys, $K_0$ and $K_1$, are sufficient to provide integrity and authenticity.

Traffic protection is maintained against active HPT adversaries for both honest parties since knowledge of $K_0$ and the secret state parameter is required to be able to forge the traffic secrets.

6, 7: The MAC keys, $K_{MAC_A}$ and $K_{MAC_B}$, are derived from $K_1$. An HPT adversary without the intermediate key $K_1$ has negligible probability to forge both MAC secrets.

8: The initiator Alice computes the long-term HPT-secure authentication tag $\tau_A$ of the public information l$_{MAC_{A2}}||ctr||$ H(m$_{QKD})$ with the derived secret key $K_{MAC_A}$. An HPT adversary without the MAC key $K_{MAC_A}$ has negligible probability to forge the MAC tag.

9: If traffic protection is enabled, the initiator Alice AEAD encrypts $\tau_A$ with the key $K_{TS_A}$ and sends the output, \{$\tau_A$\}$_{K_{TS_A}}$, to the responder Bob. Otherwise, $\tau_A$ is sent directly.

10: The responder Bob (decrypts the received output with the corresponding traffic secret and) verifies the received authentication tag with the public information l$_{MAC_{A2}}||ctr$ $||$H(m$_{QKD}$) and the derived secret key $K_{MAC_A}$. An impersonation from an HPT adversary that tried to forge the MAC tag $\tau_A$ is detected with high probability, as taken into account within the security proof.

11: The responder Bob computes the long-term HPT-secure authentication tag $\tau_B$ of the public information l$_{MAC_{B2}}||ctr||$H(m$_{QKD}||\text{m}_1)$ with the derived secret key $K_{MAC_B}$. An HPT adversary without the MAC key $K_{MAC_B}$ has negligible probability to forge the MAC tag.

12: If traffic protection is enabled, the responder Bob AEAD encrypts $\tau_B$ with the key $K_{TS_B}$ and sends the output, \{$\tau_B$\}$_{K_{TS_B}}$, to the initiator Alice. Otherwise, $\tau_B$ is sent directly.

13: The initiator Alice (decrypts the received output with the corresponding traffic secret and) verifies the received authentication tag with the public information l$_{MAC_{B2}}||ctr||$ H(m$_{QKD}||\text{m}_1$) and the derived secret key $K_{MAC_B}$. An impersonation from an HPT adversary that tried to forge the MAC tag $\tau_B$ is detected with high probability, as taken in account within the security proof.

14: Both parties update the secret state $SecState$ by deriving it from the intermediate key $K_1$ and the classical traffic traded so far.

15: The counter $ctr$ is updated with $ctr+1$.

Since the key extracted from the shared secret key pool provides randomness, uniqueness has to be included to avoid replay attacks. The counter $ctr$ provides the uniqueness required, so nonces are not necessary in this scheme.

16: Honest parties keep the now authenticated QKD bits, ss$_{rest}$, in the pre-agreed shared secret key pool.

Note that in this protocol, there is no involvement of ephemeral key. Adding an ephemeral key would add another layer of protection regarding post-compromise security. If it is assumed that the HPT adversary Eve has knowledge of the shared secret key pool and the secret state parameter $SecState$ in one iteration ``a", and if she becomes passive for at least an iteration ``b" $>$ ``a", then the parameter $SecState$ would be updated by the honest parties to a value unknown to Eve, because of the ephemeral key. Thus the adversary Eve would be detected at the MAC verification step by honest parties.

The choice of the non-ITS MAC function may depend on a security level parameter and the secret key consumption constraint, if there is any such constraint. The obvious choice is to use a MAC function with a time complexity that scales well with the length input, in addition to requiring little secret key material to operate. Possible candidates would be Hash-based MAC (HMAC)~\cite{rfc6234-hashbasedmac}, Cipher-based Message Authentication Code (CMAC)~\cite{CMAC-NIST-SP-800-38B, CMAC-AES-rfc4493} or Galois Message Authentication Code (GMAC)~\cite{GaloisMAC-NIST-SP-800-38D, Galois-MAC-AES-rfc4543}.

These MAC algorithms allow reuse of the shared secret key $K_0$ without compromising the security of the schemes. In that scenario, the key eventually expires, either because it has been too long since the last update, or because it has been used too many times. Renewal of the key is done by simply repeating step 2: a new QKD key is extracted from the shared secret key pool. For the security proof of the non-ITS MAC-based protocol, $K_0$ is assumed to be refreshed every time the protocol is run, to facilitate some post-compromise security properties and freshness. If the shared secret key pool has insufficient bits, the PQC-based protocol would need to be run again.

\section{Hybrid Authenticated Key Exchange Framework}\label{framework}

To show that the presented examples (Figures \ref{signHAKEauthFurther}, \ref{KEMHAKEauthFurther} and \ref{renewalQKDonlyAuth}) are secure, an adapted version of the HAKE framework introduced in~\cite{Muckle} and recalled in~\cite{MucklePlus2023, MuckleKEM2024} is used. This section recalls and adapts the tools required to execute the HAKE framework in the context of this work. The algorithmic description of the security model, Exp$_{\Pi,n_P,n_S,n_T}^{\text{HAKE},\text{clean},\mathcal{A}}(\kappa)$, is reused and is not recalled within this work, however it can be found in~\cite{Muckle}, figure 5 of Appendix C. To investigate security, the indistinguishability experiment is played between an HPT challenger $\mathcal{C}$ and two adversaries $\mathcal{A} = \{\mathcal{A_{HPT}}, \mathcal{A}_{unbounded}\}$, where $\mathcal{A_{HPT}}$ and $\mathcal{A}_{unbounded}$ correspond to an HPT (Definition \ref{def1.1}) and conditioned unbounded (Definition \ref{def1.2}) adversary, respectively.

The presented examples do not use classical cryptography (non-PQC), however the tools to deal with classical primitives are also presented for completeness. Protocols that combine classical cryptography with different PQC algorithms follow a practice to prevent that if one of the algorithms believed to be secure is found to be compromised, then said combined protocol's security would still hold. This is useful, especially nowadays where PQC algorithms are under intense scrutiny and testing to check their practical security.

\subsection{Secret Key Generation}\label{framework_key_gen}

\paragraph*{Recall and adaptation of the HAKE framework.}

The HAKE framework categorizes secret key generation into long-term and ephemeral. Long-term keys are generated once and reused in every execution of a protocol, while ephemeral keys are freshly generated at certain stages within each protocol execution. Furthermore, these are divided into the following sub-categories:

\begin{itemize}
    \item Post-quantum asymmetric secret generation. This includes all public-secret key pair generation given by any PQC algorithm. The algorithm that generates these secrets for a long-term manner is denoted as LQKeyGen, while EQKeyGen denotes the ephemeral version.
    
    \item Classical asymmetric secret generation. This includes all public-secret key pair generation given by any classical algorithm, known to be insecure against quantum algorithms. The algorithm that generates these secrets for a long-term manner is denoted as LCKeyGen, while ECKeyGen denotes the ephemeral version.
    
    \item Symmetric secrets. This includes pre-shared keys and keys distilled from an already working QKD network. The algorithm that generates these secrets for a long-term manner is denoted as LSKeyGen, while ESKeyGen denotes the ephemeral version.
\end{itemize}

The set of algorithms that produces the key material are denoted as:

\begin{itemize}
    \item XYKeyGen($\kappa$) $\stackrel{\$}{\rightarrow}$ (pk, sk), where X $\in \{L, E\}$, Y $\in \{Q,C\}$, pk is public key, sk is secret key and $\kappa$ is a security parameter.
    \item XSKeyGen($\kappa$) $\stackrel{\$}{\rightarrow}$ (ss, ssid), where X $\in \{L, E\}$, ss is shared secret (or symmetric secret), ssid is the identity parameter of the shared secret ss and $\kappa$ is a security parameter.
\end{itemize}

\paragraph*{Adaptations, added conditions and comments.}

Long-term Post-Quantum secret key generation is indirectly replaced by T$_{HPT}$-term Post-Quantum secret key generation, following Definition \ref{def9}. This distinction evokes that the security of the PQC algorithm  secrets has a temporal bound T$_{HPT}$ that relates to the bounded resources of the considered HPT adversaries, required to make the QKD iterations secure. Additionally, the presented protocols do not use ephemeral Post-Quantum secrets, thus EQKeyGen is never executed (or returns $\perp$). However, ephemeral asymmetric post-quantum algorithms can be implemented to achieve redundant security properties, similarly to the approach in the Muckle series~\cite{Muckle, MucklePlus2023, MucklePlusPlus, MuckleKEM2024}.

Furthermore, classical asymmetric secret generation is never used within the presented protocols, thus LCKeyGen and ECKeyGen are never executed (or return $\perp$). However, classical asymmetric algorithms can be added to provide redundant security properties, as discussed at the start of this section.

The unauthenticated QKD keys from the first steps of the PQC-based protocols are considered symmetric ephemeral keys, thus, if ephemeral QKD keys are involved, then ESKeyGen is executed via QKD.KeyGen in the presented protocols.

Since the protocols presented here are designed to require no pre-shared key, LSKeyGen is not executed (or returns $\perp$). Once any of the presented protocols (Figures \ref{fig:generalAKE}, \ref{signHAKEauthFurther}, \ref{KEMHAKEauthFurther} and \ref{renewalQKDonlyAuth}) are securely finalized, parties obtain a non empty shared secret key pool given by the unused QKD bits. These QKD bits stored inside the shared secret key pool are considered to be equivalent to session keys. Additionally, if the session keys are used to perform the MAC-based protocol, the used keys are no longer considered as session keys, but rather as a different element (shared secret key pool keys), with the condition that the session keys were securely derived.

Next, the HAKE execution environment is recalled and adapted.

\subsection{Execution Environment}

\paragraph*{Recall and adaptation of the HAKE framework.}

The environment is composed of $n_P$ parties $P_1,\dots,P_{n_P}$ capable of interacting with each other. Each party is able to run $n_S$ sessions of a general key-exchange protocol denoted by $\Pi$. Each session consists of $n_T$ stages: i.e., for a new stage, the protocol $\Pi$ is run once again while maintaining the per-session parameters from the previous stage.

The per-session parameters are denoted as $\pi$ and described as:

\begin{itemize}
    \item $\rho \in \{\text{init},\text{resp}\}$: The role of the party in the current session, i.e., initiator or responder.
    \item $pid \in \{1,\dots,n_P,\star\}$: The intended communication partner, where $\star$ denotes unspecified. For the unspecified case, the parameter can be updated once to a specified party $\in \{1,\dots,n_P\}$.
    \item $stid \in [n_T]$: The current or most recently completed stage of the session.
    \item $\alpha \in \{\text{active},\text{accept}, \text{reject},\perp\}$: The status of the session, initialized with $\perp$.
    \item $m_i[stid] \in \{0,1\}^* \cup \{\perp\}$, $i \in \{s,r\}$: Array of the concatenation of messages sent (if $i=s$) or received (if $i=r$) by the session in each stage. Initialized with $\perp$ and indexed by the stage identifier $stid$.
    \item $k[stid] \in \{0,1\}^* \cup\{\perp\}$: Array of the session keys from each stage. The used session keys $k[i]$ to run the MAC-based protocol $\Pi_{MAC}$ are removed and deleted, where the index $i$ is the closest and non-used integer to $stid$ such that $i<stid$. Initialized with $\perp$ and indexed by the stage identifier $stid$.
    \item $qk[stid] \in \{0,1\}^* \cup\{\perp\}$: Array of the T$_{HPT}$-term post-quantum secrets. Initialized with $\perp$ and indexed by the stage identifier $stid$
    \item $exk[stid] \in \{0,1\}^* \cup \{\perp\}$, $x \in \{q, c, s\}$: Array of the ephemeral post-quantum (if $x=q$), classical (if $x=c$) and symmetric (if $x=s$) secret keys used by the session in each stage. Initialized with $\perp$ and indexed by the stage identifier $stid$.
    \item $pss[stid] \in \{0,1\}^* \cup \{\perp\}$: Per-stage secret state ($SecState$) that is established during protocol execution for use in the following stage. Initialized at an arbitrary, and public, value and indexed by the stage identifier $stid$.
    \item $sskp[stid] \in \{0,1\}^* \cup \{\perp\}$: Array of the keys extracted from $k[i]$ to run the MAC-based protocol ($\Pi_{MAC}$, defined below), where the index $i$ is the closest and non-used integer to $stid$ such that $i<stid$. Initialized with $\perp$ and indexed by the stage identifier $stid$.
    \item $st[stid]\in\{0,1\}^*$: Any additional state used by the session in each stage. Indexed by the stage identifier $stid$.
\end{itemize}

The identifier of the s-th instance of the protocol $\Pi$ being run by party $P_i$, and the collection of per-session variables maintained for the s-th instance of the protocol $\Pi$ being run by party $P_i$, are both denoted by $\pi^s_i$ in the following.

In the context of this work, three examples of protocols can be run in an alternative manner. The signature-based, the KEM-based and the MAC-based protocols from figures \ref{signHAKEauthFurther}, \ref{KEMHAKEauthFurther} and \ref{renewalQKDonlyAuth} respectively, are denoted as $\Pi_{\Sigma}$, $\Pi_{KEM}$ and $\Pi_{MAC}$ respectively. In the following discussion, $\Pi$ corresponds to any element of the set $\{\Pi_\Sigma,\Pi_{KEM},\Pi_{MAC}\}$. Note that the first iteration of QKD has to be run with a PQC-based protocol, since the shared secret key pool is empty, thus rendering insecure the usage of the MAC-based protocol.

The key-exchange protocol $\Pi$ is represented as a tuple of algorithms $(f,$ $EQKeyGen,$ $ECKeyGen,$ $ESKeyGen,$ $LQKeyGen,$ $LCKeyGen,$ $LSKeyGen)$, where:

\begin{itemize}
    \item $f(\kappa$, pk$_i$, sk$_i$, ssid$_i$, ss$_i$, $\pi$, $m$) $\rightarrow$ ($m'$, $\pi'$) is a probabilistic algorithm that takes a security parameter $\kappa$, the set of long-term asymmetric key pairs pk$_i$, sk$_i$ of the party $P_i$, a collection of per-session variables $\pi$ and an arbitrary bit string $m\in\{0,1\}^* \cup \{\emptyset\}$ as input, and then outputs a response $m'\in\{0,1\}^* \cup \{\emptyset\}$ and an updated per-session state $\pi'$, behaving as an honest protocol implementation.
    \item The KeyGen algorithms are the ones described in subsection \ref{framework_key_gen}.
\end{itemize}

To investigate security, the challenger $\mathcal{C}$ runs a total number of $n_P$ times the pertinent secret key generation algorithms, to generate public-secret key pairs (pk$_i$, sk$_i$) and/or symmetric keys with the corresponding identifiers (ss, ssid), for each party $P_i \in \{P_1,\dots,P_{n_P}\}$. All the public keys and symmetric key identifiers, pk$_i$ and ssid, are shared with the adversary $\mathcal{A}$. The challenger $\mathcal{C}$ then samples a random bit $b \stackrel{\$}{\leftarrow} \{0,1\}$ and interacts with the adversary following the adversarial queries defined below. After terminating, the adversary $\mathcal{A}$ outputs a guess $d$ of the challenger bit $b$. The adversary wins the HAKE indistinguishability experiment, defined in Figure 5 of the Appendix C of~\cite{Muckle}, if $d = b$, and if the test session $\pi$ satisfies a cleanness predicate, also recalled below.

Since the PQC secrets are T$_{HPT}$-term secure, the challenger is allowed to run LQKeyGen again when required to obtain new PQC secret key material. Additionally, the adversaries are allowed to ask for the PQC secrets once the sessions are not in an active status. The challenger eventually only runs $\Pi$ with certain PQC secrets if CorruptQK, defined in subsection \ref{framework_adv_inter}, has not been issued before reaching an accept status.

Furthermore, the challenger $\mathcal{C}$ manages a set of corruption registers that keeps track of which secrets have been corrupted. Most of the registers are formally defined in Appendix C of~\cite{Muckle} and new ones would follow the same structure.

\paragraph*{Adaptations, added conditions and comments.}

In the context of the protocols presented in this paper, the session parameter $k[stid]$ (the array containing the session keys) corresponds to the QKD bits that are stored in the shared secret key pool once the protocols end without aborting. Note that once $\Pi_{MAC}$ is run, the extracted session keys, $K_0$, are moved to a different array, $sskp[stid]$ to prove some post-compromise security property and to avoid re-usage of the same key. 

The array $qk[stid]$ is also added, to express not only the idea that the PQC secrets are T$_{HPT}$-term secure, but also that the adversaries are allowed to request the PQC secrets once the test sessions reach the accept status. Eventually, the challenger creates new PQC secret material when required before making the sessions to start a new stage.

In the same way, the arrays containing the ephemeral secret keys, $exk[stid]$, update depending on which protocols have been executed for the different stages:

\begin{itemize}
    \item If the signature-based protocol (Figure \ref{signHAKEauthFurther}) is executed in stage $stid$, $esk[stid]$ is updated with the symmetric ephemeral QKD key ss$_{QKD}$.
    \item If the KEM-based protocol (Figure \ref{KEMHAKEauthFurther}) is executed in stage $stid$, $eqk[stid]$ is updated with the post-quantum ephemeral keys, k$_A$ and k$_B$, and $esk[stid]$ is updated with the symmetric ephemeral QKD key ss$_{QKD}$.
    \item If the MAC-based protocol (Figure \ref{renewalQKDonlyAuth}) is executed in stage $stid$, $exk[stid]$ is updated with some character that specifies that there is no ephemeral key computed in stage $stid$. This is because the provided example does not use ephemeral keys, however, as mentioned before, an ephemeral QKD key $ss_{QKD}$ may be derived to provide a stronger version of post-compromise security.
\end{itemize}

The session parameter $st[stid]$ is used to contain the information regarding which protocol has been run in stage $stid$; either $\Pi_\Sigma$, $\Pi_{KEM}$ or $\Pi_{MAC}$ in this case. If $\Pi_{MAC}$ is run in stage $stid$, this session parameter also provides information to determine which key $K_0$ from $k[stid]$ was extracted to execute the protocol: this could be done by storing some key identity parameter. This array is not relevant to the security proof and is assumed to be known by all adversaries $\mathcal{A}$.

Now that the execution environment has been recalled and adapted, the adversarial interactions are recalled and adapted next. 

\subsection{Adversarial Interaction}\label{framework_adv_inter}

The considered adversary able to interact with the execution environment is an HPT adversary. Additionally, the conditioned unbounded adversary is allowed to interact in a limited way with both the execution environment and the HPT adversary.

\paragraph*{Recall and adaptation of the HAKE framework.}

The adversaries $\mathcal{A} = \{\mathcal{A_{HPT}},$ $ \mathcal{A}_{unbounded}\}$ are allowed to act as stated in Definitions \ref{def1.1} and \ref{def1.2}. That is, the HPT adversary has complete control of the classical and quantum communication network, able to modify, inject, delete or delay classical or quantum messages/signals at all times.
The unbounded adversary has complete control of the quantum communication network, able to modify, inject, delete or delay quantum signals at all times. Additionally, the unbounded adversary is able to only read and delay the classical traffic. 

Following these conditions, the HPT adversary $\mathcal{A_{HPT}}$ can interact with the challenger $\mathcal{C}$ following the queries:

\begin{itemize}
    \item Create$(i,j,role) \rightarrow\{(s),\perp\}$: The adversary initializes a new session owned by party $P_i$, with role $role$, and intended partner party $P_j$. If a session $\pi^s_i$ has already been created, the challenger returns $\perp$. Otherwise $(s)$ is returned.
    \item Send$(i,s,m) \rightarrow \{m',\perp\}$: The adversary sends a message $m$ to an active session ($\pi^s_i.\alpha=$ active) for protocol execution and receives the honest output $m'$ from the challenger, computed with $\Pi.f(\kappa$, pk$_i$, sk$_i$, ssid$_i$, ss$_i$, $\pi$, $m$). The previous local (challenger side) information of the session $\pi^s_i$ is updated to $\pi^{s'}_i$.
    \item Reveal$(i,s,t)$: The adversary has access to the session keys computed in session $\pi^s_i$ ($\pi^s_i.k[t]$) where the targeted session is in the accepted state ($\pi^s_i.\alpha=$ accept). Otherwise the challenger returns $\perp$. The challenger updates the corruption registers accordingly.
    \item Test$(i,s,t) \rightarrow \{k_b,\perp\}$: The adversary is provided by the challenger with the real session key ($b = 1$) or a random session key ($b=0$), where the targeted session is in the accepted state ($\pi^s_i.\alpha=$ accept). Otherwise the challenger returns $\perp$.
    \item CorruptXY$(i, s, t) \rightarrow \{key, \perp\}$: The adversary is provided by the challenger with the long-term XY $\in$ $\{SK,$ $QK,$ $CK\}$ keys from party $P_i$. The parameters $s$ and stage $t$ are indicated only for XY $= QK$. If the keys have been corrupted previously, then $\perp$ is returned. Specifically:
    \begin{itemize}
        \item CorruptSK$(i)$: Reveals the long-term symmetric secrets (if available) of party $P_i$. This also affects the peer party of $P_i$.
        \item CorruptQK$(i, s, t)$: Reveals the post-quantum T$_{HPT}$-term keys used in stage $t$, $\pi^s_i.qk[t]$ (if available).
        \item CorruptCK$(i)$: Reveals the classical long-term keys (if available) of party $P_i$.
    \end{itemize}
    \item CompromiseXY$(i,s,t) \rightarrow \{key, \perp\}$: The adversary is provided by the challenger with the ephemeral XY $\in$ $\{QK,$ $CK,$ $SK,$ $SS,$ $KP\}$ keys created during the session $\pi^s_i$ prior to stage $t$. If the ephemeral key has already been compromised, then $\perp$ is returned. Specifically:
    \begin{itemize}
        \item CompromiseQK$(i,s,t)$: Reveals the ephemeral post-quantum keys, $\pi^s_i.eqk[t]$ (if available).
        \item CompromiseCK$(i,s,t)$: Reveals the ephemeral classical keys, $\pi^s_i.eck[t]$ (if available).
        \item CompromiseSK$(i,s,t)$: Reveals the ephemeral symmetric keys, $\pi^s_i.esk[t]$ (if available).
        \item CompromiseSS$(i,s,t)$: Reveals the per-session state, $\pi^s_i.pss[t]$ ($SecState$).
        \item CompromiseKP$(i,s,t)$: Reveals the symmetric secret extracted from the session keys (shared secret key pool) to execute $\Pi_{MAC}$, $\pi^s_i.sskp[t]$ ($K_0$).
    \end{itemize}
\end{itemize}

Additionally, when the Test query is issued, the HPT adversary $\mathcal{A}$ interacts with the conditioned unbounded adversary $\mathcal{A}_{unbounded}$ with the following query:

\begin{itemize}
    \item RevealITS$(i, s, t) \rightarrow \{\text{Reveal}(i,s,t), \perp\}$: If the conditioned unbounded adversary finds a way to compromise the QKD key $K_{QKD}$, distilled in session $\pi^s_i$ at stage $t$, the conditioned unbounded (or the HPT, does not matter) adversary $\mathcal{A}$ is allowed to issue the Reveal$(i,s,t)$ query to the challenger $\mathcal{C}$. Otherwise, $\perp$ is returned.
\end{itemize}

The algorithmic description of RevealITS is an if statement that activates the Reveal query in the case the conditioned unbounded adversary compromises the ITS security of the QKD keys; else nothing is returned. This RevealITS query gives virtual access to unbounded resources for the HPT adversary in order to test the ITS security of the final QKD keys, also named as the session keys in this context.

The query RevealITS returns the session keys if the QKD step has failed in the $\varepsilon_{QKD}(\kappa)$ sense. In other words, the probability that RevealITS returns the session keys is $\varepsilon_{QKD}(\kappa)$. Furthermore, if the unauthenticated has been bad implemented, RevealITS reveals always the session keys (QKD keys). Note that the conditioned unbounded adversary is able to interact with the quantum channel and with the public QKD traffic but not with the authentication steps.

Furthermore, the probability that the QKD keys are not ITS is equivalent to the probability that authentication is not secure (named $\varepsilon_{auth}$) or QKD fails (named $\varepsilon_{QKD}$), giving equation \ref{eq:RevealITS}.

\begin{equation}
    \text{Pr}(\text{QKD keys not ITS}) = \varepsilon_{auth} + \varepsilon_{QKD} - \varepsilon_{auth}\varepsilon_{QKD} \leq \varepsilon_{auth} + \varepsilon_{QKD}
    \label{eq:RevealITS}
\end{equation}

Where $\varepsilon_{auth}$ is indirectly estimated in the security analysis, in section \ref{security}, and $\varepsilon_{QKD}$ is defined in Definition \ref{def2.2}.

Note that there is not a single way to test the ITS security of the final QKD keys. For example, the unbounded adversary could be replaced by the challenger itself who is honest regarding the ITS security. In that case, the query RevealITS would be requested to the challenger $\mathcal{C}$, rather than the conditioned unbounded adversary.

\paragraph*{Adaptations, added conditions and comments.}

The traffic protection keys and the corresponding encryption are not effective against the unbounded adversary, because the confidentiality property of the presented protocols is not ITS, but is effective against bounded HPT adversaries. Since authentication, confidentiality and integrity are being tested against HPT adversaries, the conditioned unbounded adversary does not provide with the trivially compromised HPT-secure keys to the HPT adversary once the RevealITS query is issued.

The Corrupt and Compromise queries can be called for the three presented protocols $\Pi_{\Sigma}$, $\Pi_{KEM}$ and $\Pi_{MAC}$, yielding the following:

\begin{itemize}
    \item $\Pi_{\Sigma}:$
    \begin{itemize}
        \item CorruptQK reveals the T$_{HPT}$-term secrets of the PQC signature.
        \item CorruptSK and CorruptCK reveal nothing.
        \item CompromiseSK reveals the ephemeral QKD key.
        \item CompromiseSS reveals the SecState parameter.
        \item CompromiseQK, CompromiseSK and CompromiseKP reveal nothing.
    \end{itemize}
    \item $\Pi_{KEM}:$
    \begin{itemize}
        \item CorruptQK reveals the T$_{HPT}$-term secrets of the PQC KEMs.
        \item CorruptSK and CorruptCK reveal nothing.
        \item CompromiseSK reveals the ephemeral QKD key.
        \item CompromiseSS reveals the SecState parameter.
        \item CompromiseQK, CompromiseSK and CompromiseKP reveal nothing.
    \end{itemize}
    \item $\Pi_{MAC}:$
    \begin{itemize}
        \item CorruptQK, CorruptSK and CorruptCK reveal nothing.
        \item CompromiseKP reveals the symmetric secret $K_0$.
        \item CompromiseSS reveals the SecState parameter.
        \item CompromiseSK, CompromiseQK and CompromiseKP reveal nothing.
    \end{itemize}
\end{itemize}

When the CompromiseSK query cannot be issued to the challenger $\mathcal{C}$ caused by a cleanness predicate condition, the HPT adversary $\mathcal{A_{HPT}}$ is allowed to interact with the conditioned unbounded adversary once the unauthenticated QKD step has finalized, as stated in Definition \ref{def1.2}, and obtain the ephemeral QKD keys with $\varepsilon_{QKD}(\kappa)$ probability. 

Note that the query CorruptQK is always successfully issued after waiting the estimated time T$_{HPT}$.

If the Test and Reveal queries are issued together at any time, then the adversary trivially compromises the security of the protocol, as that is equivalent to explicit issue of the shared secret key pool. To avoid this kind of trivial compromise, the cleanness predicates are defined in subsection \ref{framework_clean_pred}, after recalling the definitions regarding partnering.

\subsection{Partnering Definition}

The matching sessions~\cite{matching_session_def} and origin session~\cite{origin_session_def} definitions are recalled. This defines how sessions are partnered.

\begin{definition}\label{def10'}\textbf{(Matching sessions).} Let $\pi^s_i$ and $\pi^r_j$ be two sessions such that $\pi^s_i.pid = j$, $\pi^r_j.pid = i$ and $\pi^s_i.\rho \neq \pi^r_j.\rho$. Session $\pi^s_i$ matches session $\pi^r_j$ in stage $t$ if $\pi^s_i.m_r[t] = \pi^r_j.m_s[t]$ and $\pi^s_i.m_s[t] = \pi^r_j.m_r[t]$, where $\pi^s_i.m_s[t]$ and $\pi^s_i.m_r[t]$ correspond to the concatenation of all messages sent and received by the session $\pi^s_i$ in stage $t$, respectively.\newline
Additionally, the session $\pi^s_i$ prefix-matches session $\pi^r_j$ in stage $t$ if $\pi^r_j.m_s[t] = \pi^s_i.m_r[t]'$, where $\pi^s_i.m_r[t]'$ is $\pi^s_i.m_r[t]$ truncated to the length of $\pi^r_j.m_s[t]$.

\end{definition}

Note that $\pi^s_i$ having a matching or prefix-matching session with $\pi^r_j$ is equivalent to $\pi^r_j$ undertaking a matching or prefix-matching session with $\pi^s_i$. If the sessions $\pi^s_i$ and $\pi^r_j$ are matching, then the sessions are also prefix-matching.

\begin{definition}\label{def11'}\textbf{(Origin sessions).} Let $\pi^s_i$ and $\pi^r_j$ be two sessions such that $\pi^s_i.pid = j$, $\pi^r_j.pid = i$ and $\pi^s_i.\rho \neq \pi^r_j.\rho$. Session $\pi^s_i$ has an origin session with $\pi^r_j$ if $\pi^s_i$ matches or prefix-matches session with $\pi^r_j$.

A session $\pi^s_i$ that has an origin session with $\pi^r_j$ implies that $\pi^s_i$ has received the messages that $\pi^r_j$ intended to send. However, that does not imply that $\pi^r_j$ is undertaking an origin session with $\pi^s_i$. Additionally, the sessions match if all messages sent and received by both sessions are identical: for this case, both sessions are undertaking an origin session with each other. The origin session distinction is necessary to include the case scenarios where an adversary starts injecting, deleting, or substituting the classical traffic for at least one of the sessions, where both sessions were matching beforehand.

\end{definition}

The cleanness predicates are defined next.

\subsection{Cleanness Predicates}\label{framework_clean_pred}

The cleanness predicates define the conditions for which an adversary $\mathcal{A}$ does not break the security of $\Pi$. The main goal is to show that the presented examples of protocols are secure against an HPT adversary. Furthermore, once any of the protocols are in an accept state (or failed state, different from active state), the HPT adversary is allowed to issue RevealITS to an unbounded adversary, to verify that the obtained QKD keys are ITS secure, where the unbounded adversary was able to interfere with the unauthenticated QKD step and only observe the authentication steps. Note that if an adversary is permanently unbounded, the presented algorithms are not secure: the only viable way of making QKD secure in this case would be to properly use ITS-MACs with pre-shared key. Thus, the unbounded adversary is defined such as being able to read all classical data and returns either the Reveal and CompromiseSK queries, or nothing, without revealing more to the HPT adversary, in function of if the sessions keys are ITS or not.

In this case, the cleanness predicates can be reduced to ``if authentication is secure, then the session keys are secure". Thus, secrecy of at least one of the secrets involving authentication has to be guaranteed for when a protocol $\Pi$ is in an active state. 

The cleanness predicates also capture perfect forward secrecy. The nature of a secure QKD protocol implies direct perfect forward secrecy, since the last post-processing step of QKD, privacy amplification, ensures that the individual QKD bits contain a min-entropy close to unity. Thus, revealing all the secrets but the actual QKD bits (and the QKD secrets such as the raw data encoded and decoded by respectively Alice and Bob) does not reveal the QKD bits distilled from previous stages, i.e. the protocols have perfect forward secrecy, secure against unbounded adversaries that interact with the public material once the protocols are finalized (in an accepted state). This is expressed in the condition 3 (below) of the cleanness predicate, by allowing the adversary to compromise all of the secret keys once the test sessions reach an $accept$ state.

Post-compromise security is also achieved for all HPT adversaries in the sense that the adversary can compromise all ephemeral secrets of a particular stage without compromising the security of the protocols, as long as there exists some previous stage that has not had the QKD ephemeral secret compromised and the adversary has been passive in all stages between the ``Test" stage and the previous ``clean" stage~\cite{post_compromise_security}, expressed in condition 3 of the cleanness predicate definition below as well.

\begin{definition}\label{def13}\textbf{(clean$_{HPT}$).} Let $\pi^s_i[t]$ be a $\Pi \in \{\Pi_\Sigma,\Pi_{KEM},\Pi_{MAC}\}$ session in stage $t$ such that $\pi^s_i.\alpha[t]=accept$ and $\pi^s_i.pid=j$. The $\Pi$ session $\pi^s_i$ in stage $t$ is clean$_{HPT}$ against an HPT adversary if all of the following conditions hold:

\begin{enumerate}
    \item Reveal$(i,s,t)$ has not been issued.
    \item For all $(j,r,t) \in n_P \times n_S \times n_T$ such that $\pi^s_i$ matches $\pi^r_j$ in stage t, Reveal$(j,r,t)$ has not been issued.    
    \item If there exists no $(j,r,t) \in n_P \times n_S \times n_T$ such that $\pi^r_j$ is an origin session of $\pi^s_i$ in stage $t$, then one of the following set of queries has not been issued before $\pi^s_i.\alpha[t] \leftarrow accept$:
    \begin{itemize}
        \item CorruptQK$(i,s,t)$ and CorruptQK$(j,r,t)$.
        \item CompromiseKP$(i,s,t)$ and CompromiseKP$(j,r,t)$.
        \item CompromiseSK$(i,s,t')$, CompromiseSK$(j,r,t')$ or CompromiseKP$(i,s,t')$,\newline CompromiseKP$(j,r,t')$ and CompromiseSS$(i,s,u)$, CompromiseSS$(j,r,u)$, with $t'\leq u < t$, and $\pi^s_i$ matches $\pi^r_j$ in stages $u$.
    \end{itemize}
    If there exists a $(j,r,t) \in n_P \times n_S \times n_T$ such that $\pi^r_j$ is an origin session of $\pi^s_i$ in stage $t$, then one of the following set of queries has not been issued before $\pi^r_j.\alpha[t] \leftarrow accept$:
    \begin{itemize}
        \item CorruptQK$(i,s,t)$ and CorruptQK$(j,r,t)$.
        \item CompromiseKP$(i,s,t)$ and CompromiseKP$(j,r,t)$.
        \item CompromiseSK$(i,s,t')$, CompromiseSK$(j,r,t')$ or CompromiseKP$(i,s,t')$,\newline CompromiseKP$(j,r,t')$ and CompromiseSS$(i,s,u)$, CompromiseSS$(j,r,u)$, with $t'\leq u < t$, and $\pi^s_i$ matches $\pi^r_j$ in stages $u$.
    \end{itemize}
    \item The protocol $\Pi_{MAC}$ is run only when $\pi^s_i.k$ has enough key material to extract $K_0$, where $\pi^s_i$ is a matching session of $\pi^r_j$ in stages $t' < t$ and $\pi^s_i.k[t'] = \pi^r_j.k[t']$. 
\end{enumerate}

\end{definition}

Condition 1. prevents the session keys being revealed to the considered adversary via the Reveal query, since this trivially compromises the security of any protocol $\Pi$. 

Condition 2. additionally prevents that the session keys of the matching session are targeted by the Reveal query as well, since matching sessions share the same session keys at the end of the accepted protocols. 

Conditions 1. and 2. impose that the challenger $\mathcal{C}$ never issues the Reveal query to the HPT adversary $\mathcal{A}$. However, note that the query RevealITS can still be issued by the HPT adversary to the unbounded adversary, bypassing conditions 1. and 2., making the session $\pi^s_i[t]$ not clean in the case that QKD keys are not ITS in the first place.

Condition 3. ensures that at least one of the secrets used to authenticate within the protocols is not known to the HPT adversary $\mathcal{A}$. This makes authentication secure against substitution and impersonation attacks. Recall that after the physical time T$_{HPT}$ has elapsed, the T$_{HPT}$-term PQC secrets are revealed by the challenger $\mathcal{C}$, which is equivalent to the HPT adversary $\mathcal{A}$ issuing a CorruptQK query. If CorruptQK is issued once the sessions reach accept status, the parties generate new PQC material in order to proceed to the next stage. 

Condition 4. ensures that the protocol $\Pi_{MAC}$ is run only when the shared secret key pool has enough key material to extract the key $K_0$. Since the shared secret key pool is empty for stage $t$, a PQC-based protocol ($\Pi_{\Sigma}$ or $\Pi_{KEM}$) has to be realized until the shared secret key pool has enough key material to realize successfully $\Pi_{MAC}$. If the adversary decides to perform Denial of Service (DoS) attacks, this could exhaust the shared secret key pool from session $\pi^s_i$, since $K_0$ is assumed to be fresh every time the protocol $\Pi_{MAC}$ is executed. In that case, the PQC-based protocol has to be run again to replenish the key pool once again. If the goal is to rely as little as possible on PQC, then ideally it would be desirable to have sufficient distilled key at stage $t=1$, i.e. at the first QKD iteration, to essentially prevent DoS attacks. Once $\Pi_{MAC}$ is finalized, more keys are stored inside the key pool that can in turn be used to execute $\Pi_{MAC}$ again. The party behind session $\pi^s_i$ would also need to communicate to its peer $\pi^r_j$ to delete the used keys from the shared secret key pool. This could be done by exchanging authenticated identity parameters of the used keys that now have to be discarded. Despite being an extensive subject, no more discussion is provided regarding the DoS attack in this paper.

Now that the cleanness predicate clean$_{HAKE}$ is defined, it is convenient to recall the advantage of an algorithm $\mathcal{A}$ in winning the HAKE key indistinguishability experiment:

\begin{definition}\label{def14}\textbf{(HAKE key indistinguishability).} Let $\Pi$ by a key-exchange protocol, and $n_P$, $n_S$, $n_T$ $\in$ $\mathbb{N}$. For a particular given predicate clean, and an algorithm $\mathcal{A}$, the advantage of $\mathcal{A}$ in the HAKE key-indistinguishability is defined as:

\begin{equation*}
    \text{Adv}^{\text{HAKE},\text{clean},\mathcal{A}}_{\Pi,n_P,n_S,n_T}(\kappa)=2  \left| \text{Pr}\left[ \text{Exp}^{\text{HAKE},\text{clean},\mathcal{A}}_{\Pi,n_P,n_S,n_T}(\kappa)=1 \right] - \frac{1}{2} \right|\;,
\end{equation*}
where the experiment $\text{Exp}^{\text{HAKE},\text{clean},\mathcal{A}}_{\Pi,n_P,n_S,n_T}(\kappa)$ is defined in Appendix C, figure 5 of the Muckle paper~\cite{Muckle} and $\kappa$ is a security parameter, and where RevealITS, defined in this work, also enters in play to test ITS security of the session keys. $\Pi$ is HPT HAKE-secure if, for all HPT adversaries $\mathcal{A}$, $\text{Adv}^{\text{HAKE},\text{clean},\mathcal{A}}_{\Pi,n_P,n_S,n_T}(\kappa)$ is negligible in the security parameter $\kappa$.

\end{definition}

As the HPT adversaries have access to the RevealITS query in the presented framework, if $\Pi$ is HPT HAKE-secure, then the session keys (QKD keys) are ITS too.

\section{Security Analysis}\label{security}

The security of the presented protocols $\Pi \in \{\Pi_\Sigma,\Pi_{KEM},\Pi_{MAC}\}$ is evaluated through the cleanness predicate clean$_{HPT}$ and Definition \ref{def14}.

Furthermore, the confidentiality property given by the traffic keys and AEAD is also taken into account. The advantages for the HAKE key-indistinguishability game and the advantages related to confidentiality are summed as given by the union bound. The following theorem \ref{theorem1} accounts for both properties.

\begin{theorem}\label{theorem1}
The multi-stage and dynamic $\Pi \in \{\Pi_\Sigma,\Pi_{KEM},\Pi_{MAC}\}$ key exchange protocol is HAKE-secure with cleanness predicate clean$_{HPT}$, and confidential secure, where unauthenticated QKD follows Definition \ref{def2.3}, PRF is a dual PRF algorithm (Definition \ref{def3}), H is a weak collision resistant hash function (Definition \ref{def4}), $\Sigma$ is a EUF-CMA secure signature algorithm (Definitions \ref{def5.1} and \ref{def5.2}), KEM is IND-CCA secure KEM algorithm (Definitions \ref{def6.1} and \ref{def6.2}) and MAC is a EUC-CMA secure MAC algorithm (Definition \ref{def7.1} and \ref{def7.2}). That is, for any adversary $\mathcal{A}$ (Definitions \ref{def1.1} and \ref{def1.2}) against the HAKE key-indistinguishability game (Definition \ref{def14}), $\text{Adv}^{\text{HAKE},\text{clean}_{HPT},\mathcal{A}}_{\Pi,n_P,n_S,n_T}(\kappa)$ is negligible in the security parameter $\kappa$, where:

\begin{eqnarray*}
    \text{Adv}^{\text{HAKE},\text{clean}_{HPT},\mathcal{A}}_{\Pi,n_P,n_S,n_T}(\kappa) &\leq& \,\,\,\,\,\,\,\,\,\,\, \\
&&2n_P^2n_Sn_{T_\Sigma}  \left[\text{Adv}^{\text{EUF-CMA}}_{\Sigma,\mathcal{A}}(\kappa) + 6\text{Adv}_{\text{H},\mathcal{A}}(\kappa) \right] \\
 &+&2n_P^2n_S^2n_{T_\Sigma}^2  \bigl[ \varepsilon_{QKD}(\kappa) + \text{Adv}^{\text{ind}}_{\text{PRF}^\text{dual}, \mathcal{A}}(\kappa)\\
 &+& (1+n_{T_\Sigma})\bigl\{5\text{Adv}^{\text{ind}}_{\text{PRF}, \mathcal{A}}(\kappa) + \text{Adv}_{\text{AEAD},\mathcal{A}}^\text{IND-CPA}(\kappa) \\
 &+&\text{Adv}_{\text{AEAD},\mathcal{A}}^\text{INT-CTXT}(\kappa) + \text{Adv}^{\text{EUF-CMA}}_{\text{MAC}, \mathcal{A}}(\kappa)+6 \text{Adv}_{\text{H},\mathcal{A}}(\kappa)\bigr\}\bigr] \\[3mm]
 &+&2n_P^2n_Sn_{T_{KEM}}  \bigl[\text{Adv}^{\text{IND-CCA}}_{\text{KEM}, \mathcal{A}}(\kappa) + \text{Adv}^{\text{ind}}_{\text{PRF}^\text{dual}, \mathcal{A}}(\kappa) \\
    &+& 4\text{Adv}^{\text{ind}}_{\text{PRF}, \mathcal{A}}(\kappa) + \text{Adv}_{\text{AEAD},\mathcal{A}}^\text{IND-CPA}(\kappa) +\text{Adv}_{\text{AEAD},\mathcal{A}}^\text{INT-CTXT}(\kappa)\\
    &+&\text{Adv}^{\text{EUF-CMA}}_{\text{MAC}, \mathcal{A}}(\kappa) + 5\text{Adv}_{\text{H},\mathcal{A}}(\kappa)\bigr] \\
 &+&2n_P^2n_S^2n_{T_{KEM}}^2  \bigl[ \varepsilon_{QKD}(\kappa) + \text{Adv}^{\text{ind}}_{\text{PRF}^\text{dual}, \mathcal{A}}(\kappa)\\
 &+& (1+n_{T_{KEM}})\bigl\{\text{Adv}^{\text{ind}}_{\text{PRF}, \mathcal{A}}(\kappa) + 2\text{Adv}_{\text{AEAD},\mathcal{A}}^\text{IND-CPA}(\kappa) \\
 &+&2\text{Adv}_{\text{AEAD},\mathcal{A}}^\text{INT-CTXT}(\kappa) + \text{Adv}^{\text{EUF-CMA}}_{\text{MAC}, \mathcal{A}}(\kappa)+5 \text{Adv}_{\text{H},\mathcal{A}}(\kappa)\bigr\}\bigr]\\[3mm]
 &+&2n_P^2n_Sn_{T_{MAC}}  \bigl[\text{Adv}^{\text{HAKE},\text{clean}_{HPT},\mathcal{A}}_{\Pi,n_P,n_S,(n_{T}'|\text{$K_0$ derivation stage})}(\kappa)\\
 &+&\text{Adv}^{\text{ind}}_{\text{PRF}^\text{dual}, \mathcal{A}}(\kappa) + 3 \text{Adv}^{\text{ind}}_{\text{PRF}, \mathcal{A}}(\kappa) + \text{Adv}_{\text{AEAD},\mathcal{A}}^\text{IND-CPA}(\kappa) \\
 &+&\text{Adv}_{\text{AEAD},\mathcal{A}}^\text{INT-CTXT}(\kappa) +\text{Adv}^{\text{EUF-CMA}}_{\text{MAC}, \mathcal{A}}(\kappa)\bigr]\\
    &+&2n_P^2n_S^2n_{T_{MAC}}^2  \bigl[\text{Adv}^{\text{HAKE},\text{clean}_{HPT},\mathcal{A}}_{\Pi,n_P,n_S,(n_{T}'|\text{$K_0$ derivation stage})}(\kappa)\\
    &+&\text{Adv}^{\text{ind}}_{\text{PRF}^\text{dual}, \mathcal{A}}(\kappa) + 4 \text{Adv}^{\text{ind}}_{\text{PRF}, \mathcal{A}}(\kappa) + \text{Adv}_{\text{AEAD},\mathcal{A}}^\text{IND-CPA}(\kappa) \\
    &+&\text{Adv}_{\text{AEAD},\mathcal{A}}^\text{INT-CTXT}(\kappa) +\text{Adv}^{\text{EUF-CMA}}_{\text{MAC}, \mathcal{A}}(\kappa) \\
    &+&n_{T_{MAC}}\bigl\{5 \text{Adv}^{\text{ind}}_{\text{PRF}, \mathcal{A}}(\kappa) + \text{Adv}_{\text{AEAD},\mathcal{A}}^\text{IND-CPA}(\kappa) \\
    &+&\text{Adv}_{\text{AEAD},\mathcal{A}}^\text{INT-CTXT}(\kappa) + \text{Adv}^{\text{EUF-CMA}}_{\text{MAC}, \mathcal{A}}(\kappa)\bigr\}\bigr]\\[3mm]
&+&n_P^2n_S^2n_T   \bigl[2\text{Adv}_{\text{AEAD},\mathcal{A}}^\text{IND-CPA}(\kappa) +\varepsilon_{QKD}(\kappa)\bigr]\\
&+&n_P^2n_S^2n_{T_{KEM}}   2\text{Adv}_{\text{AEAD},\mathcal{A}}^\text{IND-CPA}(\kappa)
\end{eqnarray*}

\end{theorem}

\textit{Proof.} As performed in the Muckle series, the proof regarding the HAKE key-indistinguishability game is divided into three separate cases where the query Test$(i,s,t)$ has been issued:

\begin{enumerate}
    \item The session $\pi^s_i$ (where $\pi^s_i.\rho=init$) has no origin session in stage $t$.
    \item The session $\pi^s_i$ (where $\pi^s_i.\rho=resp$) has no origin session in stage $t$.
    \item The session $\pi^s_i$ in stage $t$ has a matching session.
\end{enumerate}

Cases 1 and 2 verify that $\Pi$ remains secure against attacks that force sessions to remain unmatched. Both cases encapsulate the idea of an active adversary substituting, deleting or injecting messages. Case 3 ensures that $\Pi$ is secure for sessions where the transcripts fully match or prefix-match. This case takes into account a passive adversary that is only reading the traffic. The three cases together cover all adversarial scenarios relevant to the protocol, protecting against a wide variety of attacks, including both impersonation and substitution attacks, and thus provide comprehensive security guarantees for the scheme.

Additionally, since $\Pi$ could involve three different protocols, $\Pi_\Sigma$, $\Pi_{KEM}$ and $\Pi_{MAC}$, the security proof has to take into account the fact that a session can choose between one of the three algorithms at any time (with the conditions that at stage $t=1$ and condition 4. from clean$_{HPT}$, $\Pi_{MAC}$ is not run). This implies that the security of the three protocols have to be evaluated individually at some point of the proof. This is done by dividing the number of stages $n_T$ into $n_{T_\Sigma}$, $n_{T_{KEM}}$ and $n_{T_{MAC}}$, where $n_{T_\Sigma}+n_{T_{KEM}}+n_{T_{MAC}}=n_T$. This division takes into account the number of stages where each protocol has been executed. Hence, each case out of the three cases that decomposes the HAKE key-indistinguishability game present three more sub-cases: one for each protocol in $\Pi$.

It follows that:

\begin{eqnarray*}
\text{Adv}^{\text{HAKE},\text{clean}_{HPT},\mathcal{A}}_{\Pi,n_P,n_S,n_T}(\kappa) & =& \text{Adv}^{\text{HAKE},\text{clean}_{HPT},\mathcal{A},C_1}_{\Pi,n_P,n_S,n_T}(\kappa) \\
& +& \text{Adv}^{\text{HAKE},\text{clean}_{HPT},\mathcal{A},C_2}_{\Pi,n_P,n_S,n_T}(\kappa) \\
& + &\text{Adv}^{\text{HAKE},\text{clean}_{HPT},\mathcal{A},C_3}_{\Pi,n_P,n_S,n_T}(\kappa)
\end{eqnarray*}

Where $\text{Adv}^{\text{HAKE},\text{clean}_{HPT},\mathcal{A},C_\#}_{\Pi,n_P,n_S,n_T}(\kappa)$ corresponds to the advantage of the HPT adversary $\mathcal{A}$ in winning the key-indistinguishability game in case $\#$. Furthermore,

\begin{eqnarray*}
\text{Adv}^{\text{HAKE},\text{clean}_{HPT},\mathcal{A},C_\#}_{\Pi,n_P,n_S,n_T}(\kappa) & = &\text{Adv}^{\text{HAKE},\text{clean}_{HPT},\mathcal{A},C_\#}_{\Pi,n_P,n_S,n_{T_\Sigma}}(\kappa) \\
& +& \text{Adv}^{\text{HAKE},\text{clean}_{HPT},\mathcal{A},C_\#}_{\Pi,n_P,n_S,n_{T_{KEM}}}(\kappa)\\
& +& \text{Adv}^{\text{HAKE},\text{clean}_{HPT},\mathcal{A},C_\#}_{\Pi,n_P,n_S,n_{T_{MAC}}}(\kappa)
\end{eqnarray*}

In total, nine advantages have to be evaluated (three cases, for each of three protocols). However, since case 1 and case 2 present the same bounds given a protocol and case 3 can be easily treated in the same way for the three protocols, four total advantages are actually evaluated (one case for three protocols and one case for one protocol).

Here is an example of how to evaluate the advantages for different protocols. Assume that two parties are running a session with four stages where they decided to run the protocols $\Pi_{KEM}$, $\Pi_{MAC}$, $\Pi_{\Sigma}$ and $\Pi_{MAC}$ in stages 1, 2, 3 and 4 respectively. The total security levels for the different stages are:

\begin{itemize}
    \item Stage 1: $\text{Adv}^{\text{HAKE},\text{clean}_{HPT},\mathcal{A}}_{\Pi,n_P=2,n_S=1,n_T=1}(\kappa)= \text{Adv}^{\text{HAKE},\text{clean}_{HPT},\mathcal{A}}_{\Pi,n_P=2,n_S=1,n_{T_{KEM}}=1}(\kappa)$
    
    \item Stage 2: $\text{Adv}^{\text{HAKE},\text{clean}_{HPT},\mathcal{A}}_{\Pi,n_P=2,n_S=1,n_T=2}(\kappa)= \text{Adv}^{\text{HAKE},\text{clean}_{HPT},\mathcal{A}}_{\Pi,n_P=2,n_S=1,n_{T_{KEM}}=1}(\kappa) $\newline$+ \text{Adv}^{\text{HAKE},\text{clean}_{HPT},\mathcal{A}}_{\Pi,n_P=2,n_S=1,n_{T_{MAC}}=1}(\kappa)$
    
    \item Stage 3: $\text{Adv}^{\text{HAKE},\text{clean}_{HPT},\mathcal{A}}_{\Pi,n_P=2,n_S=1,n_T=3}(\kappa)= \text{Adv}^{\text{HAKE},\text{clean}_{HPT},\mathcal{A}}_{\Pi,n_P=2,n_S=1,n_{T_{KEM}}=1}(\kappa) $\newline$+ \text{Adv}^{\text{HAKE},\text{clean}_{HPT},\mathcal{A}}_{\Pi,n_P=2,n_S=1,n_{T_{MAC}}=1}(\kappa)+\text{Adv}^{\text{HAKE},\text{clean}_{HPT},\mathcal{A}}_{\Pi,n_P=2,n_S=1,n_{T_{\Sigma}}=1}(\kappa)$
    
    \item Stage 4: $\text{Adv}^{\text{HAKE},\text{clean}_{HPT},\mathcal{A}}_{\Pi,n_P=2,n_S=1,n_T=4}(\kappa)= \text{Adv}^{\text{HAKE},\text{clean}_{HPT},\mathcal{A}}_{\Pi,n_P=2,n_S=1,n_{T_{KEM}}=1}(\kappa) $\newline$+ \text{Adv}^{\text{HAKE},\text{clean}_{HPT},\mathcal{A}}_{\Pi,n_P=2,n_S=1,n_{T_{MAC}}=2}(\kappa)+\text{Adv}^{\text{HAKE},\text{clean}_{HPT},\mathcal{A}}_{\Pi,n_P=2,n_S=1,n_{T_{\Sigma}}=1}(\kappa)$
\end{itemize}

The security proof is performed by starting with the original HAKE security game (the presented protocols as they are) and aiming for an unwinnable game (by modifying little by little the presented protocols with sound replacements), reached by game-hopping~\cite{Hop_game_book1, Hop_game_book2}, whilst making sure that the session is clean$_{HPT}$ according to the defined cleanness predicates. This is typically done by stating what key material can be compromised and then evaluate the chance of the defined adversary to win the modified games by replacing the non-compromised key material by uniformly random bit strings; the capacity of an adversary to differentiate between Games is given by the defined cryptographic primitives advantages, given in the Definitions section \ref{definitions}.  Additionally, confidentiality of the traffic is also evaluated and included within the security parameters.

Since the unauthenticated QKD step is assumed to follow Definition \ref{def2.3}, i.e. the unauthenticated QKD protocol is the same as a regular QKD protocol that has a secure authenticated channel, it suffices to show that authentication is $\varepsilon_{auth}$-secure to make the final QKD keys ITS and $(\varepsilon_{auth} + \varepsilon_{QKD})$-secure, according to Definition \ref{def2.4}. Additionally, $\varepsilon_{auth}$ contains the security that keeps confidentiality and integrity intact, as given by the traffic protection keys and the MAC tags, respectively.

A reminder that the PQC secrets are used when the physical time is less than an estimation of T$_{HPT}$. Once the physical time has surpassed T$_{HPT}$ or the CorruptQK query has been issued when the test sessions reach the accept state, new PQC secrets have to be generated.

In all cases and sub-cases, the game-hopping proof starts with the same Game 0 defined as:

\paragraph*{Game 0:} This is the original HAKE security game.

\begin{equation*}
    \text{Adv}^{\text{HAKE},\text{clean}_{HPT},\mathcal{A}}_{\Pi,n_P,n_S,n_{T_\Sigma}}(\kappa) = \text{Pr}(break_0)
\end{equation*}

\subsection{Case 1: Test $init$ session without origin session}\label{security_case1}

As performed the in Muckle series, the game hops will converge to a game that takes all the potential vulnerabilities of the cryptographic primitives used, where the adversary $\mathcal{A}$ has a negligible chance to win, to afterwards hop to a final unwinnable game, where the challenger $\mathcal{C}$ always reject, regardless of the correctness of the cryptographic primitives.
Specifically, Case 1 (and Case 2) security analysis shows that $\mathcal{A}$ has negligible chance in causing the test sessions $\pi^s_i$ (or $\pi^r_j$ respectively) to reach an accept state without an origin session. Both Cases take into account substitution, injection and deleting attacks, meaning that the test sessions $\pi^s_i$ may or may not already have started undertaking a matching session with an intended partner session $\pi^r_j$ at stage $t'<t$, but in stage $t$ the adversary stops the matching session by interacting with the traffic such that the test sessions no longer have an origin session.

Game 1 is the same for all of the sub-cases for Case 1 and Case 2, unless mentioned otherwise:

\paragraph*{Game 1:} The indices $(i,s,t)$, corresponding to the test session $\pi^s_i$, the stage stage $t$, and the intended partner $j$ are guessed. If the query Test$(i',s',t')$ is issued to a session $\pi^{s'}_{i'}$, where $\pi^{s'}_{i'}.pid=j'$, and if $(i,s,t,j) \neq (i',s',t',j')$, then the game aborts, yielding the following in the worst case scenario:

\begin{equation*}
    \text{Pr}(break_0) \leq n_P^2n_Sn_{T}  \text{Pr}(break_{1})
\end{equation*}

This game takes into account not only the scenarios where the adversary wants to impersonate party $j$ from the start, but also replay and substitution attacks. This is because the security against the mentioned attacks relies solely on the cryptographic primitives and not on the timing of the adversarial attack. The collision resistance probability is also taken into account for all sub-cases but the MAC case, since the MAC input presents a counter that provides uniqueness for the input of the cryptographic primitives. Furthermore, entity protection and cipher protection are also taken in account, regardless of if this compromises or not the final session key (QKD keys).

\subsubsection{Case 1: Sub-case $\Sigma$, CorruptQK is not issued at stage $t$ before $accept$}\label{security_case1.1}

The security of the $\Pi_\Sigma$ protocol is evaluated in the case that the CorruptQK query has not been issued.

\paragraph*{Games 0 and 1:} Defined right before and right after subsection \ref{security_case1} respectively.

\paragraph*{Game 2:} If the targeted test sessions reach the reject state, $\pi^s_i.\alpha[t] =reject$, then this game aborts. In Game 1, if the targeted test sessions reach a reject state, then $\perp$ is returned by the Test query, making Game 1 and Game 2 equivalent. Since the adversary $\mathcal{A}$ gains no additional advantage compared to Game 1, it yields that:

\begin{equation*}
    \text{Pr}(break_1) = \text{Pr}(break_2)
\end{equation*}

\paragraph*{Game 3:} If the targeted sessions reach the accept state, $\pi^s_i.\alpha[t] =accept$, then this game aborts. Note that the probability of winning Game 3 is exactly zero since the Test query always returns $\perp$, caused by the constant abortion. Furthermore, Game 2 and Game 3 are identical when the adversary $\mathcal{A}$ fails to successfully forge or distinguish any of the cryptographic primitives or key material, thus by the Difference Lemma (Lemma 1 from~\cite{Hop_game_book1}), this yields:

\begin{eqnarray*}
&|\text{Pr}(break_2) - \text{Pr}(break_3)| \leq \text{Pr}(abort) \\
&\Rightarrow\text{Pr}(break_2) \leq \text{Pr}(abort)
\end{eqnarray*}
where $\text{Pr}(break_3) =0$ and $\text{Pr}(abort)$ is the probability that the session aborts, i.e. $\pi^s_i.\alpha[t] =accept$. The following Game(s) replace(s) the vulnerable primitives or key material by uniformly random and independent values to upper-bound Pr$(abort)$. On this occasion, only the signature tag is targeted since the rest of the primitives and tags are trivially vulnerable since the HPT adversary $\mathcal{A_{HPT}}$ is allowed to issue the pertinent queries while keeping the test session clean. 

\paragraph*{Game 4:} The T$_{HPT}$-term secret $\text{sk}_B$ is replaced with a uniformly random and independent value, $\widetilde{\text{sk}_B}$, of the same length. The signature tag $\sigma_B$ is then computed with $\Sigma.\text{Sign}(\widetilde{\text{sk}_B}$, l$_{s_B}||$H(m$_{QKD}||$m$_1$)). Since sk$_B$ is itself uniformly random and independent, and the HPT adversary $\mathcal{A}$ has not issued either CorruptQK$(i,j)$ or CorruptQK$(j,i)$ queries, this change is a sound replacement. If the session $\pi^s_i$ reaches the accept state, then the adversary must have forged a valid signature tag $\sigma_B$ (following Experiment 3 from Definition \ref{def5.2}). Additionally, the adversary is allowed to reuse requested signature tags but changing the message to be authenticated, to try to bypass the burden of finding the secret key $\widetilde{\text{sk}_B}$. In that case, if the session $\pi^s_i$ reaches the accept state, then a hash collision must has been found. Actually, the adversary can target six different digests: H(m$_{QKD}'$, m$_1$ to m$_i$), where m$_{QKD}'$ corresponds to the QKD traffic created by the authenticated interaction between the test party and the adversary and $i$ $\in$ $\{2,3,4,6,7,8\}$. To prevent any attack that targets the digests, a conservative bound is taken where all lines involving the hash function are considered. According to Definitions \ref{def4} and \ref{def5.2} and the union bound, it follows that:

\begin{equation*}
    \text{Pr}(abort) \leq \text{Adv}^{\text{EUF-CMA}}_{\Sigma,\mathcal{A}}(\kappa)
\end{equation*}

\begin{equation*}
    \text{Pr}(break_0) \leq n_P^2n_Sn_{T_\Sigma}  \bigl[\text{Adv}^{\text{EUF-CMA}}_{\Sigma,\mathcal{A}}(\kappa) + 6 \,\text{Adv}_{\text{H},\mathcal{A}}(\kappa)\bigr]
\end{equation*}

The conditioned unbounded adversary cannot interact (replace or substitute) with the nonces since this is not part of the unauthenticated QKD step. Additionally, if the nonces are concatenated to the labels, instead of putting them inside the hash function, the collision vulnerability becomes nonexistent.

Note that the adversaries $\mathcal{A}$ can issue the query CorruptQK at any time, as long as the test sessions reach the accept state. Since the PQC secrets are used to authenticate the QKD traffic without affecting the QKD step itself, issuing CorruptQK under the accept state condition does not affect the final advantage of winning the HAKE security game.

\subsubsection{Case 1: Sub-case $\Sigma$, CompromiseSK are not issued at stage $t'$ and CompromiseSS are not issued at stages $u$, $t'\leq u<t$}\label{security_case1.2}

\paragraph*{Game 0:} Defined immediately before subsection \ref{security_case1}.

\paragraph*{Game 1:} Same as Game 1 defined immediately after subsection \ref{security_case1}, but the adversary has to additionally guess the session $\pi^r_j$ in which the test session $\pi^s_i$ has matched in stages $u$, and taking into account the guessed stage $t'$ where the adversary $\mathcal{A}$ has not issued CompromiseSK$(i,s,t')$ or CompromiseSK$(j,r,t')$, and CompromiseSS$(i,s,u)$ or CompromiseSS$(j,r,u)$ have been issued, where the test session $\pi^s_i$ matches session $\pi^r_j$ in stages $u$, where $t' \leq u < t$:

\begin{equation*}
    \text{Pr}(break_0) \leq n_P^2n_S^2n_{T_\Sigma}^2  \text{Pr}(break_1)
\end{equation*}

\paragraph*{Games 2 and 3:} Same as sub-case $\Sigma$, CorruptQK is not issued at stage $t$ before $accept$, section \ref{security_case1.1}.

\begin{equation*}
    \text{Pr}(break_1) \leq \text{Pr}(break_2) \leq \text{Pr}(abort)
\end{equation*}

\paragraph*{Game 4:} The ephemeral QKD key is replaced with a uniformly random and independent value, $\widetilde{ss_{QKD}}$, of the same length. The replacement is performed with an unauthenticated QKD challenger that replaces the public traffic m$_{QKD}$ with m$_{QKD}'$, which is used to distill $\widetilde{ss_{QKD}}$. Since $\pi^s_i$ matches session $\pi^r_j$ in stage $t'$, the replaced traffic m$_{QKD}'$ is received by both sessions without modification. The detection of this replacement by $\mathcal{A}$ implies having an efficient distinguishing algorithm against the indistinguishability security of the unauthenticated QKD, where the algorithm does not target authentication.

\begin{equation*}
    \text{Pr}(abort) \leq \varepsilon_{QKD}(\kappa) + \text{Pr}(break_4)
\end{equation*}

\paragraph*{Game 5:} The intermediate key $K_0$ $\leftarrow$ PRF$(\widetilde{ss_{QKD}},l_0||\text{H(m})$ is replaced with a uniformly random and independent value, $\widetilde{K_0}$, of the same length. The replacement is performed with a PRF challenger after querying $l_0||$H(m), where m and H are the pertinent public data. Since $\widetilde{ss_{QKD}}$ is uniformly random and independent by Game 4, this is a sound replacement. The detection of this replacement by $\mathcal{A_{HPT}}$ implies having an efficient HPT distinguishing algorithm against the indistinguishability security of the PRF, thus:

\begin{equation*}
    \text{Pr}(break_4) \leq \text{Adv}^{\text{ind}}_{\text{PRF}, \mathcal{A}}(\kappa) + \text{Pr}(break_5)
\end{equation*}

\paragraph*{Game 6:} The intermediate key $K_1$ $\leftarrow$ PRF$(SecState,l_1||\widetilde{K_0})$ is replaced with a uniformly random and independent value, $\widetilde{K_1}$, of the same length. The replacement is performed with a dual PRF challenger after querying $SecState$ and $l_1$. Since $\widetilde{K_0}$ is uniformly random and independent by Game 5, this is a sound replacement. The detection of this replacement by $\mathcal{A_{HPT}}$ implies having an efficient HPT distinguishing algorithm against the indistinguishability security of the dual PRF, thus:

\begin{equation*}
    \text{Pr}(break_5) \leq \text{Adv}^{\text{ind}}_{\text{PRF}^{\text{dual}}, \mathcal{A}}(\kappa) + \text{Pr}(break_6)
\end{equation*}

\paragraph*{Games 7 and 8:} The traffic key $K_{TS_B}$ $\leftarrow$ PRF$(\widetilde{K_1},l_{TS_B}||\text{H(m})$ and the MAC key $K_{MAC_B}$ $\leftarrow$ PRF$(\widetilde{K_1},l_{MAC_B}||\text{H(m})$ are replaced with uniformly random and independent values, $\widetilde{K_{TS_B}}$ and $\widetilde{K_{MAC_B}}$, of the same length, respectively. These replacements are performed with a PRF challenger after querying $l_{TS_B}||$H(m) and $l_{MAC_B}||$H(m), where m and H are the pertinent public data. Since $\widetilde{K_1}$ is uniformly random and independent by Game 6, these are sound replacements. The detection of these replacements by $\mathcal{A_{HPT}}$ implies having an efficient HPT distinguishing algorithm against the indistinguishability security of the PRF, thus:

\begin{equation*}
    \text{Pr}(break_6) \leq 2 \text{Adv}^{\text{ind}}_{\text{PRF}, \mathcal{A}}(\kappa) + \text{Pr}(break_8)
\end{equation*}

Since confidentiality is also desired, the AEAD advantages regarding IND-CPA are taken into account. Additionally, the adversary not only has to forge what is inside the AEAD ciphertext, but also has to generate a valid AEAD ciphertext that the decryption process accepts (related to INT-CTXT security).

\paragraph*{Game 9:} Same as Game 7 but with the traffic key $K_{TS_A}$.

\begin{equation*}
    \text{Pr}(break_8) \leq \text{Adv}^{\text{ind}}_{\text{PRF}, \mathcal{A}}(\kappa) + \text{Pr}(break_9)
\end{equation*}

\paragraph*{Games 10 and 11:} The AEAD ciphers $\{\text{B}, \sigma_B,\tau_B\}_{K_{TS_B}}$ and $\{\text{A}, \sigma_A,\tau_A\}_{K_{TS_A}}$ are replaced with uniformly random and independent values, $\widetilde{\{\text{B}, \sigma_B,\tau_B\}_{K_{TS_B}}}$ and $\widetilde{\{\text{A}, \sigma_A,\tau_A\}_{K_{TS_A}}}$, of the same length, respectively. The replacement of $\{\text{A}, \sigma_A,\tau_A\}_{K_{TS_A}}$ is performed with a IND-CPA AEAD challenger with query $\text{A}, \sigma_A,\tau_A$, whereas the replacement of $\{\text{B}, \sigma_B,\tau_B\}_{K_{TS_B}}$ is performed with a INT-CTXT AEAD challenger with query $\text{B}, \sigma_B,\tau_B$. Since $\widetilde{K_{TS_B}}$ and $\widetilde{K_{TS_A}}$ are uniformly random and independent by Games 7 and 8, respectively, these are  sound replacements. The detection of these replacements by $\mathcal{A_{HPT}}$ implies having an efficient HPT distinguishing algorithm against the INT-CTXT or IND-CPA security of the AEAD algorithm, thus:

\begin{equation*}
    \text{Pr}(break_9) \leq \text{Adv}_{\text{AEAD},\mathcal{A}}^\text{IND-CPA}(\kappa) +\text{Adv}_{\text{AEAD},\mathcal{A}}^\text{INT-CTXT}(\kappa) + \text{Pr}(break_{11})
\end{equation*}

\paragraph*{Game 12:} If the targeted session $\pi^s_i$ accepts without an origin session in stage $t$, then this game aborts. The abortion condition is met once the final MAC tag given by no honest session $\pi^r_j$ is accepted by $\pi^s_i$. The accept condition occurs when a EUF-CMA MAC challenger computes the MAC tag $\tau_B$ $\leftarrow$ $\text{MAC.Auth}(\widetilde{K_{MAC_B}}, l_{MAC_B}||\text{H(m)})$ for $\pi^s_i$ by querying the pertinent message m. Since $\widetilde{K_{MAC_B}}$ is uniformly random and independent by Game 8, this is a sound replacement. The detection of this replacement by $\mathcal{A_{HPT}}$ implies having an efficient HPT forging algorithm against the EUF-CMA security of the MAC algorithm, thus:

\begin{equation*}
    \text{Pr}(break_{11}) \leq \text{Adv}^{\text{EUF-CMA}}_{\text{MAC}, \mathcal{A}}(\kappa) + \text{Pr}(break_{12})
\end{equation*}

\paragraph*{Game 13:} The secret state $SecState$ $\leftarrow$ PRF$(\widetilde{K_1},l_{SecState}||\text{H(m)})$ is replaced with a uniformly random and independent value, $\widetilde{SecState}$, of the same length. The replacement is performed with a PRF challenger after querying $l_{SecState}||$H(m), where m and H are the pertinent public data. Since $\widetilde{K_1}$ is uniformly random and independent by Game 6, this is a sound replacement. The detection of this replacement by $\mathcal{A_{HPT}}$ implies having an efficient HPT distinguishing algorithm against the indistinguishability security of the PRF, thus:

\begin{equation*}
    \text{Pr}(break_{12}) \leq \text{Adv}^{\text{ind}}_{\text{PRF}, \mathcal{A}}(\kappa) + \text{Pr}(break_{13})
\end{equation*}

The next game hops involving the secret state $SecState$ are repeated $(t-t')$ times in each consecutive stage $u$ where $t'< u \leq t$. Recall that there is a matching session $\pi^r_j$ with $\pi^s_i$ for all the stages following stage $t'$.

\paragraph*{Game 14:} The intermediate key $K_1$ $\leftarrow$ PRF$(\widetilde{SecState},l_1||K_0)$ is replaced with a uniformly random and independent value, $\widetilde{K_1}$, of the same length. The replacement is performed with a PRF challenger after querying $l_1||K_0$. Since $\widetilde{SecState}$ is uniformly random and independent by the previous game, this is a sound replacement. The detection of this replacement by $\mathcal{A_{HPT}}$ implies having an efficient HPT distinguishing algorithm against the indistinguishability security of the PRF, thus:

\begin{equation*}
    \text{Pr}(break_{13}) \leq n_T  \bigl\{\text{Adv}^{\text{ind}}_{\text{PRF}, \mathcal{A}}(\kappa) + \text{Pr}(break_{14})\bigr\}
\end{equation*}

\paragraph*{Games 15 to 21}: Same as Games 7 to 13 respectively.

\begin{equation*}
    \text{Pr}(break_{14}) \leq 4 \text{Adv}^{\text{ind}}_{\text{PRF}, \mathcal{A}}(\kappa) + \text{Adv}_{\text{AEAD},\mathcal{A}}^\text{IND-CPA}(\kappa) +\text{Adv}_{\text{AEAD},\mathcal{A}}^\text{INT-CTXT}(\kappa) + \text{Adv}^{\text{EUF-CMA}}_{\text{MAC}, \mathcal{A}}(\kappa)
\end{equation*}

To include potential attacks where collision of the digests could be exploited, as done in the sub-case $\Sigma$, CorruptQK is not issued at stage $t$ before $accept$, section \ref{security_case1.1}, the union bonds of the advantages related to the collision probability are also included. That is:

\begin{eqnarray*}
    && \text{Pr}(break_{13}) \leq 6 \text{Adv}_{\text{H},\mathcal{A}}(\kappa) +\text{Adv}^{\text{ind}}_{\text{PRF}, \mathcal{A}}(\kappa) + \text{Pr}(break_{14})\\
    &&\text{Pr}(break_{21}) \leq 6 \text{Adv}_{\text{H},\mathcal{A}}(\kappa)
\end{eqnarray*}

Since there are no more primitives or key material that the adversary can target, the final advantage is bounded. 

Through combining all the inequalities together,  inequality \ref{eq:case1-sign} follows.

\begin{eqnarray}
\text{Adv}^{\text{HAKE},\text{clean}_{HPT},\mathcal{A},C_1}_{\Pi,n_P,n_S,n_{T_\Sigma}}(\kappa) & \leq &n_P^2n_Sn_{T_\Sigma}  \bigl[\text{Adv}^{\text{EUF-CMA}}_{\Sigma,\mathcal{A}}(\kappa) + 6\text{Adv}_{\text{H},\mathcal{A}}(\kappa)\bigr] \nonumber\\
 &+&n_P^2n_S^2n_{T_\Sigma}^2  \bigl[ \varepsilon_{QKD}(\kappa) + \text{Adv}^{\text{ind}}_{\text{PRF}^\text{dual}, \mathcal{A}}(\kappa) \label{eq:case1-sign}\\
 & +& (1+n_{T_\Sigma})\bigl\{5\text{Adv}^{\text{ind}}_{\text{PRF}, \mathcal{A}}(\kappa) + \text{Adv}_{\text{AEAD},\mathcal{A}}^\text{IND-CPA}(\kappa) \nonumber\\
 &+&\text{Adv}_{\text{AEAD},\mathcal{A}}^\text{INT-CTXT}(\kappa) + \text{Adv}^{\text{EUF-CMA}}_{\text{MAC}, \mathcal{A}}(\kappa)+6 \text{Adv}_{\text{H},\mathcal{A}}(\kappa)\bigr\}\bigr] \nonumber
\end{eqnarray}

The sub-case for $\Pi_{KEM}$ is treated next.

\subsubsection{Case 1: Sub-case $KEM$, CorruptQK is not issued at stage $t$ before $accept$}\label{security_case1.3}

The security analysis follows analogously to what has been performed with the signature case.

\paragraph*{Games 0 and 1:} Defined immediately before and immediately after subsection \ref{security_case1}, respectively.

\begin{equation*}
    \text{Pr}(break_0) \leq n_P^2n_Sn_{T_{KEM}}  \text{Pr}(break_1)
\end{equation*}

\paragraph*{Games 2 and 3:} Same as sub-case $\Sigma$, CorruptQK is not issued at stage $t$ before $accept$, section \ref{security_case1.1}, respectively.

\begin{equation*}
    \text{Pr}(break_1) \leq \text{Pr}(break_2) \leq \text{Pr}(abort)
\end{equation*}

\paragraph*{Game 4:} The T$_{HPT}$-term secret $\text{k}_B$ is replaced with a uniformly random and independent value, $\widetilde{\text{k}_B}$, of the same length. The replacement is performed with a IND-CCA KEM challenger where the challenger's public key pk$_B'$ and the cipher output c$_B'$ takes the place of the public key pk$_B$ and cipher c$_B$ in messages m$_3$ and m$_4$, respectively. If the session $\pi^s_i$ reaches the accept state, then the adversary must have successfully decapsulated $\widetilde{\text{k}_B}$. As the queries CorruptQK$(i,j)$ or CorruptQK$(j,i)$ have not been issued, it follows that (following Experiment 4 from Definition \ref{def6.2}):

\begin{equation*}
    \text{Pr}(abort) \leq \text{Adv}^{\text{IND-CCA}}_{\text{KEM}, \mathcal{A}}(\kappa) + \text{Pr}(break_4)
\end{equation*}

\paragraph*{Game 5:} The intermediate key $K_2$ $\leftarrow$ PRF$(K_1,l_2||\widetilde{\text{k}_B})$ is replaced with a uniformly random and independent value, $\widetilde{K_2}$, of the same length. The replacement is performed with a dual PRF challenger after querying $K_1$ and $l_2$. Since $\widetilde{\text{k}_B}$ is uniformly random and independent by the previous game, this is a sound replacement. The detection of this replacement by $\mathcal{A_{HPT}}$ implies having an efficient HPT distinguishing algorithm against the indistinguishability security of the dual PRF, thus:

\begin{equation*}
    \text{Pr}(break_{4}) \leq \text{Adv}^{\text{ind}}_{\text{PRF}^\text{dual}, \mathcal{A}}(\kappa) + \text{Pr}(break_{5})
\end{equation*}

\paragraph*{Games 6 and 7:} The traffic keys $K_{TS_{A2}}$ $\leftarrow$ PRF$(\widetilde{K_1},l_{TS_{A2}}||\text{H(m)})$ and $K_{TS_{B2}}$ $\leftarrow$ PRF$(\widetilde{K_1},$ $l_{TS_{B2}}||\text{H(m)})$ are replaced with uniformly random and independent value, $\widetilde{K_{TS_{A2}}}$ and $\widetilde{K_{TS_{B2}}}$, of the same length, respectively. The replacements are performed with a PRF challenger after querying $l_{TS_{A2}}||\text{H(m)}$ and $l_{TS_{B2}}||\text{H(m)}$, where m and H are the pertinent public data. Since $\widetilde{K_1}$ is uniformly random and independent by the previous game, this is a sound replacement. The detection of this replacement by $\mathcal{A_{HPT}}$ implies having an efficient HPT distinguishing algorithm against the indistinguishability security of the PRF, thus:

\begin{equation*}
    \text{Pr}(break_{5}) \leq 2\text{Adv}^{\text{ind}}_{\text{PRF}, \mathcal{A}}(\kappa) + \text{Pr}(break_{7})
\end{equation*}

\paragraph*{Games 8 and 9:} The AEAD ciphers $\{ \text{c}_A,\tau_B\}_{K_{TS_{B2}}}$ and $\{ \text{c}_B,\tau_A\}_{K_{TS_{A2}}}$ are replaced with uniformly random and independent values, $\widetilde{\{\text{c}_A,\tau_B\}_{K_{TS_{B2}}}}$ and $\widetilde{\{\text{c}_B,\tau_A\}_{K_{TS_{A2}}}}$, of the same length, respectively. The replacement of $\{\text{c}_B,\tau_A\}_{K_{TS_{A2}}}$ is performed with a IND-CPA AEAD challenger with query $\text{c}_B,\tau_A$, whereas the replacement of $\{\text{c}_A,\tau_B\}_{K_{TS_{B2}}}$ is performed with a INT-CTXT AEAD challenger with query $\text{c}_A,\tau_B$. Since $\widetilde{K_{TS_{B2}}}$ and $\widetilde{K_{TS_{A2}}}$ are uniformly random and independent by Games 6 and 7, this is a sound replacement. The detection of these replacements by $\mathcal{A_{HPT}}$ implies having an efficient HPT distinguishing algorithm against the INT-CTXT or IND-CPA security of the AEAD algorithm, thus:

\begin{equation*}
    \text{Pr}(break_7) \leq \text{Adv}_{\text{AEAD},\mathcal{A}}^\text{IND-CPA}(\kappa) +\text{Adv}_{\text{AEAD},\mathcal{A}}^\text{INT-CTXT}(\kappa) + \text{Pr}(break_{9})
\end{equation*}

\paragraph*{Game 10:} The intermediate key $K_3$ $\leftarrow$ PRF$(\widetilde{K_2},l_3||\text{k}_A)$ is replaced with a uniformly random and independent value, $\widetilde{K_3}$, of the same length. The replacement is performed with a PRF challenger after querying $l_3||\text{k}_A$. Since $\widetilde{K_2}$ is uniformly random and independent by the previous game, this is a sound replacement. The detection of this replacement by $\mathcal{A_{HPT}}$ implies having an efficient HPT distinguishing algorithm against the indistinguishability security of the PRF, thus:

\begin{equation*}
    \text{Pr}(break_{9}) \leq \text{Adv}^{\text{ind}}_{\text{PRF}, \mathcal{A}}(\kappa) + \text{Pr}(break_{10})
\end{equation*}

\paragraph*{Game 11:} The MAC key $K_{MAC_B}$ $\leftarrow$ PRF$(\widetilde{K_3},l_{MAC_{B1}}||\text{H(m)})$ is replaced with a uniformly random and independent value, $\widetilde{K_{MAC_B}}$, of the same length. The replacement is performed with a PRF challenger after querying $l_{MAC_{B1}}||\text{H(m)}$. Since $\widetilde{K_3}$ is uniformly random and independent by the previous game, this is a sound replacement. The detection of this replacement by $\mathcal{A_{HPT}}$ implies having an efficient HPT distinguishing algorithm against the indistinguishability security of the PRF, thus:

\begin{equation*}
    \text{Pr}(break_{10}) \leq \text{Adv}^{\text{ind}}_{\text{PRF}, \mathcal{A}}(\kappa) + \text{Pr}(break_{11})
\end{equation*}

\paragraph*{Game 12:} If the targeted session $\pi^s_i$ accepts without an origin session in stage $t$, then this game aborts. The abortion condition is met once the final MAC tag given by no honest session $\pi^r_j$ is accepted by $\pi^s_i$. The accept condition occurs when a EUF-CMA MAC challenger computes the MAC tag $\tau_B$ $\leftarrow$ $\text{MAC.Auth}(\widetilde{K_{MAC_{B2}}}, l_{MAC_B}||\text{H(m)})$ for $\pi^s_i$ by querying the pertinent message m. Since $\widetilde{K_{MAC_{B2}}}$ is uniformly random and independent by Game 8, this is a sound replacement. The detection of this replacement by $\mathcal{A_{HPT}}$ implies having an efficient HPT forging algorithm against the EUF-CMA security of the MAC algorithm, thus:

\begin{equation*}
    \text{Pr}(break_{11}) \leq \text{Adv}^{\text{EUF-CMA}}_{\text{MAC}, \mathcal{A}}(\kappa) + \text{Pr}(break_{12})
\end{equation*}

Where $\text{Pr}(break_{12})=0$ since there are no more components to target that would allow the test session to reach the accept state or abort caused by the MAC verification failure. Additionally, as performed with the signature case, the hash collision advantage is also taken into account for all traded messages, to take into account potential attacks in regard of that. Hence:

\begin{eqnarray*}
\text{Pr}(break_0) &\leq& n_P^2n_Sn_{T_{KEM}}  \bigl[\text{Adv}^{\text{IND-CCA}}_{\text{KEM}, \mathcal{A}}(\kappa) + \text{Adv}^{\text{ind}}_{\text{PRF}^\text{dual}, \mathcal{A}}(\kappa) \\
&+& 4\text{Adv}^{\text{ind}}_{\text{PRF}, \mathcal{A}}(\kappa) + \text{Adv}_{\text{AEAD},\mathcal{A}}^\text{IND-CPA}(\kappa) +\text{Adv}_{\text{AEAD},\mathcal{A}}^\text{INT-CTXT}(\kappa)\\
&+&\text{Adv}^{\text{EUF-CMA}}_{\text{MAC}, \mathcal{A}}(\kappa) + 5\text{Adv}_{\text{H},\mathcal{A}}(\kappa)\bigr]
\end{eqnarray*}

\subsubsection{Case 1: Sub-case $KEM$, CompromiseSK are not issued at stage $t'$ and CompromiseSS are not issued at stages $u$, $t'\leq u<t$}

\paragraph*{Game 0:} Defined immediately before subsection \ref{security_case1}.

\paragraph*{Game 1:} Same as the sub-case $\Sigma$, CompromiseSK is not issued at stage $t'$ and CompromiseSS is not issued at stages $u$, $t'\leq u<t$, section \ref{security_case1.2}.

\begin{equation*}
    \text{Pr}(break_0) \leq n_P^2n_S^2n_{T_{KEM}}^2  \text{Pr}(break_1)
\end{equation*}

\paragraph*{Games 2 and 3:} Same as Sub-case $KEM$, CorruptQK is not issued at stage $t$ before $accept$, section \ref{security_case1.3}.

\begin{equation*}
    \text{Pr}(break_1) \leq \text{Pr}(break_2) \leq \text{Pr}(abort)
\end{equation*}

\paragraph*{Games 4 to 6:} Same as Games 4 to 6, respectively, from the sub-case $\Sigma$, CompromiseSK is not issued at stage $t'$ and CompromiseSS is not issued at stages $u$, $t'\leq u<t$, section \ref{security_case1.2} (targets $ss_{QKD}$, $K_0$ and $K_1$).

\begin{equation*}
    \text{Pr}(abort) \leq \varepsilon_{QKD}(\kappa) + \text{Adv}^{\text{ind}}_{\text{PRF}, \mathcal{A}}(\kappa) + \text{Adv}^{\text{ind}}_{\text{PRF}^\text{dual}, \mathcal{A}}(\kappa) + \text{Pr}(break_6)
\end{equation*}

\paragraph*{Games 7 and 8:} Same as Games 6 and 7, respectively, from the sub-case $KEM$, CorruptQK is not issued at stage $t$ before $accept$, section \ref{security_case1.3}, but targeting $K_{TS_{A1}}$ and $K_{TS_{B1}}$ with $\widetilde{K_1}$.

\begin{equation*}
    \text{Pr}(break_6) \leq 2\,\text{Adv}^{\text{ind}}_{\text{PRF}, \mathcal{A}}(\kappa) + \text{Pr}(break_8)
\end{equation*}

\paragraph*{Games 9 and 10:} Same as Games 8 and 9, respectively, from the sub-case $KEM$, CorruptQK is not issued at stage $t$ before $accept$, section \ref{security_case1.3}, but targeting $\{B\}_{K_{TS_{B1}}}$ and $\{\text{c}_B\}_{K_{TS_{A1}}}$ with $\widetilde{K_{TS_{A1}}}$ and $\widetilde{K_{TS_{B1}}}$.

\begin{equation*}
    \text{Pr}(break_8) \leq \text{Adv}_{\text{AEAD},\mathcal{A}}^\text{IND-CPA}(\kappa) +\text{Adv}_{\text{AEAD},\mathcal{A}}^\text{INT-CTXT}(\kappa) + \text{Pr}(break_{10})
\end{equation*}

\paragraph*{Games 11 to 18:} Same as Games 5 to 12, respectively, from the sub-case $KEM$, CorruptQK is not issued at stage $t$ before $accept$, section \ref{security_case1.3}, but targeting $K_2$, $K_{TS_{A2}}$, $K_{TS_{B2}}$, $\{B\}_{K_{TS_{B1}}}$, $\{ \text{c}_A,\tau_B\}_{K_{TS_{B2}}}$, $\{ \text{c}_B,\tau_A\}_{K_{TS_{A2}}}$, $K_3$, $K_{MAC_B}$ and $\tau_B$ with the respective uniformly random and independent parameters and security challengers.

\begin{eqnarray*}
\text{Pr}(break_{10}) &\leq& 5\text{Adv}^{\text{ind}}_{\text{PRF}, \mathcal{A}}(\kappa)  +\text{Adv}_{\text{AEAD},\mathcal{A}}^\text{IND-CPA}(\kappa)+\text{Adv}_{\text{AEAD},\mathcal{A}}^\text{INT-CTXT}(\kappa) \\
& +& \text{Adv}^{\text{EUF-CMA}}_{\text{MAC}, \mathcal{A}}(\kappa)+ \text{Pr}(break_{18})
\end{eqnarray*}

\paragraph*{Game 19:} Same as Game 13 from the sub-case $\Sigma$, CompromiseSK is not issued at stage $t'$ and CompromiseSS is not issued at stages $u$, $t'\leq u<t$, section \ref{security_case1.2}, but targeting $SecState$ with $\widetilde{K_3}$.

\begin{equation*}
    \text{Pr}(break_{18}) \leq \text{Adv}^{\text{ind}}_{\text{PRF}, \mathcal{A}}(\kappa) + \text{Pr}(break_{19})
\end{equation*}

The next game hops involving the secret state $SecState$ are repeated $(t-t')$ times in each consecutive stage $u$ where $t'< u \leq t$. Recall that there is a matching session $\pi^r_j$ with $\pi^s_i$ for all the stages following stage $t'$.

\paragraph*{Game 20:} Same as Game 6 but targeting $K_1$ with uniformly random and independent parameter $\widetilde{SecState}$, and with a PRF challenger.

\begin{equation*}
    \text{Pr}(break_{19}) \leq n_T \bigl\{\text{Adv}^{\text{ind}}_{\text{PRF}, \mathcal{A}}(\kappa) + \text{Pr}(break_{20})\bigr\}
\end{equation*}

\paragraph*{Games 21 to 33:} Same as Games 7 to 19

\begin{eqnarray*}
\text{Pr}(break_{20}) &\leq& 8 \text{Adv}^{\text{ind}}_{\text{PRF}, \mathcal{A}}(\kappa) + 2\text{Adv}_{\text{AEAD},\mathcal{A}}^\text{IND-CPA}(\kappa) +2\text{Adv}_{\text{AEAD},\mathcal{A}}^\text{INT-CTXT}(\kappa) \\
&+& \text{Adv}^{\text{EUF-CMA}}_{\text{MAC}, \mathcal{A}}(\kappa)
\end{eqnarray*}

To include potential attacks where collision of the digests could be exploited, as done in all of the precedent cases, the union bonds of the advantages related to the collision probability are also included. That is:

\begin{eqnarray*}
&& \text{Pr}(break_{18}) \leq 5\text{Adv}_{\text{H},\mathcal{A}}(\kappa) + \text{Adv}^{\text{ind}}_{\text{PRF}, \mathcal{A}}(\kappa) + \text{Pr}(break_{19}) \\
&&\text{Pr}(break_{33}) \leq 5 \text{Adv}_{\text{H},\mathcal{A}}(\kappa)
\end{eqnarray*}

Since there are no more primitives or key material that the adversary can target, the final advantage can be estimated. 

By combining all the inequalities together, the inequality \ref{eq:case1-KEM} follows.

\begin{eqnarray}
\text{Adv}^{\text{HAKE},\text{clean}_{HPT},\mathcal{A},C_1}_{\Pi,n_P,n_S,n_{T_{KEM}}}(\kappa) & \leq &n_P^2n_Sn_{T_{KEM}}  \bigl[\text{Adv}^{\text{IND-CCA}}_{\text{KEM}, \mathcal{A}}(\kappa) + \text{Adv}^{\text{ind}}_{\text{PRF}^\text{dual}, \mathcal{A}}(\kappa) \nonumber\\
&+& 4\text{Adv}^{\text{ind}}_{\text{PRF}, \mathcal{A}}(\kappa) + \text{Adv}_{\text{AEAD},\mathcal{A}}^\text{IND-CPA}(\kappa) +\text{Adv}_{\text{AEAD},\mathcal{A}}^\text{INT-CTXT}(\kappa)\nonumber\\
&+&\text{Adv}^{\text{EUF-CMA}}_{\text{MAC}, \mathcal{A}}(\kappa) + 5\text{Adv}_{\text{H},\mathcal{A}}(\kappa)\bigr] \label{eq:case1-KEM}\\
&+&n_P^2n_S^2n_{T_{KEM}}^2  \bigl[ \varepsilon_{QKD}(\kappa) + \text{Adv}^{\text{ind}}_{\text{PRF}^\text{dual}, \mathcal{A}}(\kappa)\nonumber\\
& +& (1+n_{T_{KEM}})\bigl\{9\text{Adv}^{\text{ind}}_{\text{PRF}, \mathcal{A}}(\kappa) + 2\text{Adv}_{\text{AEAD},\mathcal{A}}^\text{IND-CPA}(\kappa) \nonumber\\
&+&2\text{Adv}_{\text{AEAD},\mathcal{A}}^\text{INT-CTXT}(\kappa) + \text{Adv}^{\text{EUF-CMA}}_{\text{MAC}, \mathcal{A}}(\kappa)+5 \text{Adv}_{\text{H},\mathcal{A}}(\kappa)\bigr\}\bigr] \nonumber
\end{eqnarray}

The sub-case for $\Pi_{MAC}$ is treated next.

\subsubsection{Case 1: Sub-case $MAC$, CompromiseKP is not issued at stage $t$ before $accept$}\label{security_case1.5}

\paragraph*{Games 0 and 1:} Defined immediately before and immediately after subsection \ref{security_case1}, respectively.

\begin{equation*}
    \text{Pr}(break_0) \leq n_P^2n_Sn_{T_{MAC}}  \text{Pr}(break_1)
\end{equation*}

For this sub-case, the adversary does not have to guess the stage from which the keys $K_0$ are being taken, since honest parties follow a public convention. For example, parties could use the last derived key session from the $K_0$ derivation stage, or even extract sufficient keys in the first QKD iteration in order not to rely on asymmetric cryptography anymore in further QKD iterations. 

\paragraph*{Games 2 and 3:} Same as sub-cases $\Sigma$ (and $KEM$), CorruptQK is not issued at stage $t$ before $accept$, section \ref{security_case1.1} (and \ref{security_case1.3}), respectively.

\begin{equation*}
    \text{Pr}(break_1) \leq \text{Pr}(break_2) \leq \text{Pr}(abort)
\end{equation*}

\paragraph*{Game 4:} The shared secret $K_0$ is replaced with a uniformly random and independent value, $\widetilde{K_0}$, of the same length. The replacement is performed with a challenger, which has a shared secret key pool that stores session keys from past protocols $\Pi$ run by the challenger themselves, and $\widetilde{K_0}$ is a key from the challenger's shared secret key pool. If the session $\pi^s_i$ reaches the accept state, then the adversary must have compromised or guessed $K_0$ correctly. As the queries CompromiseKP$(i,s,t)$ or CompromiseKP$(j,r,t)$ have not been issued, it follows that:

\begin{equation*}
    \text{Pr}(abort) \leq \text{Adv}^{\text{HAKE},\text{clean}_{HPT},\mathcal{A}}_{\Pi,n_P,n_S,(n_{T}'|\text{$K_0$ derivation stage})}(\kappa) + \text{Pr}(break_4)
\end{equation*}

\paragraph*{Games 5 to 11:} Same as Games 6 to 12, respectively, from the sub-case $\Sigma$, CompromiseSK is not issued at stage $t'$ and CompromiseSS is not issued at stages $u$, $t'\leq u<t$, section \ref{security_case1.2}.

\begin{eqnarray*}
\text{Pr}(break_{4}) &\leq& \text{Adv}^{\text{ind}}_{\text{PRF}^\text{dual}, \mathcal{A}}(\kappa) + 3 \text{Adv}^{\text{ind}}_{\text{PRF}, \mathcal{A}}(\kappa) + \text{Adv}_{\text{AEAD},\mathcal{A}}^\text{IND-CPA}(\kappa) \\
&+&\text{Adv}_{\text{AEAD},\mathcal{A}}^\text{INT-CTXT}(\kappa) +\text{Adv}^{\text{EUF-CMA}}_{\text{MAC}, \mathcal{A}}(\kappa)+\text{Pr}(break_{11})
\end{eqnarray*}

For this long-term HPT-secure MAC authentication protocol, the hash function cannot be used to find vulnerabilities - the attacks where an adversary plays with the digest go back to trying to compromise the primitives that take a hash digest as input, which are already taken into account in the described previous Games.

By combining everything together, this sub-sub case yields:

\begin{eqnarray*}
\text{Pr}(break_0) &\leq& n_P^2n_Sn_{T_{MAC}}  \bigl[\text{Adv}^{\text{HAKE},\text{clean}_{HPT},\mathcal{A}}_{\Pi,n_P,n_S,(n_{T}'|\text{$K_0$ derivation stage})}(\kappa)\\
&+&\text{Adv}^{\text{ind}}_{\text{PRF}^\text{dual}, \mathcal{A}}(\kappa) + 3 \text{Adv}^{\text{ind}}_{\text{PRF}, \mathcal{A}}(\kappa) + \text{Adv}_{\text{AEAD},\mathcal{A}}^\text{IND-CPA}(\kappa) \\
&+&\text{Adv}_{\text{AEAD},\mathcal{A}}^\text{INT-CTXT}(\kappa) +\text{Adv}^{\text{EUF-CMA}}_{\text{MAC}, \mathcal{A}}(\kappa)\bigr]
\end{eqnarray*}

\subsubsection{Case 1: Sub-case $MAC$, CompromiseKP are not issued at stage $t'$ and CompromiseSS are not issued at stages $u$, $t'\leq u<t$}

\paragraph*{Game 0:} Defined immediately before subsection \ref{security_case1}.

\paragraph*{Game 1:} Same as previous Game 1 (sub-case $MAC$, CompromiseKP is not issued at stage $t$ before $accept$, section \ref{security_case1.5}), where the adversary has to additionally guess the session $\pi^r_j$ in which the test session $\pi^s_i$ has matched in stages $u$, and taking into account the guessed stage $t'$ where the adversary $\mathcal{A}$ has not issued CompromiseKP$(i,s,t')$ or CompromiseKP$(j,r,t')$, and CompromiseSS$(i,s,u)$ or CompromiseSS$(j,r,u)$ have been issued, where the test session $\pi^s_i$ matches session $\pi^r_j$ in stages $u$, where $t' \leq u < t$:

\begin{equation*}
    \text{Pr}(break_0) \leq n_P^2n_S^2n_{T_{MAC}}^2  \text{Pr}(break_1)
\end{equation*}

\paragraph*{Games 2 to 11:} Same as Games 2 to 11, respectively, of the sub-case $MAC$, CompromiseKP is not issued at stage $t$ before $accept$, section \ref{security_case1.5}:

\begin{eqnarray*}
\text{Pr}(break_0) &\leq& n_P^2n_S^2n_{T_{MAC}}^2  \bigl[\text{Adv}^{\text{HAKE},\text{clean}_{HPT},\mathcal{A}}_{\Pi,n_P,n_S,(n_{T}'|\text{$K_0$ derivation stage})}(\kappa)\\
&+&\text{Adv}^{\text{ind}}_{\text{PRF}^\text{dual}, \mathcal{A}}(\kappa) + 3 \text{Adv}^{\text{ind}}_{\text{PRF}, \mathcal{A}}(\kappa) + \text{Adv}_{\text{AEAD},\mathcal{A}}^\text{IND-CPA}(\kappa) \\
&+&\text{Adv}_{\text{AEAD},\mathcal{A}}^\text{INT-CTXT}(\kappa) +\text{Adv}^{\text{EUF-CMA}}_{\text{MAC}, \mathcal{A}}(\kappa) + \text{Pr}(break_{11})\bigr]
\end{eqnarray*}

\paragraph*{Game 12:} Same as Game 13 from the sub-case $\Sigma$, CompromiseSK is not issued at stage $t'$ and CompromiseSS is not issued at stages $u$, $t'\leq u<t$, section \ref{security_case1.2}, but targeting $SecState$ using $\widetilde{K_1}$).

\begin{equation*}
    \text{Pr}(break_{11}) \leq \text{Adv}^{\text{ind}}_{\text{PRF}, \mathcal{A}}(\kappa) + \text{Pr}(break_{12})
\end{equation*}

The next game hops involving the secret state $SecState$ are repeated $(t-t')$ times in each consecutive stage $u$ where $t'< u \leq t$. Recall that there is a matching session $\pi^r_j$ with $\pi^s_i$ for all the stages following stage $t'$.

\paragraph*{Game 13:} Same as Game 5 from sub-case $MAC$, CompromiseKP is not issued at stage $t$ before $accept$, section \ref{security_case1.5}, but targeting $K_1$ using a uniformly random and independent parameter $\widetilde{SecState}$ and with a PRF challenger.

\begin{equation*}
    \text{Pr}(break_{12}) \leq n_{T_{MAC}} (\text{Adv}^{\text{ind}}_{\text{PRF}, \mathcal{A}}(\kappa) + \text{Pr}(break_{13}))
\end{equation*}

\paragraph*{Games 14 to 20:} Same as Games 6 to 12 but targeting $K_{TS_A}$, $K_{TS_B}$, $K_{MAC_B}$, $\{ \tau_A \}_{K_{TS_A}}$, $\{ \tau_B \}_{K_{TS_B}}$, $\tau_B$ and $SecState$) with the corresponding uniformly random material:

\begin{eqnarray*}
\text{Pr}(break_{13}) &\leq& 4 \text{Adv}^{\text{ind}}_{\text{PRF}, \mathcal{A}}(\kappa) + \text{Adv}_{\text{AEAD},\mathcal{A}}^\text{IND-CPA}(\kappa) +\text{Adv}_{\text{AEAD},\mathcal{A}}^\text{INT-CTXT}(\kappa) \\
&+& \text{Adv}^{\text{EUF-CMA}}_{\text{MAC}, \mathcal{A}}(\kappa) + \text{Pr}(break_{20})
\end{eqnarray*}

For this long-term HPT-secure MAC authentication protocol, the hash function cannot be used to find vulnerabilities - the attacks where an adversary plays with the digest go back to trying to compromise the primitives that take a hash digest as input, which are already taken into account.

Since there are no more primitives or key material that the adversary can target, the final advantage can be estimated. 

By combining all the inequalities together, inequality \ref{eq:case1-MAC} follows.

\begin{eqnarray}
\text{Adv}^{\text{HAKE},\text{clean}_{HPT},\mathcal{A},C_1}_{\Pi,n_P,n_S,n_{T_{MAC}}}(\kappa) &\leq& n_P^2n_Sn_{T_{MAC}}  \bigl[\text{Adv}^{\text{HAKE},\text{clean}_{HPT},\mathcal{A}}_{\Pi,n_P,n_S,(n_{T}'|\text{$K_0$ derivation stage})}(\kappa)\nonumber\\ 
&+&\text{Adv}^{\text{ind}}_{\text{PRF}^\text{dual}, \mathcal{A}}(\kappa) + 3 \text{Adv}^{\text{ind}}_{\text{PRF}, \mathcal{A}}(\kappa) + \text{Adv}_{\text{AEAD},\mathcal{A}}^\text{IND-CPA}(\kappa) \nonumber\\
&+&\text{Adv}_{\text{AEAD},\mathcal{A}}^\text{INT-CTXT}(\kappa) +\text{Adv}^{\text{EUF-CMA}}_{\text{MAC}, \mathcal{A}}(\kappa)\bigr]\nonumber\\
&+&n_P^2n_S^2n_{T_{MAC}}^2  \bigl[\text{Adv}^{\text{HAKE},\text{clean}_{HPT},\mathcal{A}}_{\Pi,n_P,n_S,(n_{T}'|\text{$K_0$ derivation stage})}(\kappa)\nonumber\\
&+&\text{Adv}^{\text{ind}}_{\text{PRF}^\text{dual}, \mathcal{A}}(\kappa) + 4 \text{Adv}^{\text{ind}}_{\text{PRF}, \mathcal{A}}(\kappa) + \text{Adv}_{\text{AEAD},\mathcal{A}}^\text{IND-CPA}(\kappa) \nonumber\\
&+&\text{Adv}_{\text{AEAD},\mathcal{A}}^\text{INT-CTXT}(\kappa) +\text{Adv}^{\text{EUF-CMA}}_{\text{MAC}, \mathcal{A}}(\kappa) \nonumber\\
&+&n_{T_{MAC}}\bigl\{5 \text{Adv}^{\text{ind}}_{\text{PRF}, \mathcal{A}}(\kappa) + \text{Adv}_{\text{AEAD},\mathcal{A}}^\text{IND-CPA}(\kappa) \nonumber\\
&+&\text{Adv}_{\text{AEAD},\mathcal{A}}^\text{INT-CTXT}(\kappa) + \text{Adv}^{\text{EUF-CMA}}_{\text{MAC}, \mathcal{A}}(\kappa)\bigr\}\bigr]
\label{eq:case1-MAC}
\end{eqnarray}

\subsection{Case 2: Test $resp$ session without origin session}

Case 2 follows analogously to Case 1 and is the same regarding the signature-based and MAC-based protocols, whereas a small change is introduced for the KEM-based protocol.

\subsubsection{Case 2: Sub-case $\Sigma$}

Analogously, the advantage of Case 2 is exactly the same as the advantage of Case 1 for the signature-based protocol, giving the inequality \ref{eq:case2-sign}.

\begin{eqnarray}
\text{Adv}^{\text{HAKE},\text{clean}_{HPT},\mathcal{A},C_2}_{\Pi,n_P,n_S,n_{T_\Sigma}}(\kappa) &  \leq &n_P^2n_Sn_{T_\Sigma}  \bigl[\text{Adv}^{\text{EUF-CMA}}_{\Sigma,\mathcal{A}}(\kappa) + 6\text{Adv}_{\text{H},\mathcal{A}}(\kappa)) \nonumber\\
 &+&n_P^2n_S^2n_{T_\Sigma}^2  ( \varepsilon_{QKD}(\kappa) + \text{Adv}^{\text{ind}}_{\text{PRF}^\text{dual}, \mathcal{A}}(\kappa) \label{eq:case2-sign}\\
 & +& (1+n_{T_\Sigma})\bigl\{5\text{Adv}^{\text{ind}}_{\text{PRF}, \mathcal{A}}(\kappa) + \text{Adv}_{\text{AEAD},\mathcal{A}}^\text{IND-CPA}(\kappa) \nonumber\\
 &+&\text{Adv}_{\text{AEAD},\mathcal{A}}^\text{INT-CTXT}(\kappa) + \text{Adv}^{\text{EUF-CMA}}_{\text{MAC}, \mathcal{A}}(\kappa)+6 \text{Adv}_{\text{H},\mathcal{A}}(\kappa)\bigr\}\bigr] \nonumber
\end{eqnarray}

\subsubsection{Case 2: Sub-case $KEM$}

Analogously, the advantage of Case 2 is exactly the same as the advantage of Case 1 for the KEM-based protocol, giving the inequality \ref{eq:case2-KEM}.

\begin{eqnarray}
\text{Adv}^{\text{HAKE},\text{clean}_{HPT},\mathcal{A},C_2}_{\Pi,n_P,n_S,n_{T_{KEM}}}(\kappa) & \leq &n_P^2n_Sn_{T_{KEM}}  \bigl[\text{Adv}^{\text{IND-CCA}}_{\text{KEM}, \mathcal{A}}(\kappa) + \text{Adv}^{\text{ind}}_{\text{PRF}^\text{dual}, \mathcal{A}}(\kappa) \nonumber\\
&+& 4\text{Adv}^{\text{ind}}_{\text{PRF}, \mathcal{A}}(\kappa) + \text{Adv}_{\text{AEAD},\mathcal{A}}^\text{IND-CPA}(\kappa) +\text{Adv}_{\text{AEAD},\mathcal{A}}^\text{INT-CTXT}(\kappa)\nonumber\\
&+&\text{Adv}^{\text{EUF-CMA}}_{\text{MAC}, \mathcal{A}}(\kappa) + 5\text{Adv}_{\text{H},\mathcal{A}}(\kappa)\bigr] \label{eq:case2-KEM}\\
&+&n_P^2n_S^2n_{T_{KEM}}^2  \bigl[ \varepsilon_{QKD}(\kappa) + \text{Adv}^{\text{ind}}_{\text{PRF}^\text{dual}, \mathcal{A}}(\kappa)\nonumber\\
& +& (1+n_{T_{KEM}})\bigl\{9\text{Adv}^{\text{ind}}_{\text{PRF}, \mathcal{A}}(\kappa) + 2\text{Adv}_{\text{AEAD},\mathcal{A}}^\text{IND-CPA}(\kappa) \nonumber\\
&+&2\text{Adv}_{\text{AEAD},\mathcal{A}}^\text{INT-CTXT}(\kappa) + \text{Adv}^{\text{EUF-CMA}}_{\text{MAC}, \mathcal{A}}(\kappa)+5 \text{Adv}_{\text{H},\mathcal{A}}(\kappa)\bigr\}\bigr] \nonumber
\end{eqnarray}

\subsubsection{Case 2: Sub-case $MAC$}

Analogously, the advantage of Case 2 is exactly the same as the advantage of Case 1 for the MAC-based protocol, giving the inequality \ref{eq:case2-MAC}.

\begin{eqnarray}
\text{Adv}^{\text{HAKE},\text{clean}_{HPT},\mathcal{A},C_2}_{\Pi,n_P,n_S,n_{T_{MAC}}}(\kappa) &\leq &n_P^2n_Sn_{T_{MAC}}  \bigl[\text{Adv}^{\text{HAKE},\text{clean}_{HPT},\mathcal{A}}_{\Pi,n_P,n_S,(n_{T}'|\text{$K_0$ derivation stage})}(\kappa)\nonumber\\
&+&\text{Adv}^{\text{ind}}_{\text{PRF}^\text{dual}, \mathcal{A}}(\kappa) + 3 \text{Adv}^{\text{ind}}_{\text{PRF}, \mathcal{A}}(\kappa) + \text{Adv}_{\text{AEAD},\mathcal{A}}^\text{IND-CPA}(\kappa) \nonumber\\
&+&\text{Adv}_{\text{AEAD},\mathcal{A}}^\text{INT-CTXT}(\kappa) +\text{Adv}^{\text{EUF-CMA}}_{\text{MAC}, \mathcal{A}}(\kappa)\bigr]\nonumber\\
&+&n_P^2n_S^2n_{T_{MAC}}^2  \bigl[\text{Adv}^{\text{HAKE},\text{clean}_{HPT},\mathcal{A}}_{\Pi,n_P,n_S,(n_{T}'|\text{$K_0$ derivation stage})}(\kappa)\nonumber\\
&+&\text{Adv}^{\text{ind}}_{\text{PRF}^\text{dual}, \mathcal{A}}(\kappa) + 4 \text{Adv}^{\text{ind}}_{\text{PRF}, \mathcal{A}}(\kappa) + \text{Adv}_{\text{AEAD},\mathcal{A}}^\text{IND-CPA}(\kappa) \nonumber\\
&+&\text{Adv}_{\text{AEAD},\mathcal{A}}^\text{INT-CTXT}(\kappa) +\text{Adv}^{\text{EUF-CMA}}_{\text{MAC}, \mathcal{A}}(\kappa) \nonumber\\
&+&n_{T_{MAC}}\bigl\{5 \text{Adv}^{\text{ind}}_{\text{PRF}, \mathcal{A}}(\kappa) + \text{Adv}_{\text{AEAD},\mathcal{A}}^\text{IND-CPA}(\kappa)\nonumber\\
&+&\text{Adv}_{\text{AEAD},\mathcal{A}}^\text{INT-CTXT}(\kappa) + \text{Adv}^{\text{EUF-CMA}}_{\text{MAC}, \mathcal{A}}(\kappa)\bigr\}\bigr]
\label{eq:case2-MAC}
\end{eqnarray}

\subsection{Case 3: Test session with matching session}

All sub-cases of Case 3 have almost the same upper bound, given by the nature of QKD and the lack of derivation of a proper session key. If the test session $\pi^s_i$ has a matching session $\pi^r_j$, then the public information that both sessions intended to send to each other arrived at the intended destination without modification by the adversary.

The adversary can try to perform unauthenticated QKD with the test session $\pi^s_i$ and the matching session $\pi^r_j$, such that the public traffic to authenticate in both cases is the same, or the digests given by different traffics collide. In this way, the adversary would be able to share session keys with honest parties without having to deal with the authentication step. Due to the no-cloning theorem and the rejection given a high enough QBER, the adversary has a negligible probability to realize such a person-in-the-middle attack without being detected, which is taken into account within $\varepsilon_{QKD}$: the negligible chance of success for related attacks that target QKD itself rather than the authentication process are all included within the security of QKD, inside the parameter $\varepsilon_{QKD}(\kappa)$.

\paragraph*{Game 1:} The indices $(i,s,t)$ and $(j,r,t)$, corresponding to the targeted matching sessions $\pi^s_i$ and $\pi^r_j$, are guessed. If the query Test$(i',s',t')$ is issued to a session $\pi^{s'}_{i'}$, where $\pi^{r'}_{j'}$ matches $\pi^{s'}_{i'}$ in stage $t'$, and if $(i,s,j,r,t)$ $\neq$ $(i',s',j',r',t')$, then the game aborts, yielding for the worst-case scenario:

\begin{equation*}
    \text{Pr}(break_0) \leq n_P^2n_S^2n_T \text{Pr}(break_1)
\end{equation*}

The following games target the confidentiality security of the exchanged AEAD ciphers. Since the adversary has no flexibility to forge and only targets learning what is inside the AEAD ciphers, only the IND-CPA security for the AEAD algorithm is taken into account. 

\paragraph*{Games 2 to 5 for $\Pi_{\Sigma}$ and $\Pi_{MAC}$, Games 2 to 5 and $2_{\text{bis}}$
 to $5_{\text{bis}}$ for $\Pi_{KEM}$:} These indistinguishability games target the traffic keys ($K_{TS_A}$, $K_{TS_B}$) or ($K_{TS_{A1}}$, $K_{TS_{B1}}$, $K_{TS_{A2}}$, $K_{TS_{B2}}$) and the AEAD ciphers ($\{B,$ $\sigma_B,$ $\tau_B\}_{K_{TS_B}}$, $\{A,$ $\sigma_A,$ $\tau_A\}_{K_{TS_A}}$), ($\{B\}_{K_{TS_{B1}}}$, $\{A\}_{K_{TS_{A1}}}$, $\{\text{c}_A$, $\tau_B\}_{K_{TS_{B2}}}$, $\{\text{c}_B$, $\tau_A\}_{K_{TS_{A2}}}$) or ($\{\tau_A\}_{K_{TS_A}}$, $\{\tau_B\}_{K_{TS_B}}$) for $\Pi_\Sigma$, $\Pi_{KEM}$ and $\Pi_{MAC}$, respectively, in same way as undertaken in previous Games (for example Games 7, 8, 10 and 11 from the sub-case $\Sigma$, CompromiseSK is not issued at stage $t'$ and CompromiseSS is not issued at stages $u$, $t'\leq u<t$, section \ref{security_case1.2}).

For $\Pi_\Sigma$ and $\Pi_{MAC}$:

\begin{equation*}
    \text{Pr}(break_1) \leq 2\text{Adv}_{\text{AEAD},\mathcal{A}}^\text{IND-CPA}(\kappa) + \text{Pr}(break_5)
\end{equation*}

For $\Pi_{KEM}$:

\begin{equation*}
    \text{Pr}(break_1) \leq 4\text{Adv}_{\text{AEAD},\mathcal{A}}^\text{IND-CPA}(\kappa)+ \text{Pr}(break_5)
\end{equation*}

\paragraph*{Game 6:} The session key $ss_{rest}$ is replaced with a uniformly random and independent value, $\widetilde{ss_{rest}}$, of the same length. The replacement is performed with an unauthenticated QKD challenger that replaces the public traffic m$_{QKD}$ with m$_{QKD}'$, which is used to distill $\widetilde{ss_{rest}}$. Since $\pi^s_i$ matches session $\pi^r_j$ in stage $t'$, the replaced traffic m$_{QKD}'$ is received by both sessions without modification. The detection of this replacement by $\mathcal{A}$ implies having an efficient distinguishing algorithm against the indistinguishability security of the unauthenticated QKD, where the algorithm does not target authentication. This relates to the RevealITS query, in which the conditioned unbounded adversary does not compromise the QKD keys with a small probability $\varepsilon_{QKD}(\kappa)$, for a security parameter $\kappa$:

\begin{equation*}
    \text{Pr}(break_5) \leq \varepsilon_{QKD}(\kappa) + \text{Pr}(break_6)
\end{equation*}

Since the session keys are uniformly random and independent, Pr($break_6$) is equal to zero. Hence, the security analysis of Case 3 yields the inequality \ref{eq:case3-general}.

\begin{eqnarray}
\text{Adv}^{\text{HAKE},\text{clean}_{HPT},\mathcal{A},C_3}_{\Pi,n_P,n_S,n_T}(\kappa) & \leq &n_P^2n_S^2n_{T_\Sigma}   \bigl[2\text{Adv}_{\text{AEAD},\mathcal{A}}^\text{IND-CPA}(\kappa) +\varepsilon_{QKD}(\kappa)\bigr] \nonumber\\
&+& n_P^2n_S^2n_{T_{KEM}}   \bigl[4\text{Adv}_{\text{AEAD},\mathcal{A}}^\text{IND-CPA}(\kappa) +\varepsilon_{QKD}(\kappa)\bigr] \nonumber\\
&+& n_P^2n_S^2n_{T_{MAC}}   \bigl[2\text{Adv}_{\text{AEAD},\mathcal{A}}^\text{IND-CPA}(\kappa) +\varepsilon_{QKD}(\kappa)\bigr] \nonumber\\
&&\,\,\,\,\,\,\,= n_P^2n_S^2n_T   \bigl[2\text{Adv}_{\text{AEAD},\mathcal{A}}^\text{IND-CPA}(\kappa) +\varepsilon_{QKD}(\kappa)\bigr]\nonumber\\
&&\,\,\,\,\,\,\,+n_P^2n_S^2n_{T_{KEM}}   2\text{Adv}_{\text{AEAD},\mathcal{A}}^\text{IND-CPA}(\kappa)
\label{eq:case3-general}
\end{eqnarray}

\section{Discussion}\label{discussion}

The presented protocols in this work are comparable to those presented within the Muckle works. Specifically, the presented signature-based, KEM-based and MAC-based protocols are comparable with the Muckle+~\cite{MucklePlus2023}, Muckle\#~\cite{MuckleKEM2024} and Muckle~\cite{Muckle} works, respectively.

\subsection{Comparison with the Muckle protocols}

The difference of the advantages given by the HAKE key-indistinguishability games for the protocols presented here and those in the Muckle works can be explained through the different cleanness predicates used for this work and the Muckle approaches. Furthermore, this work takes into account the confidentiality aspect, which can be ignored in order to compare this work with the Muckle series. 

There are two main differences between the cleanness predicates presented in this work compared to those defined in the Muckle works. 

First, the cleanness predicates defined in this work regarding the existence, or not, of an origin session $\pi^r_j$ of $\pi^s_i$ (in stage $t$) have more queries that are not allowed to be issued before sessions get to an $accept$ state. This set of prohibited queries covers every possibility regarding authentication. Specifically, the adversary is allowed to obtain all but one of the key materials used to authenticate and the authentication still holds. This difference in the cleanness predicate is the cause of the additional MAC EUF-CMA advantages that this work presents for Cases 1 and 2 - the final key confirmation step is used as a probe to detect fraudulent attempts at authentication. The ephemeral key advantages that the protocols from the Muckle series present are replaced in this work with the advantage given by the ephemeral QKD keys, namely $\varepsilon_{QKD}$. In respect to the MAC-based protocol, the advantage of the shared material is taken into account, which is not the case for the protocol presented within Muckle, where the shared key is assumed to be perfectly secure. And finally, the different number of PRF and dual PRF advantages is natural, since the number of uses of those functions depends on how the protocols are constructed. 

The second main difference is the cleanness predicate regarding the matching session scenario, because the session keys do not depend on the intermediate keys derived in the protocol, nor the authentication material. The QKD keys are shown to be ITS and $\varepsilon_{QKD}$ secure, regardless of the authentication method. This makes sense, since if the adversary is only interacting passively with the QKD traffic (no impersonation nor substitution attacks), QKD is secure within its security parameter. 

The idea of using PQC to authenticate QKD is to only rely on PQC for the very first QKD iteration. Afterwards, authentication can rely on non-ITS MAC, or ITS MAC. The following part provides an example of how to manipulate the advantage inequality formula given in theorem \ref{theorem1}, in the case scenario where KEM-based PQC authentication is performed for stage 1 (the first iteration of QKD) and the non-ITS MAC is used for the subsequent stages (2nd and beyond iterations of QKD). Since the security of the QKD keys decreases as the number of iterations increases, the first distilled QKD keys in iteration 1 may be preferred to perform the MAC-based protocol. Nevertheless, QKD keys distilled in other iterations can be used, as long as the security parameter given by the MAC-based protocol does not surpass an arbitrary threshold. 

\subsection{Example: KEM-based in stage 1, MAC-based for rest}

Let two parties perform one session to derive session keys during $n$ stages, $n_P=2$ and $n_S=1$. They perform the KEM-based protocol in stage 1, afterwards performing the MAC-based protocol in stages 2 to n. Since the signature-based protocol is not applied, $n_{T_\Sigma}$ = 0 at all times.

The security level at stage $n_T=1$ is:

\begin{eqnarray*}
\text{Adv}^{\text{HAKE},\text{clean}_{HPT},\mathcal{A}}_{\Pi,n_P=2,n_S=1,n_T=1}&(\kappa)& = \text{Adv}^{\text{HAKE},\text{clean}_{HPT},\mathcal{A}}_{\Pi,n_P=2,n_S=1,n_{T_{KEM}}=1}(\kappa) \\
&\leq& \, 8  \bigl[\text{Adv}^{\text{IND-CCA}}_{\text{KEM}, \mathcal{A}}(\kappa) + \varepsilon_{QKD}(\kappa) + 2\text{Adv}^{\text{ind}}_{\text{PRF}^\text{dual}, \mathcal{A}}(\kappa)\\
&& + 22\text{Adv}^{\text{ind}}_{\text{PRF}, \mathcal{A}}(\kappa) + 5\text{Adv}_{\text{AEAD},\mathcal{A}}^\text{IND-CPA}(\kappa) \\
&&+5\text{Adv}_{\text{AEAD},\mathcal{A}}^\text{INT-CTXT}(\kappa) + 3\text{Adv}^{\text{EUF-CMA}}_{\text{MAC}, \mathcal{A}}(\kappa)+15 \text{Adv}_{\text{H},\mathcal{A}}(\kappa)\bigr]  \\
&&+4   \bigl[4\text{Adv}_{\text{AEAD},\mathcal{A}}^\text{IND-CPA}(\kappa) +\varepsilon_{QKD}(\kappa)\bigr]\\
\end{eqnarray*}

For the following stages, it is assumed that parties have stored sufficient keys in their shared secret key pool from stage 1 to run the rest of the stages. The MAC-based advantage, for stage 2 and the subsequent stages, takes as input the advantage where the keys were derived, i.e. in stage $n_T=1$ in this case.

The security level at stage $t$, for $2 \leq t$, is:

\begin{eqnarray*}
\text{Adv}^{\text{HAKE},\text{clean}_{HPT},\mathcal{A}}_{\Pi,n_P=2,n_S=1,n_T=t}&(\kappa)& = \text{Adv}^{\text{HAKE},\text{clean}_{HPT},\mathcal{A}}_{\Pi,n_P=2,n_S=1,n_{T_{KEM}}=1}(\kappa) + \text{Adv}^{\text{HAKE},\text{clean}_{HPT},\mathcal{A}}_{\Pi,n_P=2,n_S=1,n_{T_{MAC}}=t-1}(\kappa) \\
&\leq&\,\text{Adv}^{\text{HAKE},\text{clean}_{HPT},\mathcal{A}}_{\Pi,n_P=2,n_S=1,n_T=1}(\kappa) \\
&&+4(t-1)\big[ \text{Adv}^{\text{HAKE},\text{clean}_{HPT},\mathcal{A}}_{\Pi,n_P=2,n_S=1,n_T=1}(\kappa) + \text{Adv}^{\text{ind}}_{\text{PRF}^\text{dual}, \mathcal{A}}(\kappa) \\
&&+3\text{Adv}^{\text{ind}}_{\text{PRF}, \mathcal{A}}(\kappa) + \text{Adv}_{\text{AEAD},\mathcal{A}}^\text{IND-CPA}(\kappa) + \text{Adv}_{\text{AEAD},\mathcal{A}}^\text{INT-CTXT}(\kappa) \\
&& + \text{Adv}^{\text{EUF-CMA}}_{\text{MAC}, \mathcal{A}}(\kappa) \big]\\
&&+4(t-1)^2\big[ \text{Adv}^{\text{HAKE},\text{clean}_{HPT},\mathcal{A}}_{\Pi,n_P=2,n_S=1,n_T=1}(\kappa) + \text{Adv}^{\text{ind}}_{\text{PRF}^\text{dual}, \mathcal{A}}(\kappa) \\
&&+4\text{Adv}^{\text{ind}}_{\text{PRF}, \mathcal{A}}(\kappa) + \text{Adv}_{\text{AEAD},\mathcal{A}}^\text{IND-CPA}(\kappa) + \text{Adv}_{\text{AEAD},\mathcal{A}}^\text{INT-CTXT}(\kappa) \\
&&+ \text{Adv}^{\text{EUF-CMA}}_{\text{MAC}, \mathcal{A}}(\kappa) \big]\\
&&+4(t-1)^3\big[ 5\text{Adv}^{\text{ind}}_{\text{PRF}, \mathcal{A}}(\kappa) + \text{Adv}_{\text{AEAD},\mathcal{A}}^\text{IND-CPA}(\kappa) \\
&& + \text{Adv}_{\text{AEAD},\mathcal{A}}^\text{INT-CTXT}(\kappa) + \text{Adv}^{\text{EUF-CMA}}_{\text{MAC}, \mathcal{A}}(\kappa) \big]\\
&&+4(t   \left[2\text{Adv}_{\text{AEAD},\mathcal{A}}^\text{IND-CPA}(\kappa) +\varepsilon_{QKD}(\kappa)\right]+2\text{Adv}_{\text{AEAD},\mathcal{A}}^\text{IND-CPA}(\kappa))\\
\end{eqnarray*}

Once the corresponding advantage terms are known, the security level at stage $t$, which takes into account confidentiality, integrity and secrecy, is upper bounded by a known term.

\section{Conclusion}\label{conclusion}

The motivating scenario for this work is authentication of QKD, when parties have no initial shared secret material to leverage. A solution for this scenario enables new QKD users to securely join a new (to them) already-secure network remotely. This is clearly a very relevant and important scenario in the modern mobile world.

To address this scenario, two protocols of PQC-based authentication (signature and KEM-based) for QKD and a non-ITS long-term secure MAC-based authentication have been analyzed and proven to be secure, according to defined security parameters and bounded adversaries. The $T_{HPT}$-term security has been defined, along with the HPT and conditioned unbounded adversaries. Given the security framework, honest parties can distill the ITS QKD keys with the underlying security provided by the security analysis. 

After comparison between the protocols presented in this work and those given in the Muckle series~\cite{Muckle, MucklePlus2023, MuckleKEM2024}, a practical example of the advantages to take into account, when the first QKD iteration is authenticated using PQC KEM algorithms and authentication for posterior QKD iterations relays only in non-ITS instead of PQC, is presented. 

Neither the form of QKD, nor the PQC algorithms, are specified in this work, so it is widely applicable across the QKD and PQC spectra. Any PQC algorithms that are secure following the definitions from section \ref{definitions} and \ref{assumptions} are valid PQC algorithms for use. However, using standardized PQC algorithms, such as those from NIST, is identified as best practice~\cite{nist_pqc}. The security proofs presented here also take into account the confidentiality aspect. Additionally, honest parties are free to switch between the three presented protocols. However, it is always possible to instead adopt an ITS authentication protocol starting from stage 2, provided that the parties have derived sufficient key material in the first stage 1 iteration. 

PQC can offer remote and secure authentication to QKD for the very first iteration between new correspondents and can also be used to provide redundant security once a QKD link is established, as shown in the Muckle series. An open question is to determine if there are other methods to combine PQC with QKD to complement each other, or combine securities.

\paragraph*{Acknowledgements} JAVG thanks Christopher Battarbee, Ludovic Perret, Delaram Kahrobaei and Panagiotis Papanastasiou for valuable discussions. JAVG has conducted this work with the support of EPSRC PhD studentship Grant EP/W524657/1. JAVG and TS have conducted this work partially with the support of ONR Grant 62909-24-1-2002.

\printbibliography
\end{document}